\documentclass[aps,twocolumn,showpacs,preprintnumbers,floatfix]{revtex4-1}

\usepackage{amsmath, amssymb, amsfonts, graphicx}
\usepackage[subfigure]{graphfig}
\usepackage{subfigure}
\usepackage{epsfig}
\usepackage{epstopdf}

\usepackage[colorinlistoftodos]{todonotes}
\usepackage[colorlinks=true, allcolors=blue]{hyperref}

\usepackage{color}      

\usepackage{dcolumn}
\usepackage{bm}
\usepackage{braket}
\usepackage{slashed}
\usepackage{enumitem}
\usepackage{bbm}
\usepackage{float}

\def\be{\begin{equation}}
\def\ee{\end{equation}}
\def\bea{\begin{eqnarray}}
\def\eea{\end{eqnarray}}

\def\path{./pdf/}

\begin{document}
\title{Lattice Calculation of Pion Form Factor with Overlap Fermions}

\author{
Gen Wang$^{1}$
}
\email{genwang27@uky.edu}
\author{Jian Liang$^{1,2,3}$}
\author{Terrence Draper$^{1}$}
\author{Keh-Fei Liu$^{1}$}
\author{Yi-Bo Yang$^{4,5,6}$
\vspace*{-0.5cm}
\begin{center}
\large{
\includegraphics[scale=0.2]{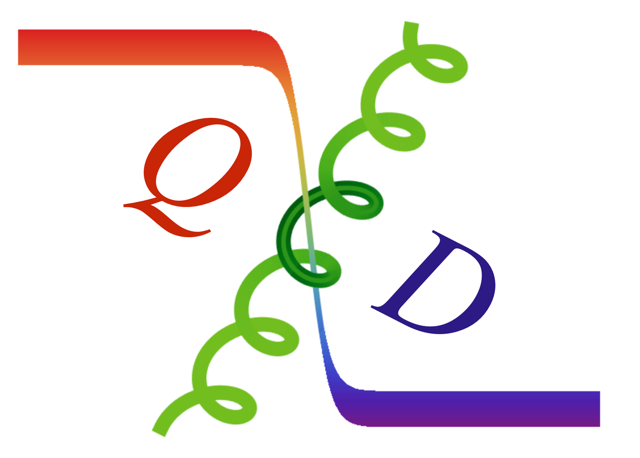}\\
\vspace*{0.0cm}
($\chi$QCD Collaboration)
}
\end{center}
}

\affiliation{
$^{1}$\mbox{Dept. of Physics and Astronomy, University of Kentucky, Lexington, KY 40506, USA}
$^{2}$\mbox{Guangdong Provincial Key Laboratory of Nuclear Science,}\\ \mbox{Institute of Quantum Matter, South China Normal University, Guangzhou 510006, China}
$^{3}${Guangdong-Hong Kong Joint Laboratory of Quantum Matter, }\\ \mbox{Southern Nuclear Science Computing Center, South China Normal University, Guangzhou 510006, China}
$^{4}$\mbox{CAS Key Laboratory of Theoretical Physics, Institute of Theoretical Physics,}\\ \mbox{Chinese Academy of Sciences, Beijing 100190, China} \\
$^{5}$\mbox{School of Fundamental Physics and Mathematical Sciences,}\\ \mbox{Hangzhou Institute for Advanced Study, UCAS, Hangzhou 310024, China}
$^{6}$\mbox{International Centre for Theoretical Physics Asia-Pacific, Beijing/Hangzhou, China}
}

\begin{abstract}
We present a precise calculation of the pion form factor using overlap fermions on seven ensembles of 2+1-flavor domain-wall configurations 
with pion masses varying from 139 to 340 ${\rm{MeV}}$.
Taking advantage of the fast Fourier transform and other techniques to access many combinations of source and sink momenta, we find the pion mean square charge radius to be $\braket{r_\pi^2}= 0.430(5)({13})\ {\rm{fm^2}}$, which agrees well with the experimental result, and includes the systematic uncertainties from chiral extrapolation, lattice spacing and finite-volume dependence.
We also {find} that $\braket{r_\pi^2}$ depends on both the valence and sea quark masses strongly and predict the pion form factor up to $Q^2 = 1.0 \ {\rm{GeV^2}}$ which agrees with experiments very well.
\end{abstract}

\maketitle

{
\section{Introduction}\label{sec:intro}
{The space-like pion electric form factor $f_{\pi\pi}(Q^2)$ is defined from the pionic matrix element and its slope at
$Q^2 =0$ gives the mean square charge radius}
\begin{eqnarray}
&&\langle \pi^i (p')| V^j_{\mu} (0) | \pi^k (p)\rangle = i \epsilon^{ijk} (p_{\mu} + p'_{\mu}) f_{\pi\pi}(Q^2), \\
&&\langle r_{\pi}^2 \rangle \equiv -6 \frac{d\,\! f_{\pi \pi} (Q^2)}{d Q^2}|_{Q^2 = 0}, \label{eq:radius}
\end{eqnarray}
where $V^j_{\mu} = \bar{\psi} \frac{1}{2}\tau^j \gamma_{\mu} \psi$ is the isovector vector current, $\tau^i$ are the Pauli matrices in flavor space, and $|\pi^i \rangle$ are the pion triplet states.
$\langle r_{\pi}^2 \rangle$ has been determined precisely based on the existing $\pi e$ scattering data~\cite{Dally:1982zk,Amendolia:1986wj,GoughEschrich:2001ji} and \mbox{$e^+ e^- \rightarrow \pi^+ \pi^-$} data~\cite{Ananthanarayan:2017efc,Colangelo:2018mtw} averaged by the Particle Data Group (PDG)~\cite{Tanabashi:2018oca} as $\langle r^2_{\pi} \rangle =0.434(5) \ {\rm{fm^2}}$.
Phenomenologically, $f_{\pi\pi}(Q^2)$ is fitted quite well over the range $0 < Q^2/m_{\rho}^2<0.4$ with the single monopole form $(1 + Q^2/\Lambda^2)^{-1}$, with $\Lambda \sim m_{\rho}$. This gives credence to the idea of vector dominance~\cite{Frazer:1960zzb,Holladay:1956zz}. In chiral perturbation theory,  $\langle r_{\pi}^2 \rangle$ has been calculated with $SU$(2) Chiral Perturbation Theory~\cite{Gasser:1984ux} at NNLO and also at NLO with $SU$(3) formula~\cite{Bijnens:1998fm}, 
{which entails the uncertainties of the low-energy constants.}

Since lattice QCD is an {\it ab initio} calculation and the experimental determination of $\langle r_{\pi}^2 \rangle$ from the $\pi e$ scattering is very precise,
it provides a stringent test for lattice QCD calculations to demonstrate complete control over the statistical and systematic errors in estimates of the relevant pionic matrix element
in order to enhance confidence in their reliability to calculate other hadronic matrix elements where further technical complications  occur.
Over the years, the pion form factor has been calculated with the quenched approximation~\cite{Martinelli:1987bh,Draper:1988bp}, and for the $N_{\textrm{f}} =2$~\cite{Brommel:2006ww,Frezzotti:2008dr,Aoki:2009qn,Brandt:2013dua,Alexandrou:2017blh}, $N_{\textrm{f}} = 2+1$~\cite{Bonnet:2004fr,Boyle:2008yd,Nguyen:2011ek,Fukaya:2014jka,Aoki:2015pba,Feng:2019geu} and $N_{\textrm{f}} = 2+1+1$~\cite{Koponen:2015tkr} cases.

In this work, we use valence overlap fermions to calculate the pion form factor
on seven ensembles of domain-wall fermion configurations with different sea pion masses, including three at the physical pion mass, four lattice spacings and different volumes to control the systematic errors.
Due to the multi-mass algorithm available for overlap fermions, we can effectively calculate several valence quark masses on each ensemble~\cite{Li:2010pw,Yang:2018nqn,Sufian:2016pex,Yang:2016plb} and also {\cal O}(100) combinations of the initial and final pion momenta with little overhead with the use of the fast Fourier transform (FFT) algorithm~\cite{cooleyAlgorithmMachineCalculation1965} in the three-point function contraction.
This allows us to study both the sea and the valence quark mass dependence of $\langle r^2_\pi \rangle$ in terms of partially quenched chiral perturbative theory, besides giving an accurate result at the physical pion mass. 
{This work is based on Ref.~\cite{Wang:2020yqv} with more statistics on the ensembles at the physical pion mass.}

The paper is organized as follows:
In Section \ref{sec:numeric}, we present the numerical details of this calculation and a brief description of the FFT on stochastic-sandwich method.
Fits and extrapolations are discussed in Section \ref{sec:analysis} with results compared with other studies.
A brief summary is given in Sec.~\ref{sec:summary}.

\begin{table}[!htb]
  \centering{
  \begin{tabular}{| c | c | c | c | c | c | c |  }
    \hline
    Lattice & $L^3\times T$ & $a\ ({\rm{fm}})$ &  $La\ ({\rm{fm}})$ & $m_\pi ({\rm{MeV}})$ & $m_\pi L$ & $n_{\rm cfg}$ \\
    \hline
    24IDc     & $24^3\times 64$ & $0.195$ & $4.66 $ & $141$ & $3.33$  & $231$ \\
    \hline
    32IDc     & $32^3\times 64$ & $0.195$ & $6.24 $ & $141$ & $4.45$  & $ 53$ \\
    \hline
    32ID      & $32^3\times 64$ & $0.143$ & $4.58 $ & $172$ & $3.99$  & $199$ \\
    \hline
    32IDh     & $32^3\times 64$ & $0.143$ & $4.58 $ & $250$ & $5.80$  & $100$ \\
    \hline
    48I       & $48^3\times 96$ & $0.114$ & $5.48 $ & $139$ & $3.86$  & $158$ \\
    \hline
    24I       & $24^3\times 64$ & $0.111 $ & $2.65 $ & $340$ & $4.56$ & $202$ \\
    \hline
    32I       & $32^3\times 64$ & $0.083$ & $2.65 $ & $302$ & $4.05$  & $309$ \\
    \hline
  \end{tabular}
  \caption{The ensembles and their respective lattice size $L^3\times T$, lattice spacing $a$, pion mass $m_{\pi}$ and number of configurations $n_{\rm cfg}$.}
  \label{tab:lattice_para0}
  }
\end{table}
}

{
\section{Numerical details}\label{sec:numeric}
We use overlap fermions on seven ensembles of HYP smeared 2+1-flavor domain-wall fermion configurations with Iwasaki gauge action (labeled with I)~\cite{Aoki:2010dy,Blum:2014tka} and Iwasaki with Dislocation Suppressing Determinant Ratio (DSDR) gauge action (labeled with ID)~\cite{Arthur:2012opa,Boyle:2015exm} as listed in Table~\ref{tab:lattice_para0}. {The effective quark propagator of the massive
overlap fermions is the inverse of the operator $(D_c + m)$~\cite{Chiu:1998eu,Liu:2002qu}, where $D_c$ is chiral, i.e., $\{D_c, \gamma_5\} = 0$~\cite{Chiu:1998gp}}. 
It can be expressed in terms of the overlap Dirac operator $D_{ov}$ as $D_c = \rho D_{ov}/(1 - D_{ov}/2)$, with $\rho = -(1/(2 \kappa) - 4)$ and $\kappa = 0.2$.
A multi-mass inverter is used to calculate the propagators with 2 to 6 valence pion masses varying from the unitary point to {$\sim$ 390 ${\rm{MeV}}$}.
On 24I, 32I and 24IDc (c stands for coarse lattice spacing), Gaussian smearing~\cite{DeGrand:1990dz} is applied with root mean square (RMS) radii 0.49 ${\rm{fm}}$, 0.49 ${\rm{fm}}$ and 0.53 ${\rm{fm}}$, respectively, {for both source and sink}.
On 48I, 32ID and 32IDh (h for heavier pion mass), box-smearing~\cite{Allton:1990qg,Liang:2016fgy} with box half sizes 0.57 ${\rm{fm}}$, 1.0 ${\rm{fm}}$ and 1.0 ${\rm{fm}}$, respectively, is applied as an economical substitute for Gaussian smearing.

To extract pionic matrix elements, the three-point function (3pt) $C_{\rm 3pt}(\tau,t_{\textrm{f}},\vec{p}_{\textrm{i}},\vec{p}_{\textrm{f}})$ is computed,
\begin{eqnarray}\label{eq:3pt_S0}
\begin{aligned}
&C_ {\rm 3pt}=
 \sum_{\vec{x}_{\textrm{f}},\vec{z}} e^{-i \vec{p}_{\textrm{f}} \cdot \vec{x}_{\textrm{f}}} e^{ i \vec{q} \cdot \vec{z}}
 \braket{{\rm{T}}[\chi_{\pi^+}{(x_{\textrm{f}}}) V_4^3({z}) \chi_{\pi^+}^{\dagger}({\mathcal{G}})] }\\
=& \sum_{\vec{x}_{\textrm{f}},\vec{z}} e^{-i \vec{p}_{\textrm{f}} \cdot \vec{x}_{\textrm{f}}} e^{ i \vec{q} \cdot \vec{z}}\braket{{\rm{Tr}}\big[\gamma_5 S({\mathcal{G}}|{z}) \gamma_4 S({z}|{x_{\textrm{f}}}) \gamma_5 S({x_{\textrm{f}}}|{\mathcal{G}})\big]}, \\
\end{aligned}
\end{eqnarray} 
where $\chi_{\pi^+}(\vec{x},t) = \bar{d} (\vec{x},t) \gamma_{5} u (\vec{x},t)$ is the interpolating field of the pion
 with $u$ and $d$ the up and down quark spinors,
 $S({y}|{x})$ is the quark propagator from ${x}$ to ${y}$, ${z}\equiv \{\tau,\vec{z}\}$, ${x_{\textrm{f}}}\equiv \{t_{\textrm{f}},\vec{x}_{\textrm{f}}\}$, 
 $\vec{p}_{\textrm{i}}$ and $\vec{p}_{\textrm{f}}$ are the initial and final momenta of the pion, respectively,
 $\vec{q} = \vec{p}_{\textrm{f}} - \vec{p}_{\textrm{i}}$ is the momentum transfer,
and $\mathcal{G}$ is the smeared $Z_3$-noise grid source~\cite{Dong:1993pk}. The disconnected insertions in~Eq.(\ref{eq:3pt_S0}) vanish in the ensemble average~\cite{Draper:1988bp}.

In practice, $S(\mathcal{G}|z)$ in Eq.~(\ref{eq:3pt_S0}) is calculated using $\gamma_5$ hermiticity, i.e., $S(\mathcal{G}|z) = \gamma_5 S^\dagger({z}|{\mathcal{G}}) \gamma_5$, and $S({z}|{x_{\textrm{f}}})$ is usually obtained in the sequential source method with 
$\gamma_5 S({x_{\textrm{f}}}|{\mathcal{G}})$ as the source~\cite{Bernard:1985tm,Martinelli:1988rr}. The calculation of the sequential propagators would need to be repeated for different $\vec{p}_{\textrm{f}}$ and different quark mass $m$,
so that the cost would be very high when dozens of momenta and multiple quark masses are calculated.
Instead, we use the stochastic-sandwich method~\cite{Yang:2015zja, Liang:2018pis}, but without low-mode substitution (LMS) for $S({x_{\textrm{f}}}|{\mathcal{G}})$ since it is not efficient for pseudoscalar mesons~\cite{Li:2010pw}.
{However, the separation of sink position $x_{\rm f}$ and current position $z$ in splitting the low and high modes for the propagator $S({z}|{x_{\textrm{f}}})$ between the current and sink can facilitate FFT along with LMS which is still useful here.}
More specifically, the propagator from the sink at $x_{\textrm{f}}$ to the current at $z$,
$S({z}|{x_{\textrm{f}}})$, can be split into the exact low-mode part based on the low lying overlap eigenvalues $\lambda_i$ and eigenvectors $v_i$ of the {$i$th} eigenmode of $D_c$, plus the noise-source estimate $S^{H}_{\rm noi}$ of the high-mode part,
\begin{eqnarray}\label{eq:S_LH}
\begin{aligned}
S({z}|{x_{\textrm{f}}})   &= S^L({z}|{x_{\textrm{f}}}) + S^H({z}|{x_{\textrm{f}}}) , \\    
S^L({z}|{x_{\textrm{f}}}) &=\sum_{\lambda_i \leq \lambda_c} \frac{1}{\lambda_i +m} v_i({z}) v_i^{\dagger}({x_{\textrm{f}}}),    \\      
S^H({z}|{x_{\textrm{f}}}) &= \frac{1}{n_{\textrm{f}}} \sum_{j=1}^{n_{\textrm{f}}} S^{H}_{\rm noi}({z},\eta_j)\eta_j^\dagger({x_{\textrm{f}}}),
\end{aligned}
\end{eqnarray} 
where $\lambda_c$ is the highest eigenvalue in LMS and is much larger than the quark mass $m$ with the typical number of eigenmodes $n_{\rm v} \sim 400$ on 24I and 32I, and $n_{\rm v} \sim 1800$ on 32ID, 32IDh, 24IDc, 32IDc and 48I;
and $S^{H}_{\rm noi}({z},\eta_j)$ is the noise-estimated propagator for the high modes with the low-mode deflated $Z_3$ noise 
$\eta_j({x_{\textrm{f}}})$~\cite{Yang:2015zja, Liang:2018pis}.
Sink smearing is applied on all the sink spatial points $x_{\textrm{f}}$ of noise $\eta_j({x_{\textrm{f}}})$ and eigenvectors $v_i^\dagger(x_f)$.

Thus $C_{\rm 3pt}$ can be decomposed into factorized forms within the sums of the eigenmodes for the low modes and the $n_{\rm f}$ number of noises $\eta_j$ for the high modes,
\begin{eqnarray}
&C_{\rm 3pt}(\tau,t_{\textrm{f}},\vec{p}_{\textrm{i}},\vec{p}_{\textrm{f}}) =\langle \sum_{\lambda_i \leq \lambda_c}  {\rm{Tr}} [\frac{1}{\lambda_i + m}  G^L_i(\vec{q},\tau) F^L_i(\vec{p}_{\textrm{f}},t_{\textrm{f}}) ]\nonumber\\
&\quad +\sum_{j=1}^{n_{\textrm{f}}} \frac{1}{n_{\textrm{f}}} {\rm{Tr}}[G_j^H(\vec{q},\tau) F_j^H(\vec{p}_{\textrm{f}},t_{\textrm{f}})]\rangle, 
\end{eqnarray} 
where 
\begin{eqnarray}
G^L_i(\vec{q},\tau)&=&\sum_{\vec{z}} e^{ i \vec{q} \cdot \vec{z}} \gamma_5 S({\mathcal{G}}|{z}) \gamma_4  v_i({z}),\\
F^L_i(\vec{p}_{\textrm{f}},t_{\textrm{f}})&=& \sum_{\vec{x}_{\textrm{f}}}  e^{-i \vec{p}_{\textrm{f}} \cdot \vec{x}_{\textrm{f}}} v_i^{\dagger}({x_{\textrm{f}}}) \gamma_5 S({x_{\textrm{f}}}|{\mathcal{G}}),\\
G^H_j(\vec{q},\tau)&=&\sum_{\vec{z}} e^{ i \vec{q} \cdot \vec{z}} \gamma_5 S({\mathcal{G}}|{z}) \gamma_4 S_{\rm noi}^H({z},\eta_j),\\
F^H_j(\vec{p}_{\textrm{f}},t_{\textrm{f}})&=& \sum_{\vec{x}_{\textrm{f}}} e^{-i \vec{p}_{\textrm{f}} \cdot \vec{x}_{\textrm{f}}} \eta_j^\dagger({x_{\textrm{f}}}) \gamma_5 S({x_{\textrm{f}}}|{\mathcal{G}}),
\end{eqnarray} 
which are calculated by using FFTs on the spatial points $\vec{z}$ and $\vec{x}_{\rm f}$ for each of $G^L_i$, $F^L_i$, $G^H_j$ and $F^H_j$ to obtain any $\vec{q}$ and $\vec{p}_{\textrm{f}}$ with the computational complexity $\mathcal{O}(V {\rm{log}}V)$, with $V$ the lattice spatial volume.
Compared with the stochastic-sandwich method for a fixed $\vec{p}_{\textrm{f}}$
which also includes the summation over the spatial points $\vec{z}$ and $\vec{x}_{\rm f}$, eigenvectors $v_i$ and noises $\eta_j$, the additional cost factor of using FFTs, {namely} $\mathcal{O}({\rm{log}}V)$, is only of order $\sim 7$ for our largest 48I lattice.
This allows us to calculate any combination of $\vec{q}$ and $\vec{p_{\textrm{f}}}$ without much additional cost compared to the traditional stochastic-sandwich method; this is of order $\sim 10$ times less expensive if we calculate more than seven different sink momenta $\vec{p_{{f}}}$ and average over different directions.
{In practice, since larger $\vec{p_{\textrm{i}}}$ or $\vec{p_{\textrm{f}}}$ will lead to worse signals,
we choose three cases
so that for a given $Q^2$ we use as small $\vec{p}_{\textrm{f}}$ and $\vec{p_{\textrm{i}}}$ as possible:
(1) $\vec{p}_{\rm i}=0$ with $\vec{q} = \vec{p}_{\rm f}$ or $\vec{p}_{\rm f}=0$ with $\vec{q} = - \vec{p}_{\rm i}$ which probes small $Q^2$;
(2) $\vec{p}_{\rm f} = -\vec{p}_{\rm i}$ with $\vec{q} = 2 \vec{p}_{\rm f}$  which probes reasonably high $Q^2$;
(3) For a given $\vec{q}$, we calculate $\vec{q}/2$ and choose lattice momenta $\vec{p}_{\rm f}$ and $-\vec{p}_{\rm i}$ which are close to $\vec{q}/2$. This can also probe high $Q^2$ which fills $Q^2$ between the previous two cases.
}

{
We use the lattice dispersion relation $\hat{E}^2 = \hat{m}^2 + \sum_i {\hat{p}}_i^2$ with $a \hat{E} = 2 {\rm{sinh}}(a E /2)$, $a \hat{m} = 2 {\rm{sinh}}(a m /2)$ and $a \hat{p}_i = 2 {\rm{sin}}(ap_i/2)$
to define $Q^2$ for all ensembles so that there is a well defined physical limit.
This is also used in Ref.\cite{Feng:2019geu} to calculate the pion charge radius.
More details about checking the dispersion relation are included in Appendix~\ref{Appendix:dispersion}. }

\begin{table}[htbp]
  \centering{
  \begin{tabular}{| c | c | c | c | c | c | c | c |}
    \hline
    Lattice & $n_{\textrm{i}}$ & $n_t$ & $n_s$& $t_{\textrm{f}}/a$ &  $n_{\textrm{f}}$ &$n_{\rm{meas}}$  \\
    \hline
    24IDc    & $32$ & $1$  & $3$ & $6,7,8,9,10$ & $4,4,6,4,4$ & 199584 \\
    \hline
    32IDc    & $16$ & $2$  & $4$ & $6,7,8,9,10$ & $4,4,4,4,4$ & 108544 \\
    \hline
    32ID     & $6$  & $2$  & $2$ & $9,10,11$ & $4,5,12$ &19104         \\
    \hline
    32IDh    & $6$  & $2$  & $2$ & $9,10,11$ & $4,5,12$ &\  9600       \\
    \hline
    48I      & $16$ & $3$  & $4$ & $8,10,12,14$ & $4,4,4,4$ & 485376   \\
    \hline
    24I      & $8$  & $1$  & $2$ & $10,11,12$ & $3,5,5$  &  12928      \\
    \hline
    32I      & $8$  & $1$  & $2$ & $8,12,15$ & $4,8,12$ & 19776        \\
    \hline
  \end{tabular}
  \caption{The lattice setup of this calculation. The $n_{\textrm{i}}$ sets of smeared noise-grid sources with $\{n_s,n_s,n_s,n_t\}$ points in $\{x,y,z,t\}$ directions, respectively, are placed on the lattice to improve the statistics, together with $n_{\textrm{f}}$ sets of $S^H_{\rm noi}$ at {$2 n_t$ sink time slices at $i \frac{T}{n_t} t_{\textrm{f}}$ and $T - i \frac{T}{n_t} t_{\textrm{f}}$ with $i=\{1 \cdots n_t \}$}. On a given configuration, the total number of the propagators we generated is $n_{\textrm{i}}+n_{\textrm{f}}$ and $n_{\rm{meas}}=n_{\textrm{i}} n_s^3 n_t * n_{\rm{cfg}}$ is the number of measurements of 3pt.}
  \label{tab:lattice_para1}
  }
\end{table}
}

{
\section{Analysis and results}\label{sec:analysis}

{
\subsection{Three-point function fit}
{The source-sink separations $t_{\textrm{f}}$ used in this work with different ensembles are collected in Table~\ref{tab:lattice_para1}.
The largest $t_{\textrm{f}}$ is $\sim 2.0$ fm on the coarsest lattice 24IDc and the smallest one is $\sim 0.7$ fm on the finest lattice 32I.}
 
{With the use of Wick contractions and gauge invariance, the three-point function (3pt) with two coherent sources (we have put a source at each of $t=0$ and $t = T/2$ for most ensembles to increase statistics) has contributions from the three diagrams shown in Fig.~\ref{fig:dia_use}. (We assume $T/2 > t_{\textrm{f}}> \tau > 0$.) The diagram \ref{fig:dia_use}.(1) contributes}
\begin{eqnarray} \label{eq:3pt_1}
\begin{aligned}
C_{{\rm 3pt},(1)}&(\tau,t_{\textrm{f}},\vec{p}_{\textrm{i}},\vec{p}_{\textrm{f}}) \\
&= \frac{Z_{\vec{p}_{\textrm{i}}} Z_{\vec{p}_{\textrm{f}}} (E_{\textrm{i}}+E_{\textrm{f}})}{E_{\textrm{i}} E_{\textrm{f}} Z_V}  f_{\pi\pi} (Q^2)  (e^{-E_{\textrm{i}} \tau -E_{\textrm{f}}(t_{\textrm{f}}-\tau)} )  \\
&\quad + C_1 e^{-E_{\textrm{i}} \tau - E^1_{\textrm{f}} (t_{\textrm{f}} - \tau)} + C_2 e^{-E^1_{\textrm{i}} \tau -E_{\textrm{f}} (t_{\textrm{f}} - \tau)}  \\
&\quad + C_3 e^{-E^1_{\textrm{i}} \tau -E^1_{\textrm{f}} (t_{\textrm{f}} - \tau)}, \\
\end{aligned}
\end{eqnarray}
which includes the first excited-state contamination,
where $Z_{\vec{p}}$ is the spectral weight and $E$ and $E^1$ are the ground state and first excited state energies, respectively.
$Z_{\vec{p}_{\textrm{i}}}, Z_{\vec{p}_{\textrm{f}}}$, $E_{\textrm{i}}$, $E_{\textrm{f}}$, $E_{\textrm{i}}^1$ and $E_{\textrm{f}}^1$ 
are constrained by the joint fit with the corresponding two-point function (2pt).
$Z_V$ is the finite normalization constant for the local vector current and is determined from
the forward matrix element as $Z_V \equiv \frac{2 E}{\bra{\pi(p)} V_4 \ket{\pi(p)}}$.
$C_1, C_2$ and $C_3$ are free parameters for the excited-state contamination.
The diagram \ref{fig:dia_use}.(2) contributes
\begin{eqnarray} \label{eq:3pt_2}
&C_{{\rm 3pt},(2)}(\tau,t_{\textrm{f}},\vec{p}_{\textrm{i}},\vec{p}_{\textrm{f}}) = \frac{Z_{\vec{p}_{\textrm{i}}} Z_{\vec{p}_{\textrm{f}}} (E_{\textrm{i}}+E_{\textrm{f}})}{E_{\textrm{i}} E_{\textrm{f}} Z_V}  f_{\pi\pi} (Q^2) \\
& \quad \quad \quad \times ( e^{-E_{\textrm{i}} (T/2+\tau) -E_{\textrm{f}}(t_{\textrm{f}}-\tau)}  ),
\end{eqnarray}
in which we have ignored the excited-state contamination from the source at $T/2$ since such terms are suppressed by $e^{-E_i^1 T/2}$ which is of order $\sim 10^{-8}$ with $E_i^1 \approx 1.3 \ {\rm GeV}$ estimated with the experimental value of the first excited state of the pion,
and the diagram \ref{fig:dia_use}.(3) contributes
\begin{eqnarray} \label{eq:3pt_3}
C_{{\rm 3pt},(3)}(\tau,t_{\textrm{f}},\vec{p}_{\textrm{i}},\vec{p}_{\textrm{f}}) = C_4  e^{-E_{\textrm{i}} (T/2 -t_{\textrm{f}}) -E_{\rm h} (t_{\textrm{f}}- \tau)} ,
\end{eqnarray}
in which this term corresponds to the creation of a hadron state with operator $V_4 = \bar{q} \gamma_4 q$ at time slice $\tau$ with momentum $q$ as $\bra{h(q)} V_4 \ket{0}$,
an annihilation of a pion state at time slice $T/2$ with momentum $p_{\textrm{i}}$ as $\bra{0} \chi_{\pi^+}^{\dagger} \ket{\pi^-(p_{\textrm{i}})}$
and an unknown matrix element $\bra{\pi^-(p_{\textrm{i}})} \chi_{\pi^+} \ket{h(q)}$. 
The excited-state contamination from $E_i^1$ is ignored for the same reason as in the previous discussion and the excited-state contamination from $E_h^1$ is ignored under current statistics.

\begin{figure*}[htbp]
\centering
\includegraphics[scale=0.45]{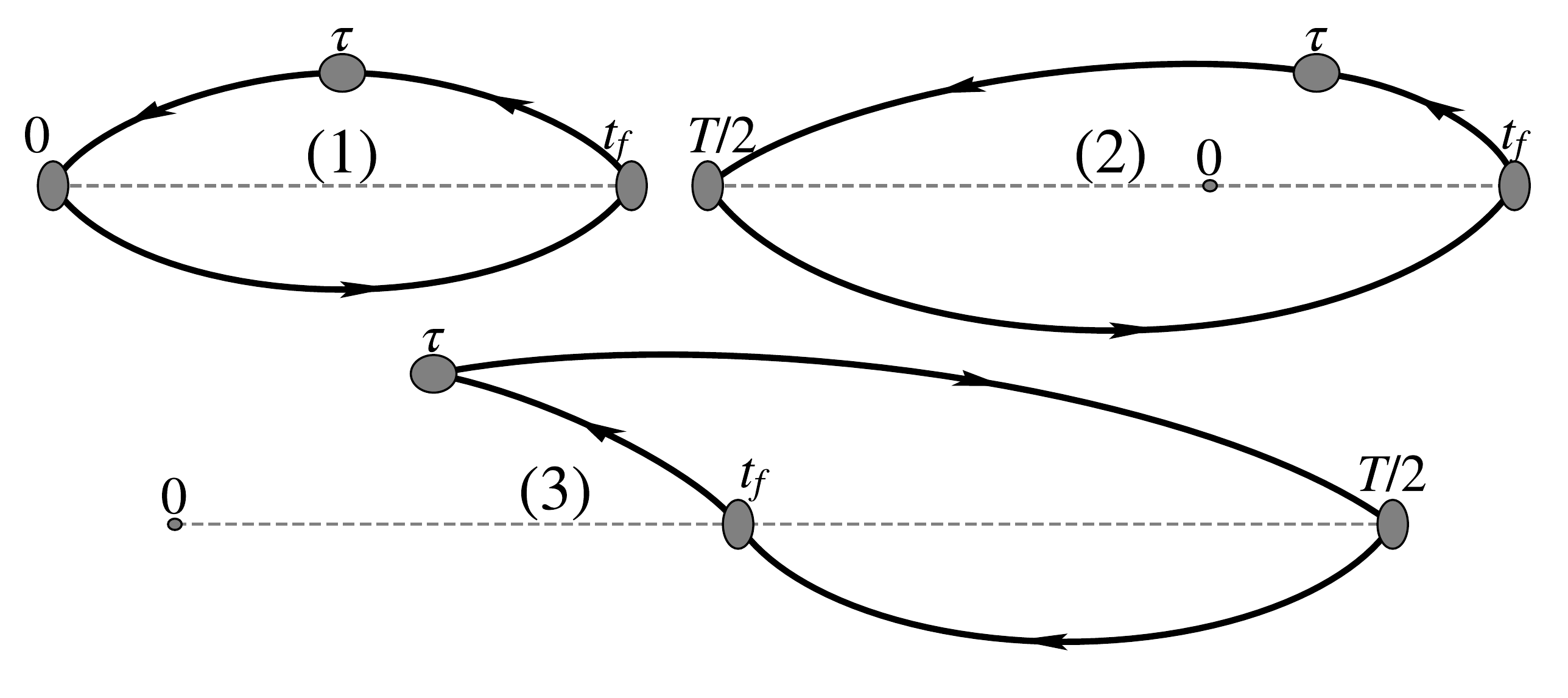} \\
\caption{Diagrams of pion three-point functions with sources at time slices $0$ and $T/2$.}
\label{fig:dia_use}
\end{figure*}

In order to test the functional form of $C_{{\rm 3pt},(3)}(\tau,t_{\textrm{f}},\vec{p}_{\textrm{i}},\vec{p}_{\textrm{f}})$, we construct 3pt with one source at time slice $T/2 = 32$ and sink time $t_{\textrm{f}}$ at $20,21,22$ with $\vec{p}_{\textrm{i}}=\{0,0,0\}$ and $\vec{p}_{\textrm{f}}=\{0,0,\frac{2 \pi}{L} \}$. Then we can evaluate the effective mass $E^{\textrm{eff}}_{\rm h}$ and $E^{\textrm{eff}}_{\textrm{i}}$ from $C_{{\rm 3pt},(3)}(\tau,t_{\textrm{f}},\vec{p}_{\textrm{i}},\vec{p}_{\textrm{f}})$ with 
\begin{eqnarray}\label{eq:CEi}
\begin{aligned}
&E^{\textrm{eff}}_{\rm h}(\tau,t_{\textrm{f}}) = {\rm{ln}}\left(\frac{C_{{\rm 3pt},(3)}(\tau+1,t_{\textrm{f}},\vec{p}_{\textrm{i}},\vec{p}_{\textrm{f}})}{C_{{\rm 3pt},(3)}(\tau,t_{\textrm{f}},\vec{p}_{\textrm{i}},\vec{p}_{\textrm{f}})}\right), \\
&E^{\textrm{eff}}_{\textrm{i}}(\tau,t_{\textrm{f}}) = {\rm{ln}}\left(\frac{C_{{\rm 3pt},(3)}(\tau+1,t_{\textrm{f}},\vec{p}_{\textrm{i}},\vec{p}_{\textrm{f}})}{C_{{\rm 3pt},(3)}(\tau,t_{\textrm{f}}-1,\vec{p}_{\textrm{i}},\vec{p}_{\textrm{f}})}\right),
\end{aligned}
\end{eqnarray} 
in which $E_{\rm i}^{\rm eff}$ is evaluated by a simultaneous change of $\tau$ and $t_{\rm f}$ to single out $E_{\rm i}$ from the exponential $e^{-E_{\textrm{i}} (T/2 -t_{\textrm{f}}) -E_{\rm h} (t_{\textrm{f}}- \tau)}$.
They should equal to $E_{h}=\sqrt{m^2_{h}+(\vec{p}_{\textrm{f}}-\vec{p}_{\textrm{i}})^2}$ and $E_{\rm i}=\sqrt{m^2_{\pi}+\vec{p}_{\textrm{i}}^2}=m_{\pi}$ in the $t_{\textrm{f}} \gg \tau$ limit, as confirmed in Fig.~\ref{fig:dia_corr} and the fit results in Fig.~\ref{fig:dia_fit}. 

\begin{figure*}[!htb]
\centering
\includegraphics[scale=0.38]{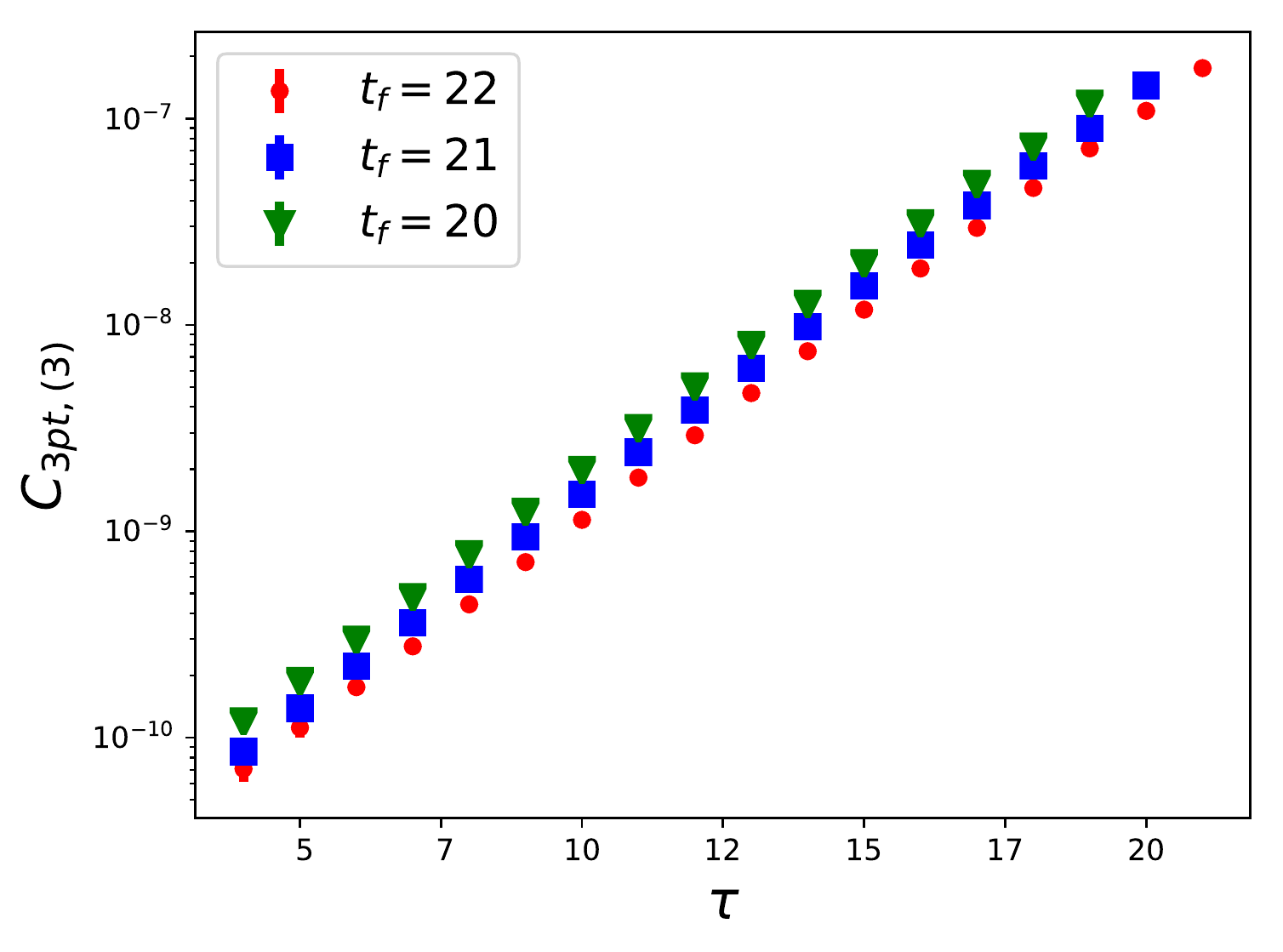}
\includegraphics[scale=0.38]{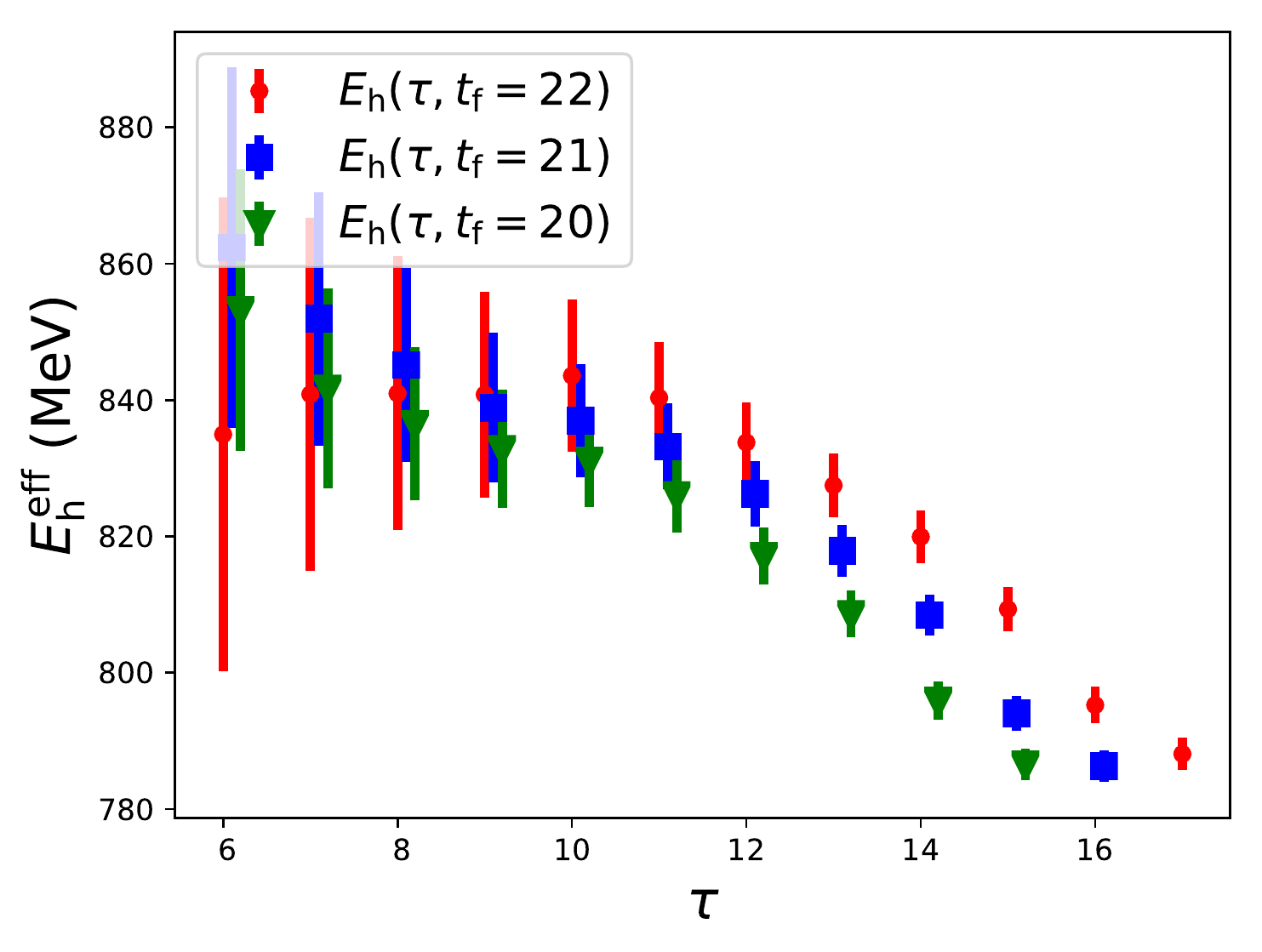}
\includegraphics[scale=0.38]{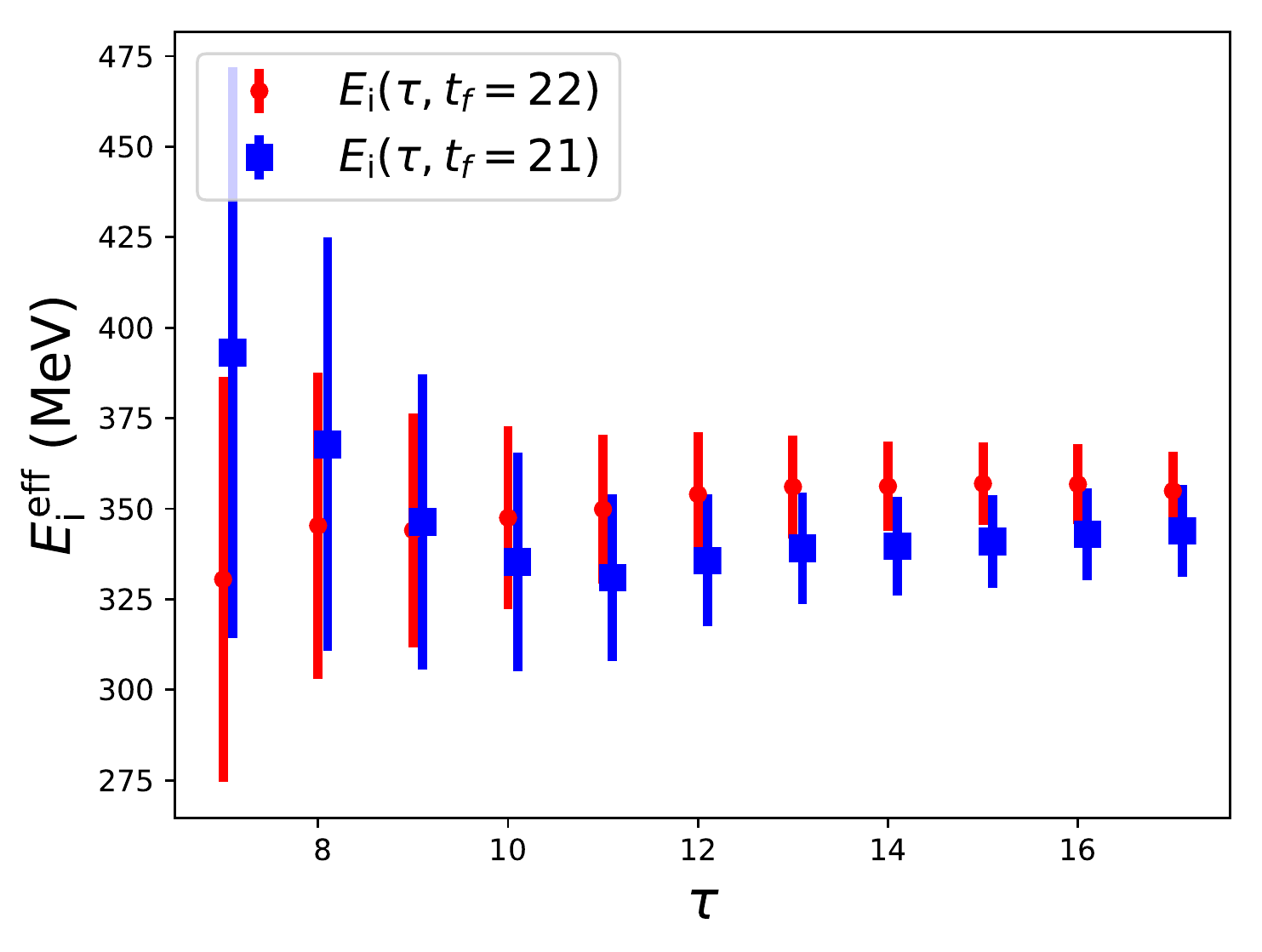}
\caption{The plot on the left is of $C_{{\rm 3pt},(3)}$ on 24I with $m_\pi = 347\ {\rm{MeV}}$, one source at time slice $T/2$, $\vec{p}_{\textrm{i}}=\{0,0,0\}$ and $\vec{p}_{\textrm{f}}=\{0,0,\frac{2 \pi}{L} \}$. The correlation function is a rising exponential which confirms that $E_h > 0$ in Eq.~(\ref{eq:3pt_3}).
The plots in the middle and right panels show the corresponding effective masses $E^{\textrm{eff}}_{\rm h}$ and $E^{\textrm{eff}}_{\textrm{i}}$, respectively, obtained with Eq.~(\ref{eq:CEi}).}
\label{fig:dia_corr}
\end{figure*}

Thus the final functional form is $C_{\rm 3pt}=C_{{\rm 3pt},(1)} + C_{{\rm 3pt},(2)} + C_{{\rm 3pt},(3)}$ as 
\begin{eqnarray} \label{eq:3pt}
\begin{aligned}
&C_{\rm 3pt}(\tau,t_{\textrm{f}},\vec{p}_{\textrm{i}},\vec{p}_{\textrm{f}}) 
= \frac{Z_{\vec{p}_{\textrm{i}}} Z_{\vec{p}_{\textrm{f}}} (E_{\textrm{i}}+E_{\textrm{f}})}{E_{\textrm{i}} E_{\textrm{f}} Z_V}  f_{\pi\pi} (Q^2)   \\
&\times (e^{-E_{\textrm{i}} \tau -E_{\textrm{f}}(t_{\textrm{f}}-\tau)} + e^{-E_{\textrm{i}} (T/2+\tau) -E_{\textrm{f}}(t_{\textrm{f}}-\tau)} ) \\
&\quad+ C_1 e^{-E_{\textrm{i}} \tau - E^1_{\textrm{f}} (t_{\textrm{f}} - \tau)} + C_2 e^{-E^1_{\textrm{i}} \tau -E_{\textrm{f}} (t_{\textrm{f}} - \tau)}  \\
&\quad + C_3 e^{-E^1_{\textrm{i}} \tau -E^1_{\textrm{f}} (t_{\textrm{f}} - \tau)} + C_4 e^{-E_{\textrm{i}} (T/2 -t_{\textrm{f}}) - E_{\rm h} (t_{\textrm{f}}- \tau)}.
\end{aligned}
\end{eqnarray} 
The associated 2pt is fitted with
\begin{eqnarray} \label{eq:fit_2pt}
\begin{aligned}
C_{2pt}(t,\vec{p}) =&  \frac{Z_{\vec{p}}^2}{E} (e^{-E t}+e^{-E (T-t)}) (1+ e^{-E (T/2-t)}) \\
& +   A_1 (e^{-E^1 t} + e^{-E^1 (T/2-t)} ) , \\
\end{aligned}
\end{eqnarray}
with $A_1$ being a free parameter for the excited-state contributions and the exponential terms with $T/2$ account for contributions from the source at $T/2$.
An example of fitted energies is shown in Fig.~\ref{fig:32ID_2pt_fit}. It can be seen that the first excited state energy $E^1$ is close to the experimental value $1.3 \ {\rm GeV}$ and it has been used to constrain that of the 3pt by the joint fit of 2pt and 3pt to extract $f_{\pi \pi}(Q^2)$.

\begin{figure}[htbp]
  \centering
   \includegraphics[width=0.40\textwidth]{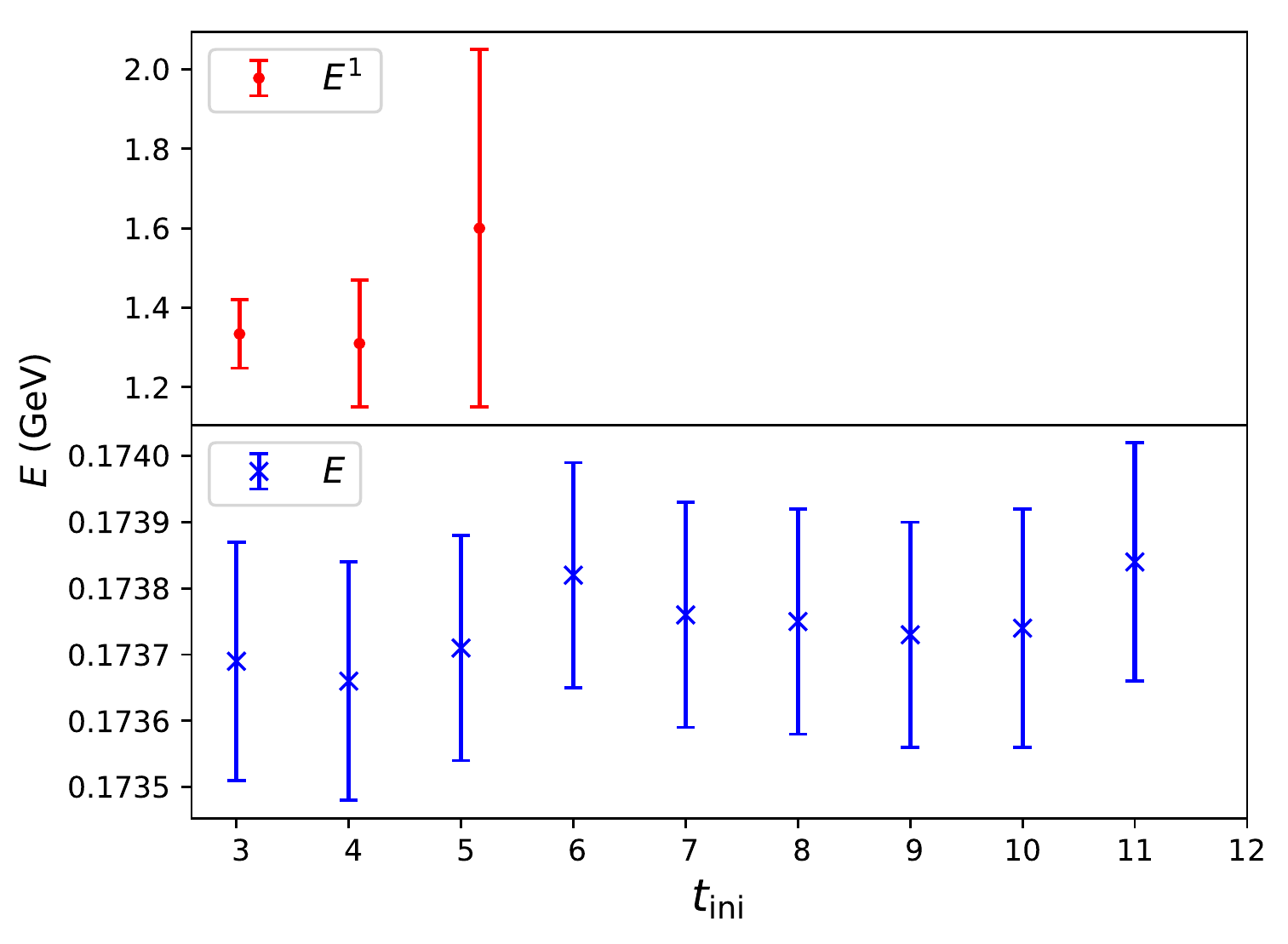}
  \caption{Pion energies as a function of $t_{\rm ini}$ with $[t_{\rm ini},15]$ the fit-range of the 2pt on 32ID with pion mass $173.7\ {\rm MeV} $ at zero momentum. The contributions from the first excited state are ignored for $t_{\rm ini} \geq 6$ under current statistics.}
  \label{fig:32ID_2pt_fit}
\end{figure}

{{For the special} $|\vec{p}_{\textrm{i}}|=|\vec{p}_{\textrm{f}}|$ case, one can simply calculate the ratio of 3pts, and obtain the pion form factor by the following parametrization of the ratio $R_1$, 
\begin{eqnarray} \label{3pt-ratio}
\begin{aligned}
&R_1(\tau,t_{\textrm{f}},\vec{p}_{\textrm{i}},\vec{p}_{\textrm{f}}) = {C_{\rm 3pt}(\tau,t_{\textrm{f}},\vec{p}_{\textrm{i}},\vec{p}_{\textrm{f}})}/{C_{\rm 3pt}(\tau,t_{\textrm{f}},\vec{p}_{\textrm{i}},\vec{p}_{\textrm{i}})}         \\
&=f_{\pi\pi}(Q^2) + B_1 (e^{-\Delta E \tau}  + e^{- \Delta E (t_{\textrm{f}} - \tau)}) + B_2 e^{-\Delta E t_{\textrm{f}}},
\end{aligned}
\end{eqnarray} 
where the terms with $B_1$ and $B_2$ are the contributions from the excited-state contamination, and $\Delta E = E^{1}(\vec{p}_{\textrm{i}})-E(\vec{p}_{\textrm{i}})$ is the energy difference between the pion energy $E(\vec{p}_{\textrm{i}})$
and that of the first excited state $E^{1}(\vec{p}_{\textrm{i}})$. 
These energies are also constrained by the joint fit with the corresponding 2pt.
Since the excited-state contamination of the forward matrix element in the denominator is known to be small and 
the contribution from $C_4$ term in Eq.~(\ref{eq:3pt}) is suppressed by $e^{-E(\vec{p}_i) T/2}$ with $\vec{p}_i \neq \vec{0}$ for both the denominator and numerator,
we have dropped them in the parametrization of the ratio and our fits can describe the data with {$\chi^2/d.o.f.\sim 1$}.
Fig.~\ref{fig:32ID_ff} shows a sample plot for 32ID with the unitary pion mass of 174 MeV at $Q^2=0.146 \ {\rm{GeV}}^2$.
In view of the fact that the data points are symmetric about  $\tau=t_{\textrm{f}}/2$, within uncertainty, it reassures that the sink smearing implemented under the FFT contraction has the same overlap with the pion state as that of the source smearing.
}

\begin{figure}[htbp]
  \centering
  \includegraphics[width=0.40\textwidth]{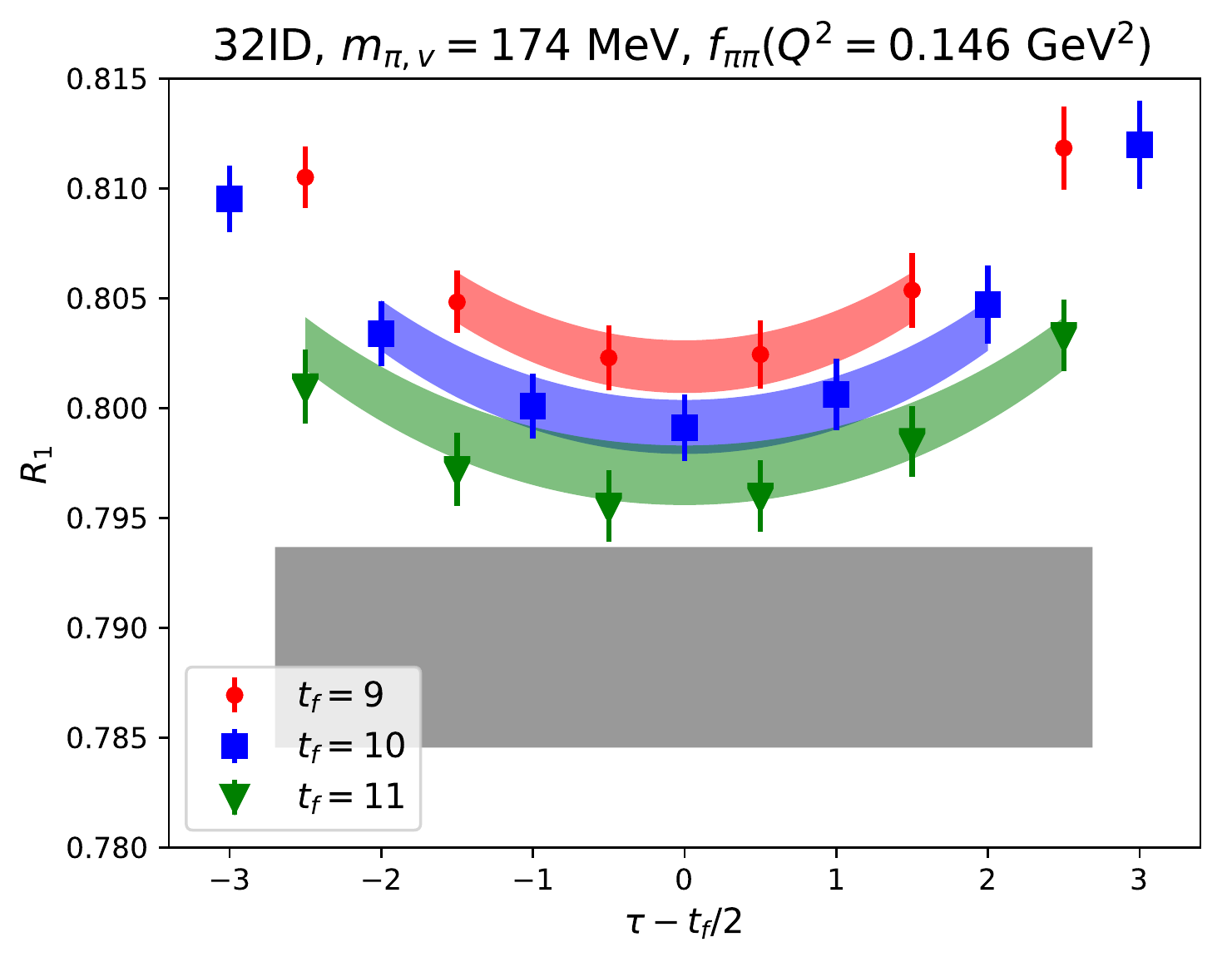} \\
  \caption{Example of the ratios for the special $|\vec{p}_{\textrm{i}}|=|\vec{p}_{\textrm{f}}|$ case on 32ID with various values of source-sink separation $t_{\textrm{f}}$ and current position $\tau$. The data points agree well with the colored bands predicted from the fit, and the gray band is for the fitted value of the ground state form factor $f_{\pi\pi}(Q^2)$.}
  \label{fig:32ID_ff}
\end{figure}

\begin{figure*}[htbp]
\centering
\includegraphics[scale=0.38]{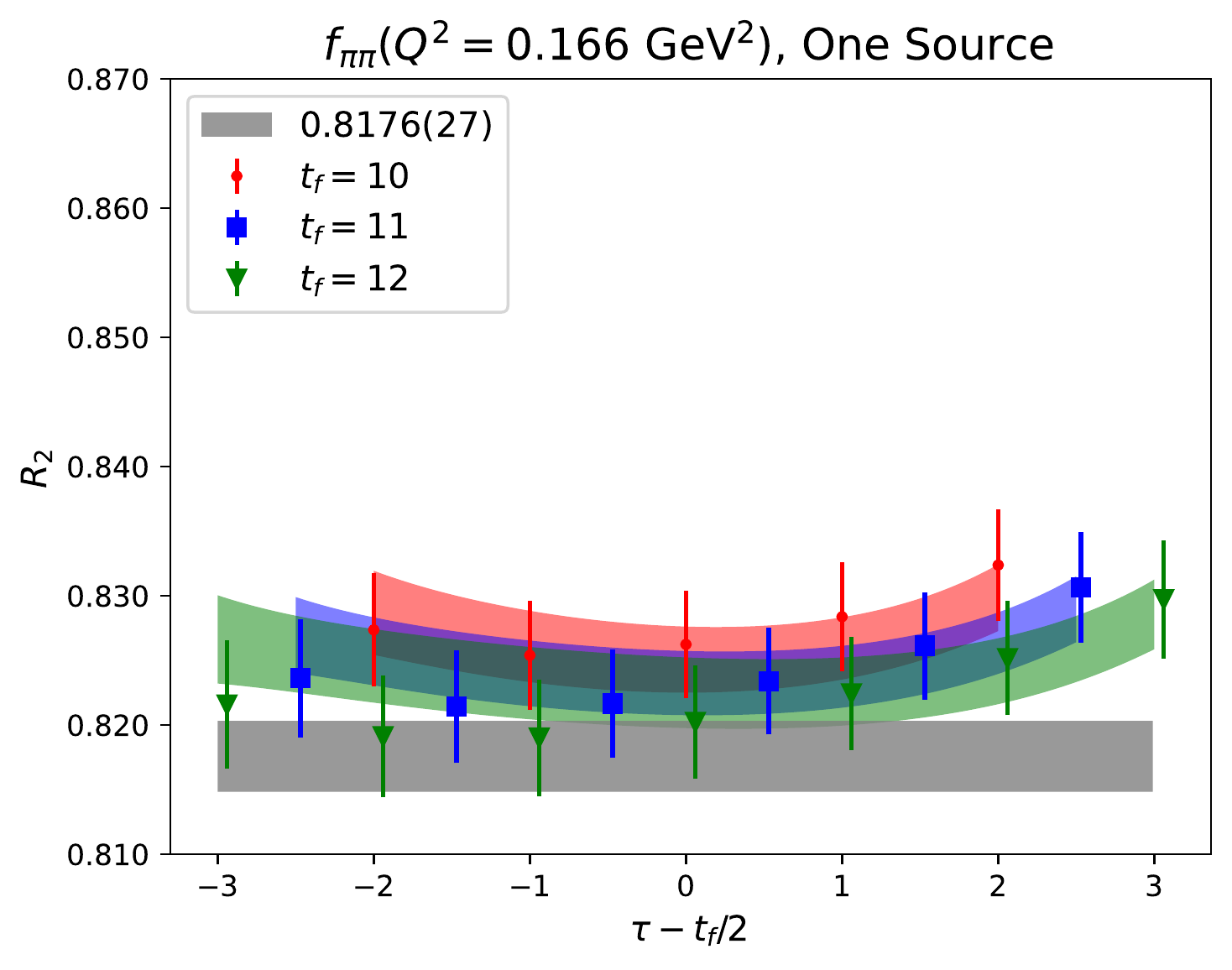}
\includegraphics[scale=0.38]{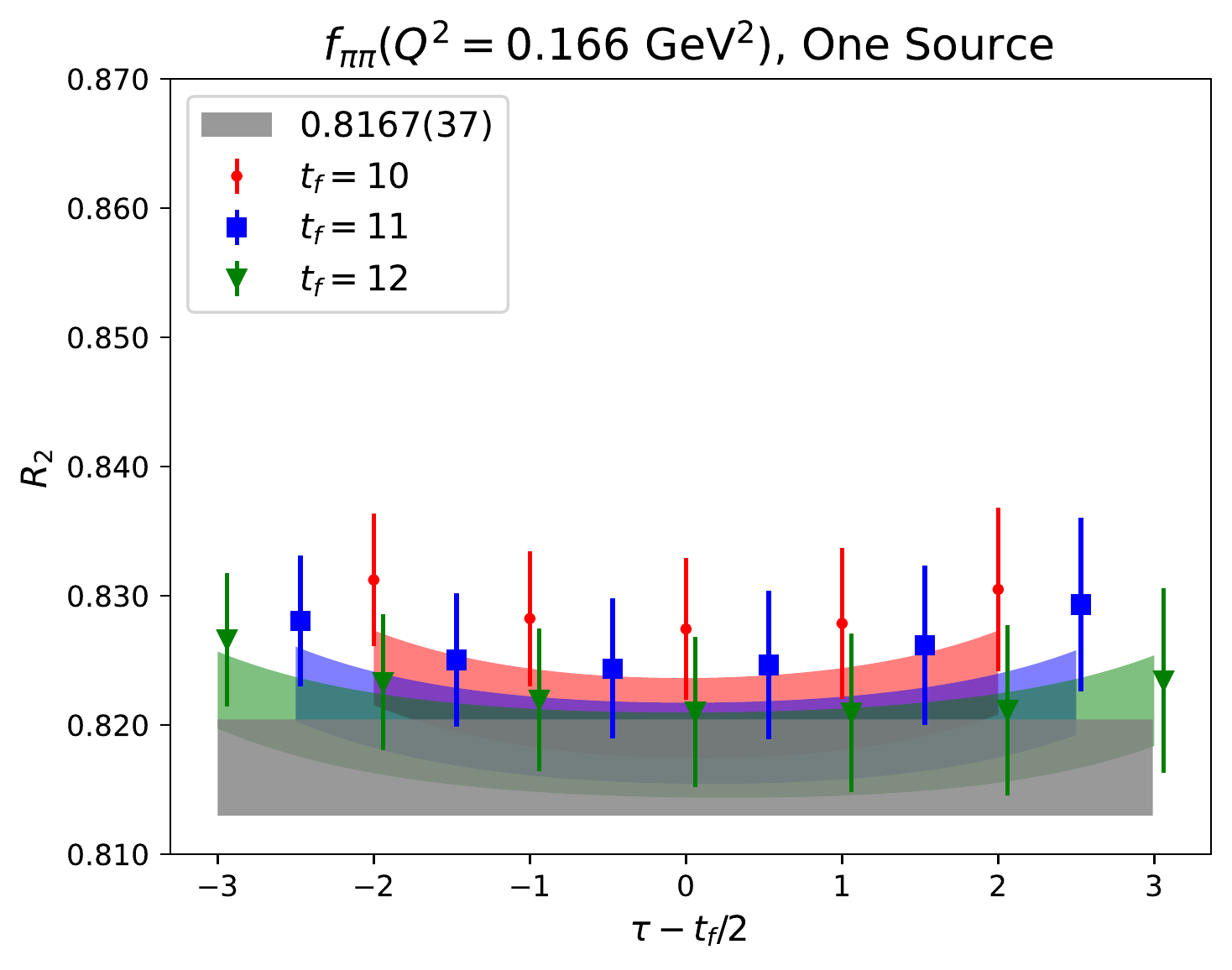}
\includegraphics[scale=0.38]{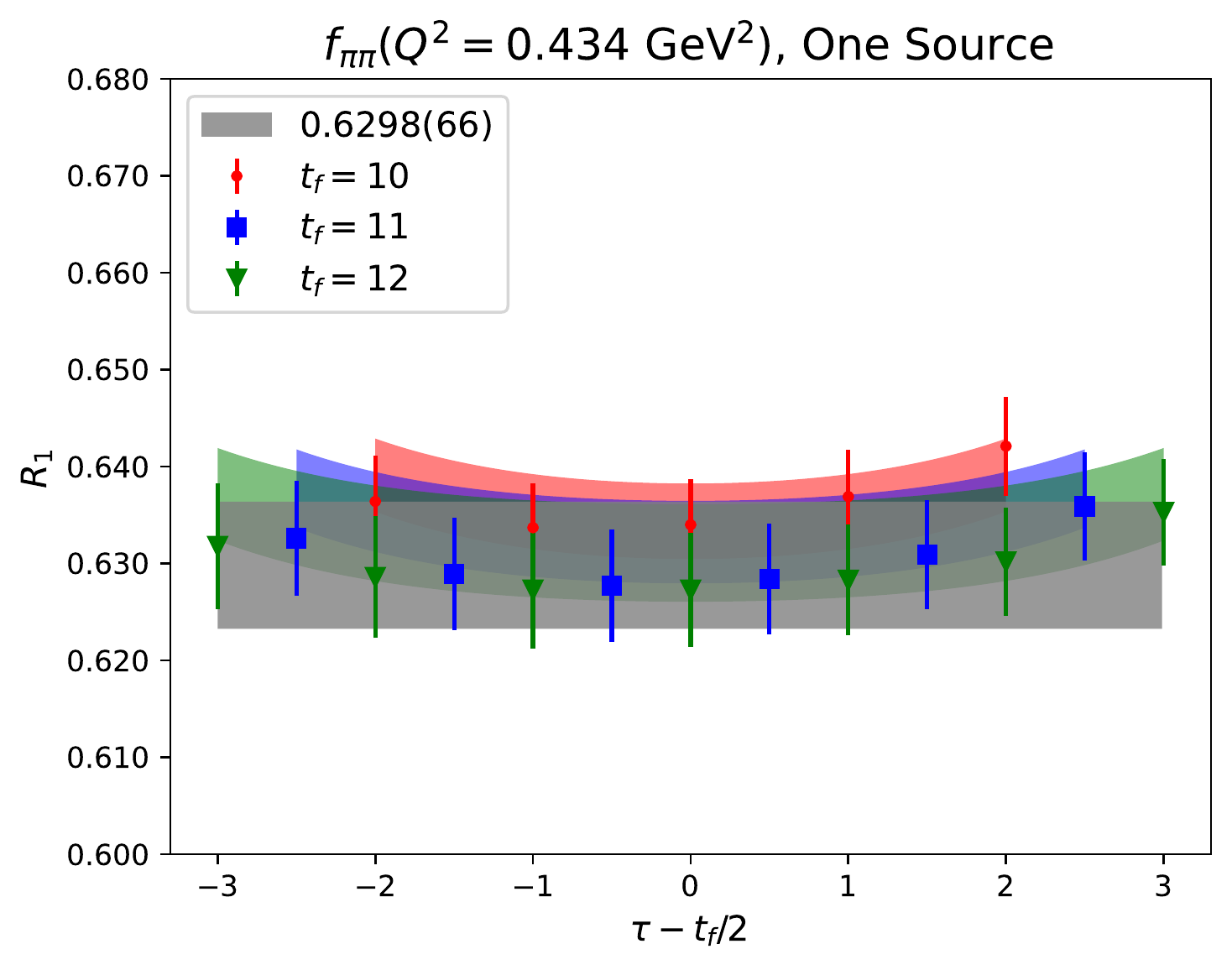} \\
\includegraphics[scale=0.38]{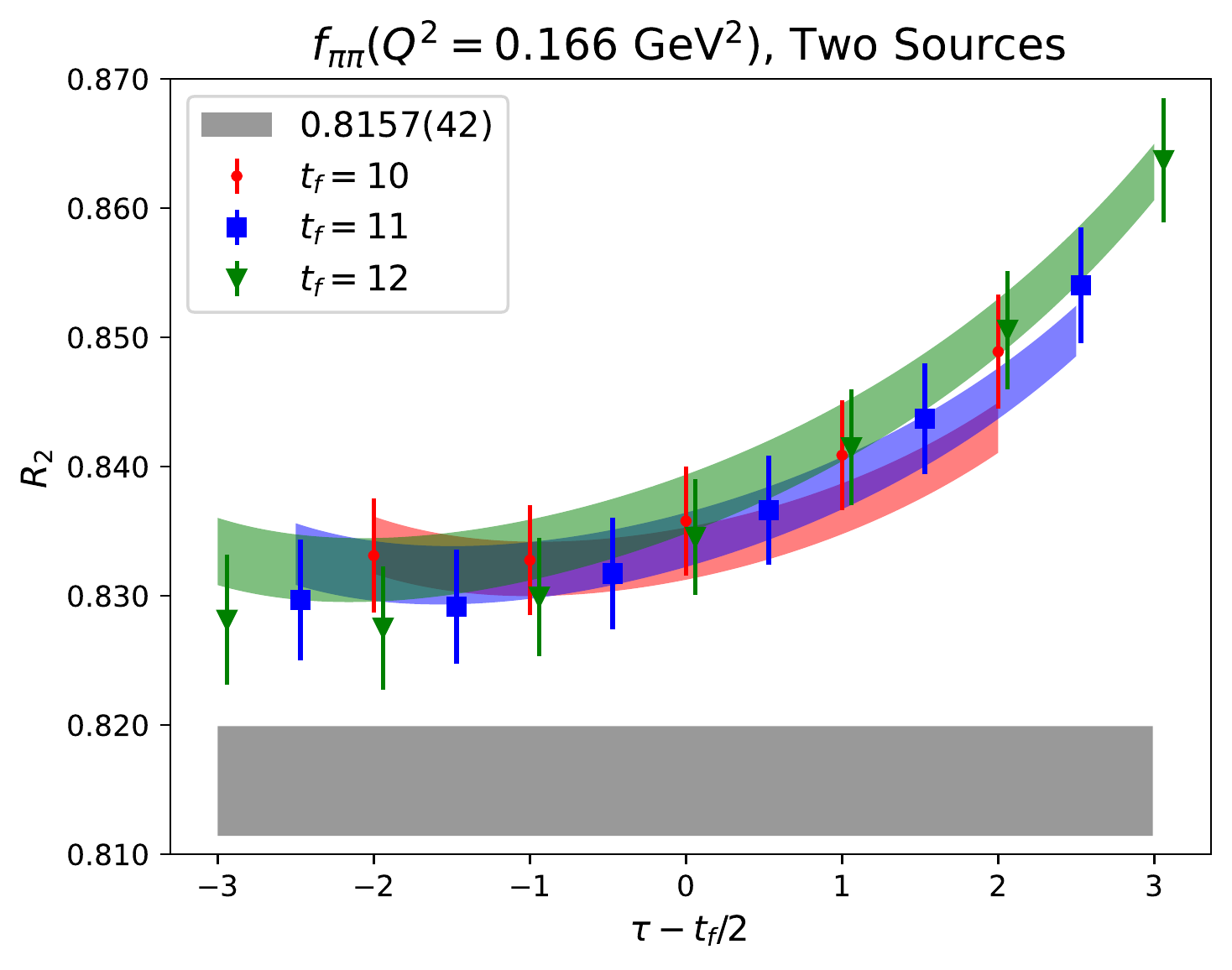}
\includegraphics[scale=0.38]{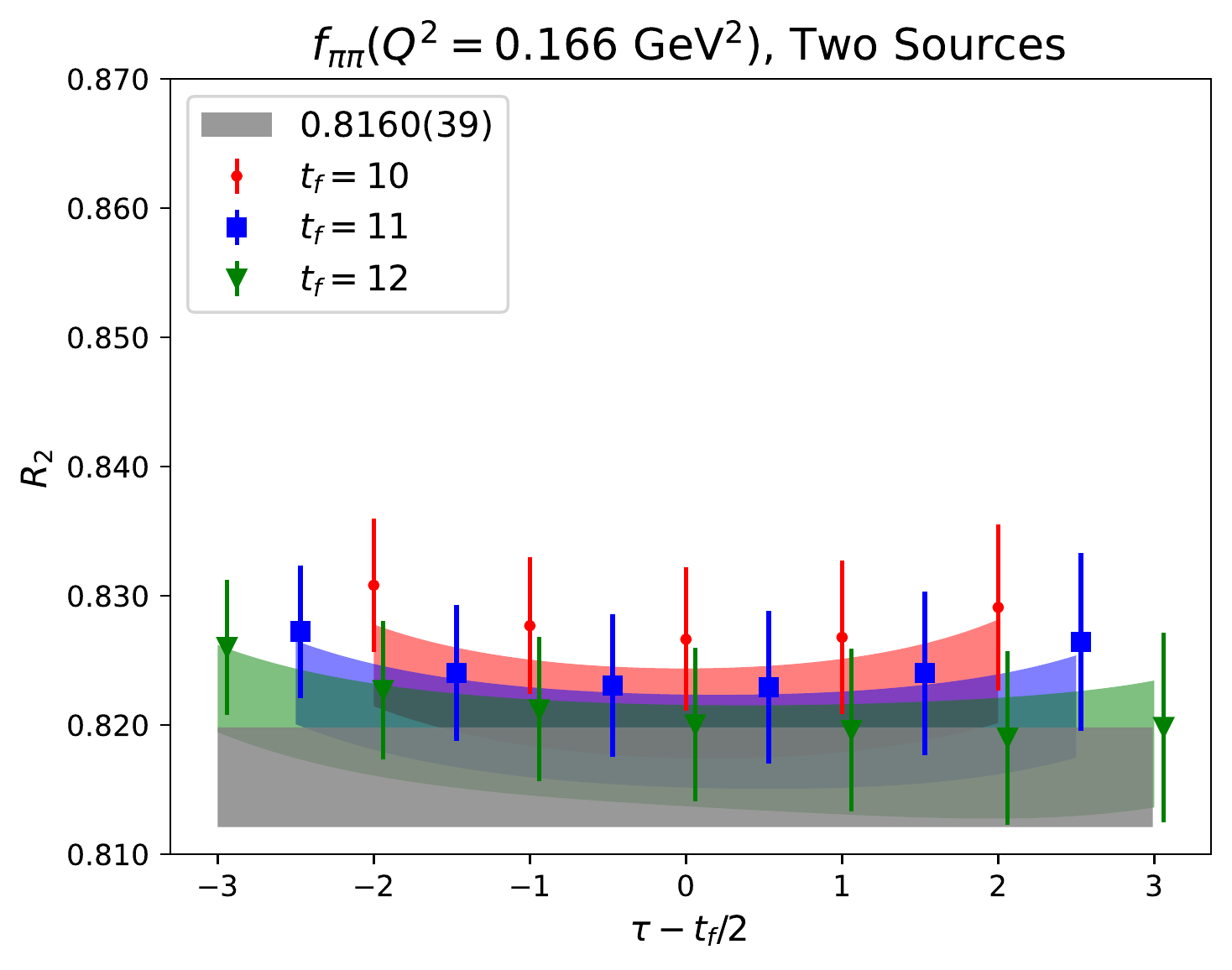}
\includegraphics[scale=0.38]{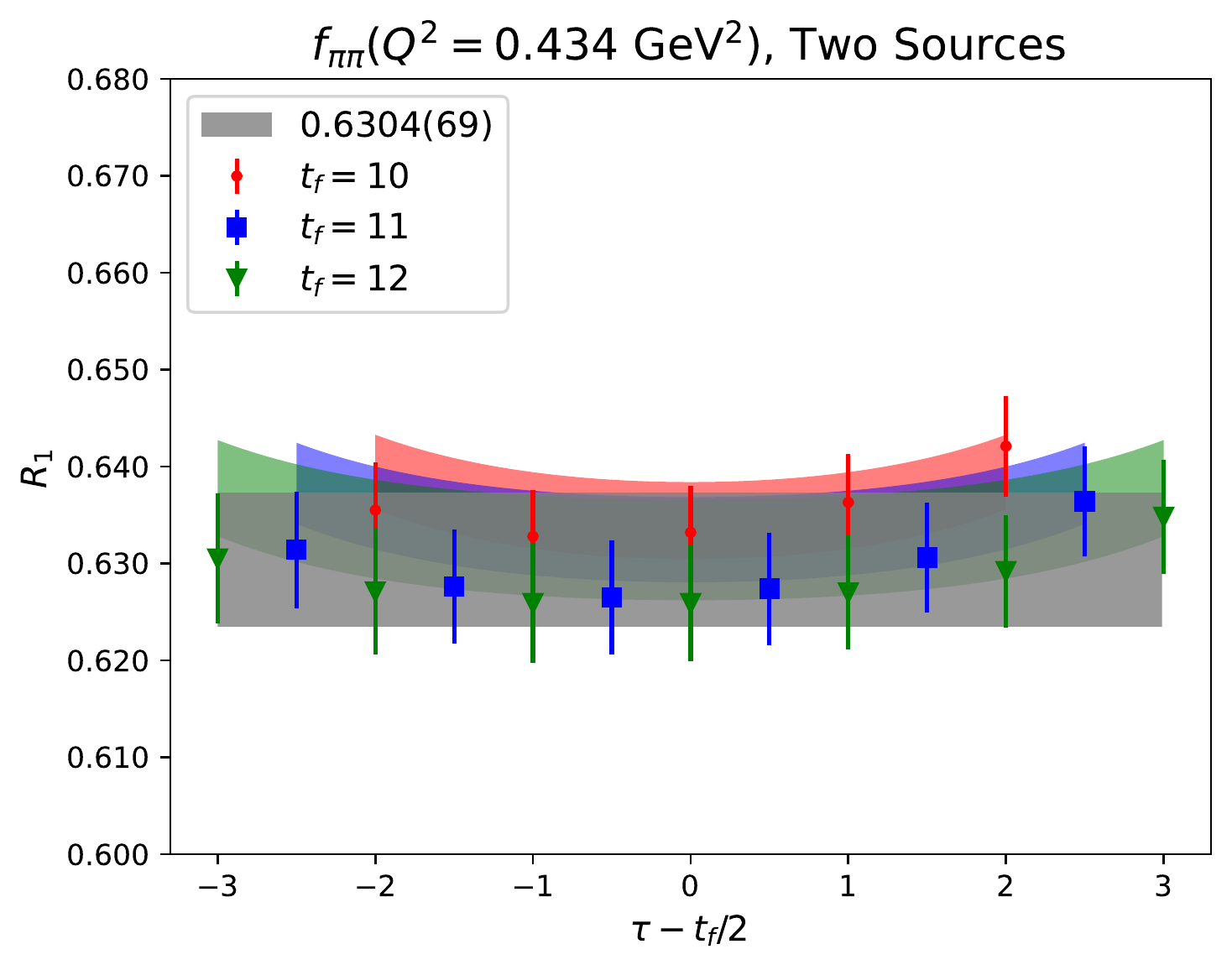} \\
\caption{Joint fit results on 24I with $m_\pi = 347\ {\rm{MeV}}$ for different source and momentum setups. The cases with $\vec{p}_i=0, \ \vec{q} = \vec{p}_f$; $\vec{p}_f=0, \ \vec{q} = -\vec{p}_i$; and $\vec{p}_f = -\vec{p}_i$,\ $\vec{q} = 2\vec{p}_f$ are shown in the left, middle and right panels, respectively. The top panels correspond to the cases of one source at time slice $0$. The lower panels correspond to the cases of a source at each of the time slices $0$ and $T/2$. Each panel's gray band is for the fitted value of the ground state form factor $f_{\pi\pi}(Q^2)$.}
\label{fig:dia_fit}
\end{figure*}

In order to test the fit function of 3pt in Eq.~(\ref{eq:3pt}), 
a comparison of the fit of the one-source result with the source at $t=0$ and that of the two-source result with a source at each of $t=0$ and $32$ in the same inversion is shown in Fig.~\ref{fig:dia_fit}.
For illustrative purpose, the data points on the left and middle panels are shown with ratio $R_2$,
\begin{eqnarray}
\begin{aligned}
&R_2(\tau,t_{\textrm{f}},\vec{p}_{\textrm{i}},\vec{p}_{\textrm{f}}) = {C_{\rm 3pt}(\tau,t_{\textrm{f}},\vec{p}_{\textrm{i}},\vec{p}_{\textrm{f}})}/ \left [\frac{Z_{\vec{p}_{\textrm{i}}} Z_{\vec{p}_{\textrm{f}}} (E_{\textrm{i}}+E_{\textrm{f}})}{4 E_{\textrm{i}} E_{\textrm{f}} Z_V} \right. \\
&\,\,\, \left. (e^{-E_{\textrm{i}} \tau -E_{\textrm{f}}(t_{\textrm{f}}-\tau)} + e^{-E_{\textrm{i}} (T/2 + \tau) -E_{\textrm{f}}(t_{\textrm{f}}-\tau)} ) \right] \\
&\,\,\, = f_{\pi \pi}(Q^2) + {\textrm{excited-state terms}} + {{C_4} \textrm{ term}} ,\\
\end{aligned}
\end{eqnarray}
in which $Z_{\vec{p}}$ and $E$ are determined from the fit of 2pt and $Z_V$ from 3pt at zero momentum transfer. 
{
The data for the top panels use $n_{{\rm{i}}}=4$ with $\{n_s,n_s,n_s,n_t\} = \{2,2,2,1\}$ and the data for the lower panels use $n_{{\rm{i}}}=2$ with $\{n_s,n_s,n_s,n_t\} = \{2,2,2,2\}$ so that their statistics are matched.
The case with one source and $\vec{p}_i=0$ and $\vec{q} = \vec{p}_f$ is shown
in the top left panel and the gray band is close to the data points due to small excited-state contamination.
The similar case with two coherent sources is shown in the lower left panel and the gray band is far away from the rising data points due to the additional $C_4$ term with fitted $E_{\rm h} = 820(110)\ {\rm{MeV}}$, which is consistent with the result of Fig.~\ref{fig:dia_corr}.
The two results agree with each other within uncertainty which confirms our fit formula,
but a comparison of the statistical errors reveals that the factor of two lower cost from using two coherent sources versus one source produces no net benefit for this case of $\vec{p}_{\rm i}=0$.
Since the contribution from the $C_4$ term is suppressed significantly for 3pts with $p_{\rm{i}} \neq 0$,
the data points and results from one source and two coherent sources agree with each other very well 
for the cases with $\vec{p}_{\rm f}=0, \ \vec{q} = -\vec{p}_{\rm i}$ and $\vec{p}_{\rm f} = -\vec{p}_{\rm i}$,\ $\vec{q} = 2\vec{p}_{\rm f}$ which are shown in the middle, and right panels, respectively.
In these kinematical cases, however, the statistical errors are the same for one source versus two coherent sources and thus the full factor of two lower cost (in computation and storage) for the latter is fully realized.
}

Thus for the general momentum setup $|\vec{p}_{\textrm{i}}|\neq |\vec{p}_{\textrm{f}}|$, we can proceed further to fit $C_{\rm 3pt}(\tau,t_{\textrm{f}},\vec{p}_{\textrm{i}},\vec{p}_{\textrm{f}})$ together with $C_{\rm 3pt}(\tau,t_{\textrm{f}},\vec{p}_{\textrm{f}},\vec{p}_{\textrm{i}})$ which corresponds to the exchange of initial and final momenta.
Fig.~\ref{fig:case0_ff_0} shows an example plot on 32ID.
The data points are fitted well ($\chi^2/d.o.f.\sim 1$) with Eq.~(\ref{eq:3pt}) and the fit results are shown with colored bands.
The data points for $C_{\rm 3pt}(\tau,t_{\textrm{f}},\vec{p},\vec{0})$ are lower and closer to the gray band since the $C_4$ term has a negative contribution with a suppression factor $e^{-E(\vec{p}) T/2}$ compared to the case of $C_{\rm 3pt}(\tau,t_{\textrm{f}},\vec{0},\vec{p})$ in which the $C_4$ term has a positive and large contribution with only a suppression factor $e^{-E(\vec{0}) T/2}$.

\begin{figure}[htbp]
  \centering
  \includegraphics[width=0.40\textwidth]{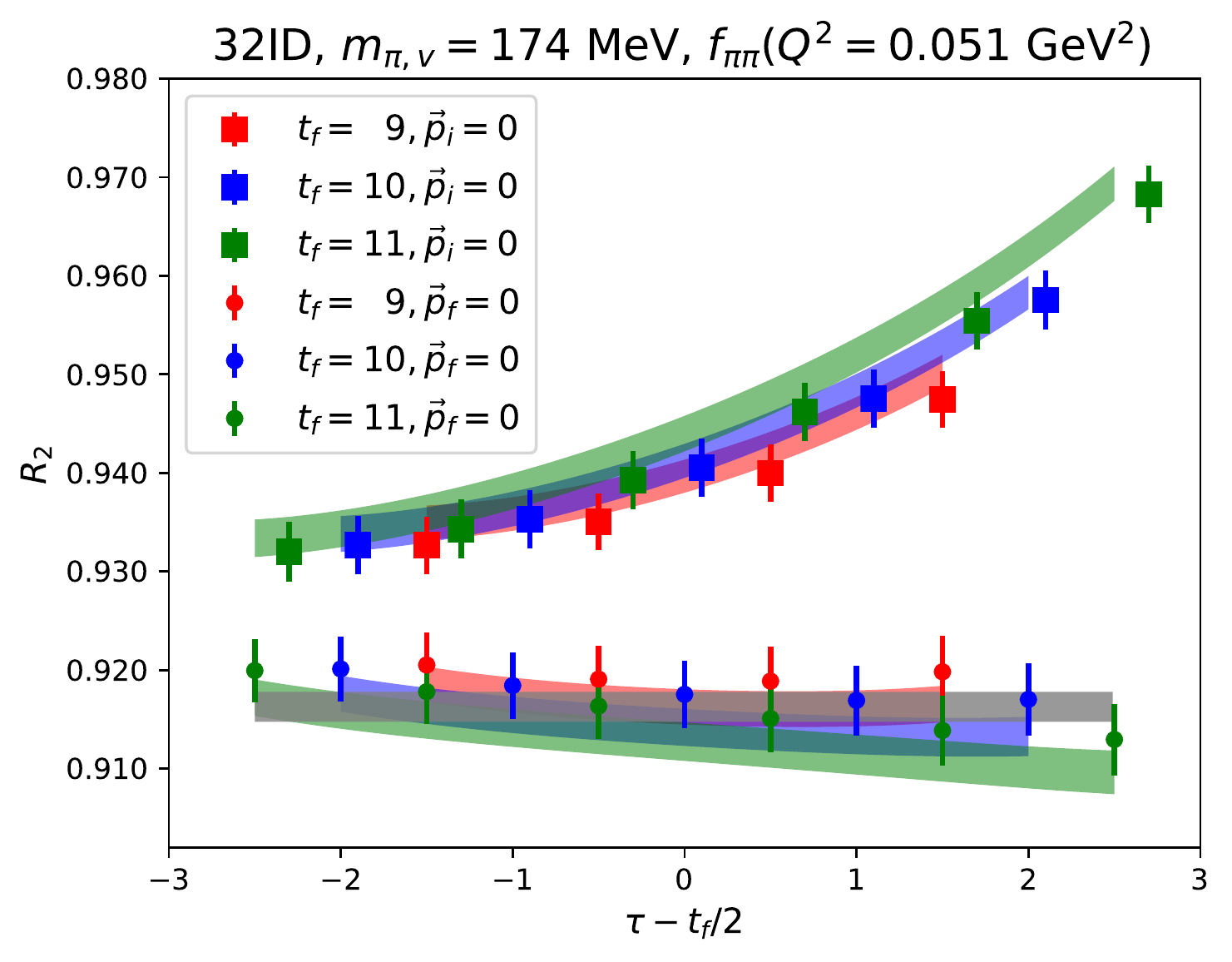}
  \caption{Examples of the ratios on 32ID with various values of source-sink separation $t_{\textrm{f}}$ and current position $\tau$ {at the valence pion mass $m_{\pi,{\rm v}} = 174 \ {\rm{MeV}}$}. The plots show the general $|\vec{p}_{\textrm{i}}|\neq |\vec{p}_{\textrm{f}}|$ case with square points $\vec{p}_{\textrm{i}}=-\vec{q},\vec{p}_{\textrm{f}}=0$ and dot points $p_{\textrm{i}}=0,\vec{p}_{\textrm{f}}=\vec{q}$.
  The data points agree well with the colored bands predicted from the fit, and the gray band is for the ground state form factor $f_{\pi\pi}(Q^2)$.
  }
  \label{fig:case0_ff_0}
\end{figure}

}

{
\subsection{$z$-expansion fit}

To obtain $f_{\pi \pi}(Q^2)$, we have done a model-independent $z$-expansion~\cite{Lee:2015jqa} fit using the following equation with $k_{\rm{max}} \geq 3$.
\begin{eqnarray}\label{eq:z-exp}
\begin{aligned}
f_{\pi \pi}(Q^2) &= \sum_{k=0}^{k_{\rm{max}}} a_k z^k \\
z(t,t_{\rm cut},t_0) &= \frac{\sqrt{t_{\rm cut}-t} - \sqrt{t_{\rm cut}-t_0}}{\sqrt{t_{\rm cut}-t} + \sqrt{t_{\rm cut}-t_0}} ,\\
\end{aligned}
\end{eqnarray}
where \mbox{$t = -Q^2$}; \mbox{$f_{\pi \pi}(0)=1$} after normalization which leads to the constraint \mbox{$a_0 = 1 - \sum_{k=1}^{k_{\rm max}} a_k z^k(t=0,t_{\rm cut},t_0)$}; \mbox{$t_{\rm cut} = 4 m_{\pi,\textrm{mix}}^2$} corresponds to the two-pion production threshold,
with $m_{\pi,\textrm{mix}}$ {the mass of the mixed valence and sea pseudoscalar meson calculated in Ref~\cite{Lujan:2012wg,Deltamix_paper} on each ensemble directly with one valence domain wall propagator and one valence overlap propagator for each valence quark mass;}
and $t_0$ is chosen to be its ``optimal" value \mbox{$t_0^{\rm opt}(Q_{\rm{max}}^2) = t_{\rm cut}(1- \sqrt{1 + Q_{\rm max}^2/t_{\rm cut}})$} to minimize the maximum value of $|z|$, with $Q_{\rm max}^2$ the maximum $Q^2$ under consideration.

In order to remove the model dependence of the $z$-expansion fit, we need to take $k_{\rm max}$ to be large enough such that the fit results are independent of the precise value of $k_{\rm max}$. 
One way of achieving this is putting a Gaussian prior on the $z$-expansion parameters $a_k$ with central value $0$.
The choice of the Gaussian prior can be investigated using the Vector Meson Dominance (VMD) model with rho meson mass $m_{\rho} = 775 \ {\rm{MeV}}$,
\begin{eqnarray}\label{eq:VMD}
\begin{aligned}
f_{\pi \pi}(Q^2) = \frac{1}{1 + Q^2/m_{\rho}^2}.
\end{aligned}
\end{eqnarray}
A non-linear least squares fit of this analytical function with $z$-expansion fit at $k_{\rm max} = 10$ gives $|a_k/a_0|_{\rm max} < 1.03$,
in which we used $t_{\rm cut} = 4 m_{\pi,{\rm phys}}^2$, $t_0^{\rm opt}(Q_{\rm max}^2) = t_{\rm cut}(1- \sqrt{1 + Q_{\rm max}^2/t_{\rm cut}})$ and $Q_{\rm max}^2 = 1.0 \  {\rm{GeV^2}} $.
Also, by investigating the $z$-expansion fits with $k_{\rm max}=3$ without priors of our data, we find $|a_k/a_0|_{\rm max} < 3.0$.
Thus we propose the use of the conservative choice of Gaussian prior~\cite{Lee:2015jqa} with $|a_k/a_0|_{\rm max}=5$ (use $``a_k/a_0 = 0(5)"$ as a Gaussian prior for all $a_k, \ k>1$, in the fits) for the pion form factor.
The $z$-expansion fitted pion form factors up to $Q^2 \ \sim 1.0 \ {\rm{GeV}}^2$ for the seven lattices with the same valence and sea pion mass are shown in Fig.~\ref{fig:ff_all} with {$\chi^2/d.o.f.\sim [0.4,0.9]$}.

\begin{figure*}[!htb]
  \centering
   \includegraphics[scale=0.45]{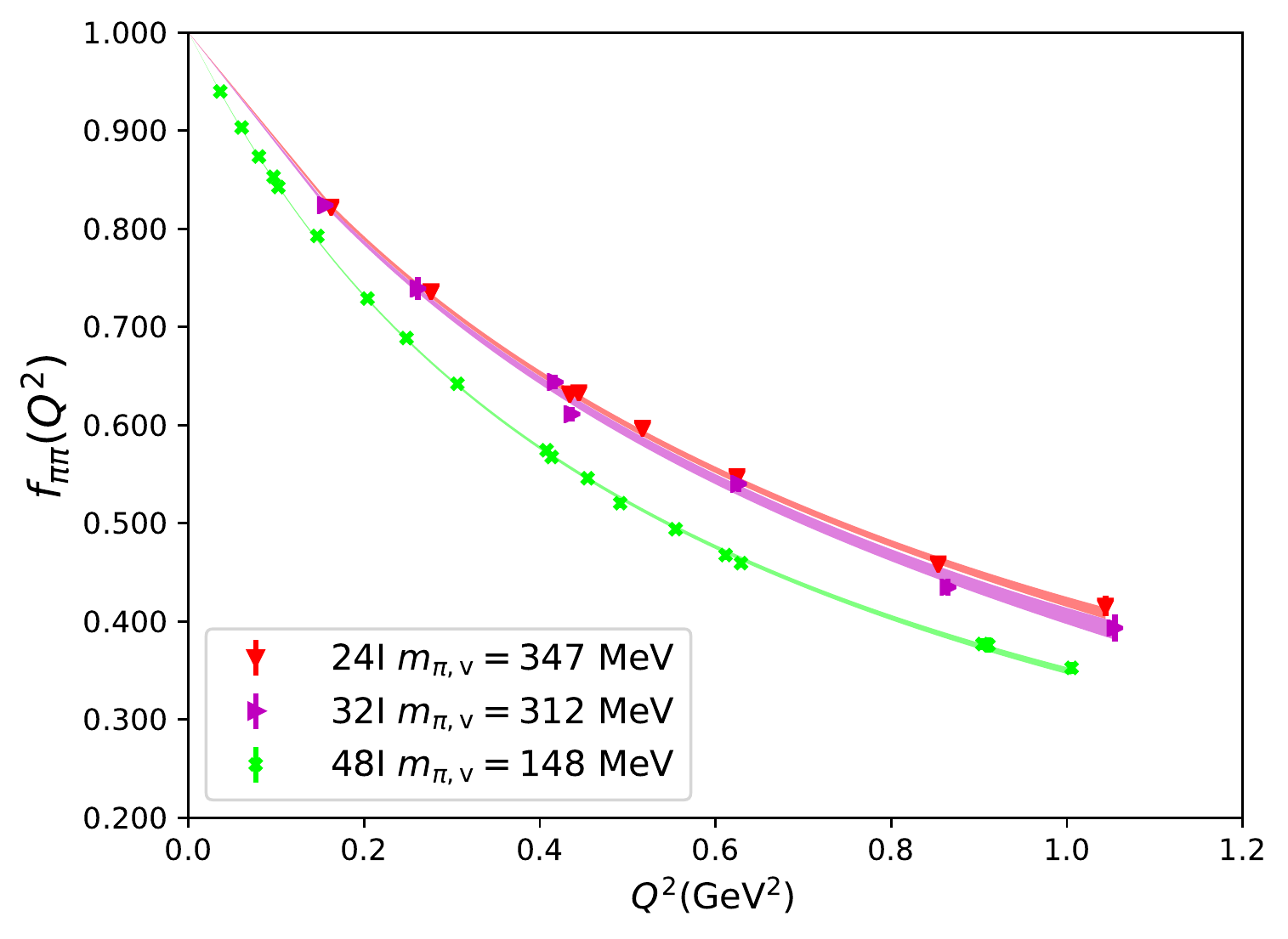}
   \includegraphics[scale=0.45]{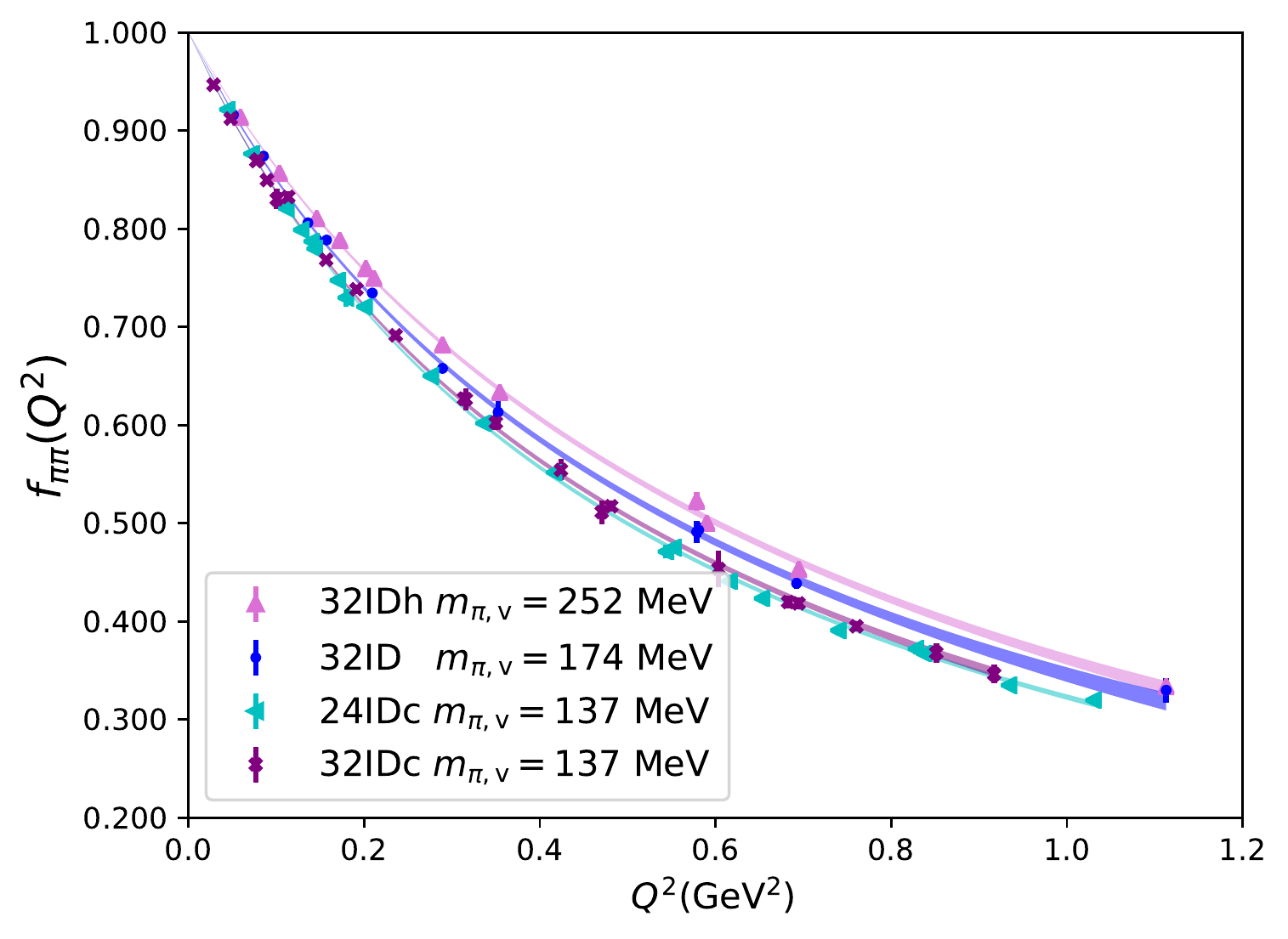}
  \caption{$z$-expansion fit of the pion form factors on seven gauge ensembles at their unitary pion mass with $k_{\rm max}=3$ and $|a_k/a_0|_{\rm max}=5$. The left panel is for the ensembles using the Iwasaki gauge action and the Iwasaki+DSDR cases are shown in the right panel.}
  \label{fig:ff_all}
\end{figure*}

Another way to reach higher $k_{\rm max}$ and control the model dependence of fits is to use the fact that at the $Q^2 \rightarrow \infty$ limit $f_{\pi \pi}(Q^2)$ falls as $1/Q^2$ up to logarithms~\cite{Lepage:1979zb,Farrar:1979aw}. Thus we have $Q^k f_{\pi \pi}(Q^2) \rightarrow 0$ for $k=0,1$ and follow the same argument in~\cite{Lee:2015jqa}, which implies
\begin{eqnarray}
{\left. \frac{d^n}{dz^n} f_{\pi \pi} \right|}_{z=1} = 0, \quad n\in \{0,1\},
\end{eqnarray}
with $z=1$ corresponding to the $Q^2 \rightarrow \infty$ limit. These equations lead to the two sum rules for pion form factors as
\begin{eqnarray}\label{eq:sum_rules}
\sum_{k=0}^{\infty} a_k = 0 ,\quad \sum_{k=1}^{\infty} k a_k = 0.
\end{eqnarray}
{We have explored this alternative results shown in Fig.~\ref{fig:radius_k_max}.}

}

{
\subsection{Chiral extrapolation of pion radius}

With the $z$-expansion fit of the form factor using Eq.~(\ref{eq:z-exp}), the charge radius of pion can be obtained through 
\begin{eqnarray}
\begin{aligned}
\langle r_{\pi}^2 \rangle \equiv -6 \frac{d\,\! f_{\pi \pi} (Q^2)}{d Q^2}|_{Q^2 = 0},
\end{aligned}
\end{eqnarray} 
for all the valence masses of each lattice.
{Fig.~\ref{fig:radius_mv} shows the results on 32ID and 32IDh as a function of valence pion mass $m_{\pi,{\rm{v}}}^2$ and mixed pion mass $m_{\pi,{\rm{mix}}}^2$ in the left panel and right panel, respectively. 
We see that there is a strong dependence on the valence pion masses from the data points on these two ensembles. 
Also, the disagreement in the left panel evinces a strong dependence on the sea pion mass.
In contrast, the right panel shows an agreement of 32ID results and 32IDh results at similar $m_{\pi,{\rm{mix}}}^2$ which guides us to use $m_{\pi,{\rm{mix}}}^2$, as proposed by Partially Quenched Chiral Perturbation Theory~\cite{Arndt:2003ww}, as a basic variable for the chiral extrapolations.}

\begin{figure*}[!htb]
  \centering
   \includegraphics[scale=0.45]{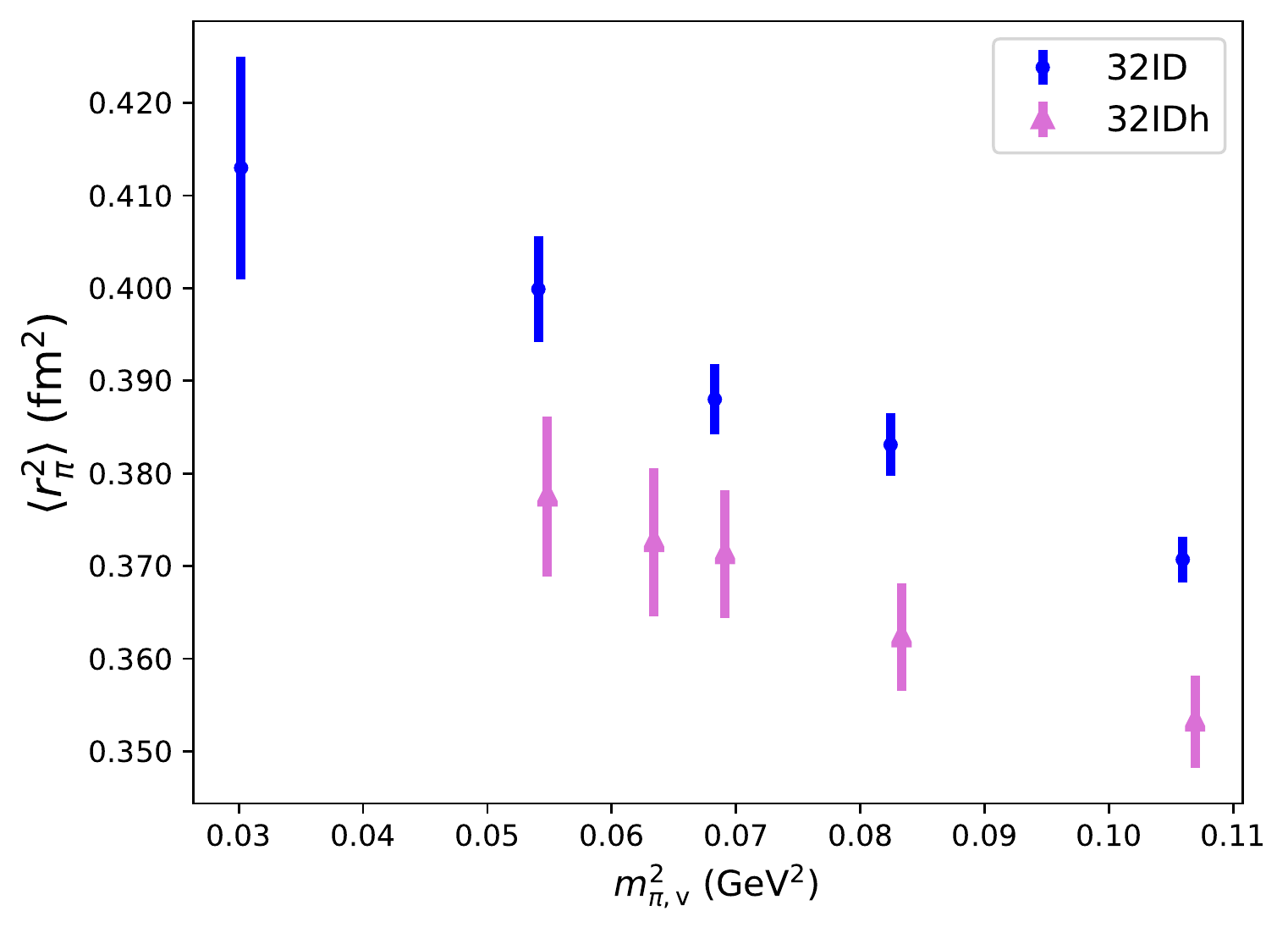}
   \includegraphics[scale=0.45]{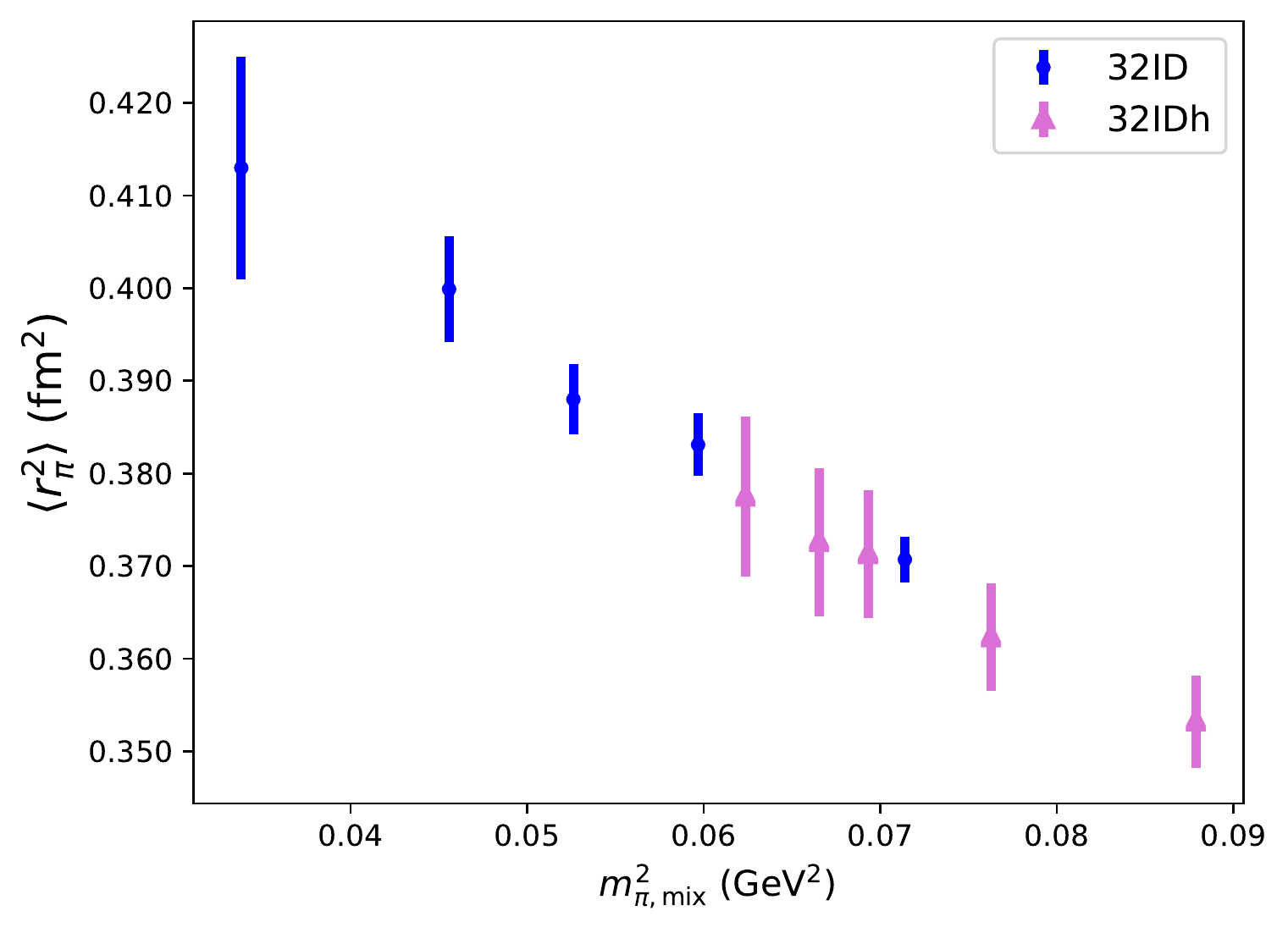}
  \caption{The left panel shows the pion radius on 32ID and 32IDh as a function of valence pion mass $m_{\pi,{\rm{v}}}^2$ and the right panel shows the same data as a function of mixed pion mass $m_{\pi,{\rm{mix}}}^2$.}
  \label{fig:radius_mv}
\end{figure*}

The $\langle r^2_\pi \rangle$ on different lattices with different valence pion masses are plotted in Fig.~\ref{fig:Radius_global}.
The following fit form as a function of $m_{\pi,\textrm{mix}}^2$ is used which includes an essential divergent log term from the $SU$(2) NLO ChPT~\cite{Bijnens:1998fm,Arndt:2003ww},
\begin{eqnarray}\label{eq:global_fit}
\begin{aligned}
\braket{r^2_{\pi}} =& 
\frac{1}{(4 \pi F_\pi)^2} (\bar{l}_6 + {\rm{ln}} \frac{M_\pi^2}{m_{\pi,\textrm{phys}}^2} - 1) \\
&\quad + {\color{black}{b_2^{I/ID}}} a^2 + \frac{ {\color{black}{b_3}}  e^{- M_\pi L} }{(4 \pi F_\pi)^4 {(M_\pi L)}^{3/2}},
\end{aligned}
\end{eqnarray} 
where $F_{\pi} = F \big(1 + \frac{M_{\pi}^2}{16 \pi^2 F_\pi^2} (\bar{l}_4 - {\rm ln} \frac{M_\pi^2}{m_{\pi,\textrm{phys}}^2}) ) \big)$ shows the pion mass dependence of the pion decay constant from partially quenched NLO $SU$(2) ChPT~\cite{Golterman:1997st} with $M_\pi^2 \equiv m_{\pi,\textrm{mix}}^2$, 
$F$ and $\bar{l}_6$ are free parameters for fitting,
$m_{\pi,\textrm{phys}} = 139.57 \ {\rm{MeV}}$ is the physical pion mass,
$L$ is the spatial size of the lattice, 
the $b_2^{I/ID}$ terms reflect the lattice spacing dependence for the two sets of ensembles with different gauge actions (Iwasaki and Iwasaki plus DSDR),
and the $b_3$ term accounts for the finite-volume effect~\cite{Bunton:2006va,Jiang:2006gna,Colangelo:2016wgs,Alexandrou:2017blh}.
{Instead of fitting both the low-energy constants $\bar{l}_4$ and $F$ as free parameters, which leads to unstable fits, we use $\bar{l}_4 = 4.40(28)$, as given by its FLAG average~\cite{Aoki:2019cca}, as a prior and treat $F$ as free parameter.}

\begin{figure}[htbp]
\centering
 \includegraphics[width=0.4\textwidth]{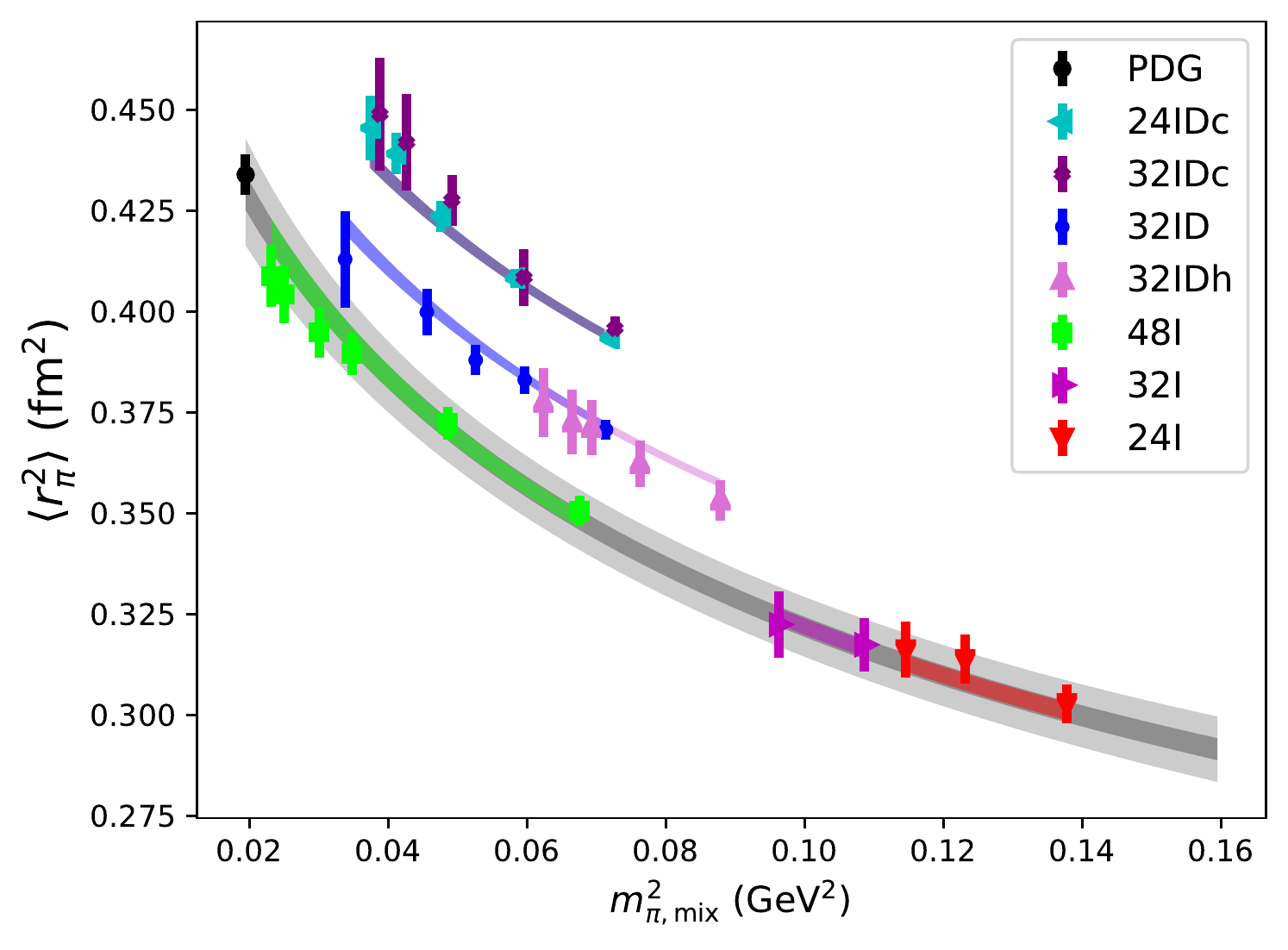}
  \caption{Pion radius squared $\braket{r^2_\pi}$ as a function of $m_{\pi,\textrm{mix}}^2$. Data points with different colors correspond to the results on the seven ensembles with different sea pion masses.}
  \label{fig:Radius_global}
\end{figure}

\begin{figure}[!htb]
  \centering
   \includegraphics[scale=0.45]{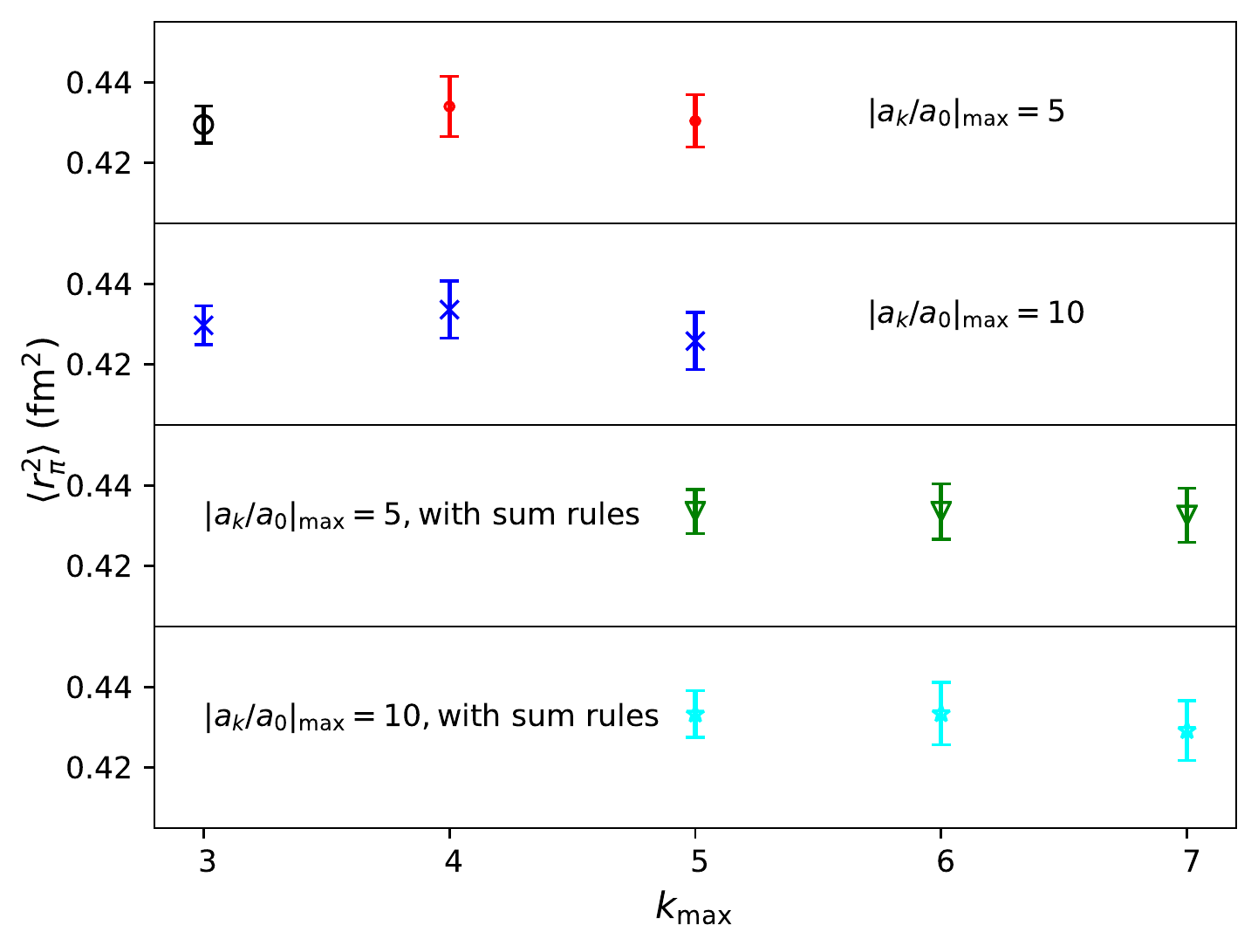}
  \caption{Comparison of extrapolated $\braket{r_{\pi}^2}$ with $z$-expansion fits with different $k_{\rm max}$. The first and second sets are the fits with priors $|a_k/a_0|_{\rm max}=5$ and $|a_k/a_0|_{\rm max}=10$, respectively. The third and fourth sets are the similar fits constrained with the sum rules in Eq.~(\ref{eq:sum_rules}).}
  \label{fig:radius_k_max}
\end{figure}

The results of the fits are shown in Fig.~\ref{fig:Radius_global}.
The colored bands show our prediction based on the global fit of $\braket{r^2_{\pi}}$ with {$\chi^2/d.o.f.= 0.85$}; 
the inner gray band shows our prediction for the unitary case of equal pion mass in the valence and the sea in the continuum and infinite volume limits and the outer band includes the systematic uncertainties from excited-state contamination, $z$-expansion fit, chiral extrapolation, lattice spacing, and finite-volume dependence.
Since the kaon mass only varies a little in the current pion mass range, we do not include the kaon log term in the fit.
The discretization errors across the Iwasaki gauge ensembles are small, while those across the Iwasaki plus DSDR gauge ensembles are obvious; this is consistent with what was found in the previous work with the DWF valence quark on similar RBC ensembles~\cite{Feng:2019geu}.
The fit gives $F_{\pi} = 96.2(4.3) \, {\rm{MeV}}$, which is consistent with $92.2(1) \, {\rm{MeV}}$,
and $\bar{l}_6 = 17.1(1.4)$, which is also  consistent with the FLAG average~\cite{Aoki:2019cca} value $\bar{l}_6 = 15.1(1.2)$.
The systematic uncertainties considered are listed as follows:
\begin{itemize}
\item
{
{Fit results for the radius from different $z$-expansion fits using Eq.~(\ref{eq:global_fit}) are shown in Fig.~\ref{fig:radius_k_max}.}
Since $b_2^I$ and $b_3$ have no statistical significance, we use only three free parameters $F$, $\bar{l}_6$ and $b_2^{ID}$ in these fits and treat the low-energy constant $\bar{l}_4 = 4.40(28)$ appearing in $F_\pi$ as a prior.
All the fits have good {$\chi^2/d.o.f.\sim 0.85$} with the central values and error values varying a little.
Thus we take the result shown in black, namely $\braket{r_{\pi}^2}=  0.4298(45) \ {\rm{fm^2}}$, which corresponds to $k_{\rm max}=3$ and $|a_k/a_0|_{\rm max}=5$ as our fit result.
{The central values and correlations of the fit parameters $F$, $\bar{l}_6$, $b_2^{ID}$, and $\bar{l}_4$ are listed in Table~\ref{tab:radius_corr}.}
The maximum difference between the result shown in black in Fig.~\ref{fig:radius_k_max} and those of the other fitted cases is treated as the systematic uncertainty from the $z$-expansion fit.
}
\item
{
The systematic uncertainty from the excited-state contamination is estimated by changing the fit ranges of 2pt and 3pt on 32ID with pion mass $174\ {\rm MeV} $ at the smallest momentum transfer which results in $f_{\pi\pi}(Q^2 = 0.051 \ {\rm GeV^2}) = 0.9158(14)(13)$; the second error corresponds to the systematic uncertainty from excited-state contamination.
This case is chosen because of its good signal to noise ratio which has the most control of the final result at close to the physical pion mass, and the smallest momentum transfer is chosen due to its largest influence on the radius.
{In order to estimate the systematic uncertainty of the radius from the form factor at only one small momentum transfer, we solve the VMD model in Eq.~(\ref{eq:VMD}),
\begin{eqnarray}
\frac{1}{1 + (0.051 \ {\rm GeV^2})/m^2} = 0.9158(14)(13)
\end{eqnarray} 
with $m$ as a free parameter. The predicted radius is $\braket{r_{\pi}^2} = {6.0}/m^2 = 0.4190(74)(68) \ {\rm{fm^2}}$.
The second error $0.0068  \ {\rm{fm^2}}$, which propagates from the systematic uncertainty of the form factor, is treated as the systematic uncertainty from the change of fit ranges for the extrapolated charge radius.}
}

\item
{

We added a linear dependence term between the charge radius of the pion and the pion mass squared as $b_4 M_\pi^2$ to Eq.~(\ref{eq:global_fit}) proposed by $SU$(2) NNLO ChPT~\cite{Gasser:1984ux} and repeated the fit with four free parameters $\braket{r^2_{\pi}}_{\textrm{phys}}$, $b_1$, $b_2^{ID}$ and $b_4$.
The coefficient $b_4$ is consistent with zero and the prediction changes by $0.0017 \ {\rm{fm^2}}$ which is treated as a chiral extrapolation systematic uncertainty.

Another source of the chiral extrapolation systematic uncertainty is the lack of a kaon log term in Eq.~(\ref{eq:global_fit}).
On 24I, the valence pion masses ranging from $323 \ {\rm{MeV}}$ to $391 \ {\rm{MeV}}$ give a range of kaon mass from $532 \ {\rm{MeV}}$ to $554 \ {\rm{MeV}}$.
Thus we estimate the maximum kaon mass for the pion mass range in consideration to be $M_{K,{\rm max}}=554 \ {\rm{MeV}}$.
With the use of $SU$(3) NLO ChPT~\cite{Bijnens:1998fm}, the systematic uncertainty from the kaon log term can be given by {$\frac{1}{32 \pi^2 F_{\pi}^2} {\rm{ln}} \frac{M_{K,{\rm{max}}}^2}{m_{K,{\rm{phys}}}^2} = 0.0034  \ {\rm{fm^2}}$}, in which $F_{\pi}=92.2 \ {\rm{MeV}}$ and $m_{K,{\rm{phys}}}=493 \ {\rm{MeV}}$ is the physical kaon mass.
}
\item 
{
We repeated the fit with four free parameters $F$, $\bar{l}_6$, $b_2^{ID}$ and $b_2^{I}$ which includes the discretization error from the Iwasaki gauge action and the prediction changes by $0.0052 \ {\rm{fm^2}}$.
With this fit, we get a difference between the fit predictions in the continuum limit with those from the smallest lattice spacing (32I) to be $0.0017 \ {\rm{fm^2}}$. We combined these two as the systematic uncertainty of finite lattice spacing.
}
\item 
{
We repeated the fit with four free parameters $F$, $\bar{l}_6$, $b_2^{ID}$ and $b_4$ which includes the finite-volume term and the prediction changes by {$0.00019 \ {\rm{fm^2}}$}.
With the inclusion of the finite-volume term, the difference of the predictions for 24IDc (which has the smallest $m_\pi L$) and 32IDc is {$0.005 \ {\rm{fm^2}}$}.
We combined these two as the systematic uncertainty of finite-volume effects.
}
\end{itemize}

Thus, the final result of the mean square charge radius of the pion at the physical pion mass in the physical limit reads
\begin{eqnarray*}\label{eq:radius_sys}
\begin{aligned}
\braket{r_{\pi}^2} &= 0.4298(45)_{\rm stat} (66)_{z{\textrm{-exp}}} (68)_{\textrm{fit-range}} ({37})_{\chi} (55)_{\rm a} (50)_{\rm V} \\
& = 0.4298(45)({126}) \ {\rm{fm^2}}, \\
\end{aligned}
\end{eqnarray*} 
with statistical error $({\rm stat})$ and systematic uncertainty from $z$-expansion fit $(z{\textrm{-exp}})$, fit-range dependence $({\textrm{fit-range}})$, chiral extrapolation $(\chi)$, finite lattice spacing $({\rm a})$, and finite-volume $({\rm V})$. The total uncertainties at heavier pion masses are estimated from the scale of the total/statistical ratio at the physical pion mass.

\begin{figure}[htbp]
\centering
 \includegraphics[width=0.5\textwidth]{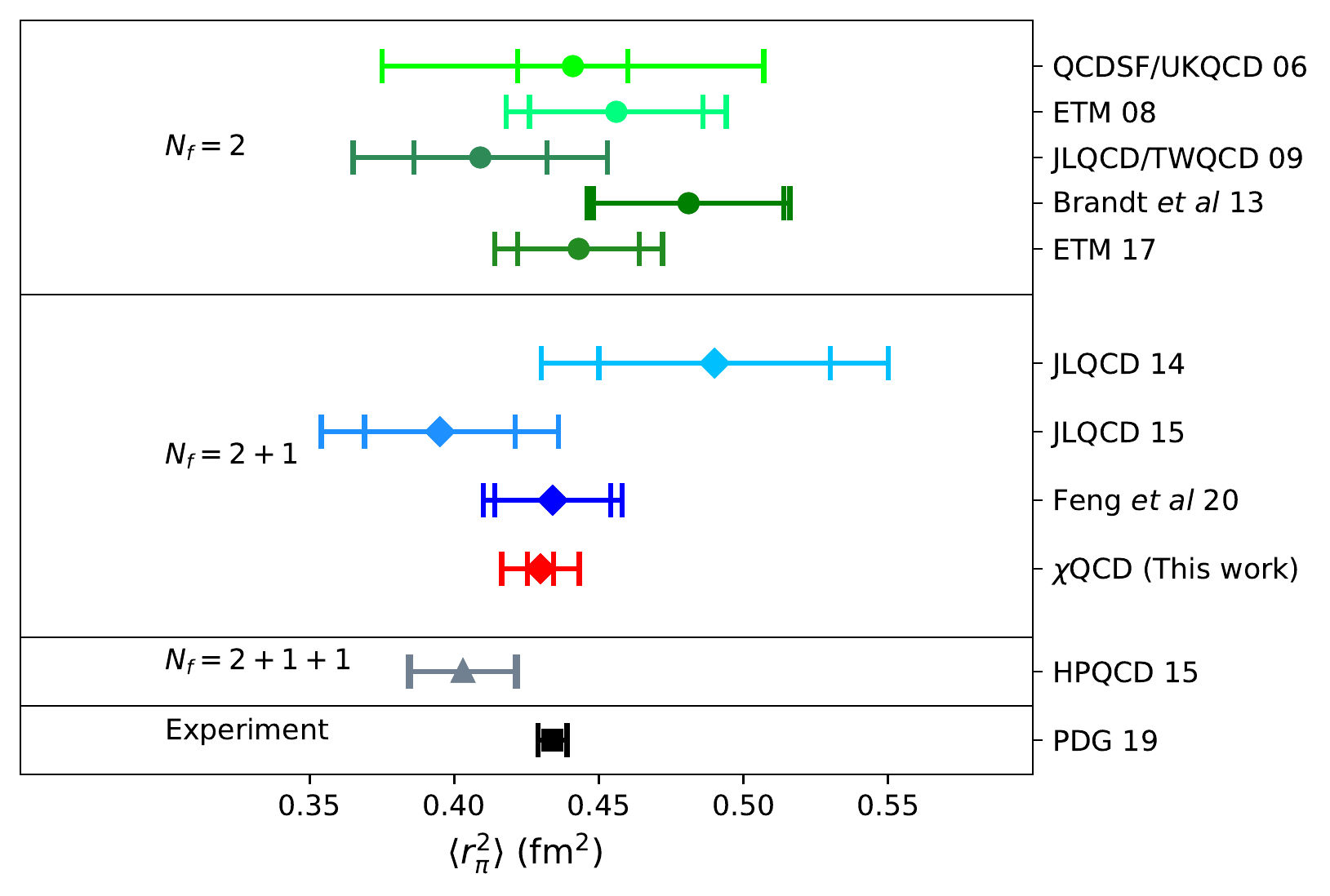}
  \caption{Summary of the pion radius results at the physical point. The lattice QCD results with different sea flavors are collected in different blocks, while all the results are consistent with each other within uncertainties. Numbers are from (QCDSF/UKQCD)~\cite{Brommel:2006ww}, (ETM)~\cite{Frezzotti:2008dr}, (JLQCD/TWQCD)~\cite{Aoki:2009qn}, (Brandt et al.)~\cite{Brandt:2013dua}, (ETM)~\cite{Alexandrou:2017blh}, (JLQCD)~\cite{Fukaya:2014jka,Aoki:2015pba}, (Feng et al.)~\cite{Feng:2019geu}, (HPQCD)~\cite{Koponen:2015tkr}, and (PDG)~\cite{Tanabashi:2018oca}.}
  \label{fig:Radius_lattice}
\end{figure}

}

{
\subsection{Chiral extrapolation of the pion form factor}

In order to make a prediction of the form factor at the continuum and infinite volume limits, we fit the {inverse} of the $f_{\pi \pi}(Q^2)$ data on different lattices with different valence pion masses, as inspired from the NLO $SU$(2) ChPT expansion~\cite{Gasser:1984ux,Bijnens:1998fm},
\begin{eqnarray}\label{eq:ff_SU2}
\begin{aligned}
&\frac{1}{f_{\pi \pi}(Q^2)} = 1 + \frac{Q^2}{6 (4 \pi F_\pi )^2} \left[ \bar{l}_6 - {\rm{ln}} \frac{M_\pi^2}{m_{\pi,{\rm phys}}^2} - 1 + R(s) \right]  \\
&\;  + \frac{Q^2 M_\pi^2}{F_\pi^4} (c_1 + c_2 \frac{Q^2}{M_\pi^2})  + {\color{black}{c_3^{I/ID}}} a^2 Q^2 + {\color{black}{c_4^{I/ID}}} a^2 Q^4 \\
&\;  + \frac{Q^2}{F_\pi^4  (M_\pi L)^{3/2}}({\color{black}{c_5}} + {\color{black}{c_6}} \frac{Q^2}{M_\pi^2}) e^{- M_\pi L}, \\
\end{aligned}
\end{eqnarray} 
in which $F$ and $\bar{l}_6$ are free parameters for fitting,
$c_1$ and $c_2$ account for possible NNLO effects, $c_3^{I/ID}$ and $c_4^{I/ID}$ reflect the lattice spacing dependence terms,
$c_5$ and $c_6$ account for the finite-volume effect,
and 
\mbox{$R(s)=\frac{2}{3} + \left(1+\frac{4}{s} \right) \left[\sqrt{1+\frac{4}{s}} {\rm{ln}} \frac{\sqrt{1+\frac{4}{s}}-1}{\sqrt{1+\frac{4}{s}}+1} + 2 \right]$}.
$F_\pi$ was defined previously with $\bar{l}_4=4.40(28)$ treated as a prior here as well.
Since the inverse of $f_{\pi \pi}(Q^2)$ is mainly dominated by the NLO contributions considering the vector dominace of the pion form factor, fitting the inverse helps avoid the need of too many low-energy constants from NNLO corrections~\cite{Alexandrou:2017blh}.
{The fit result is shown in Fig.~\ref{fig:ff_global} with the central values and correlations of the fit parameters are listed in Table~\ref{tab:pff_NLO_corr}.}
This fit (with {$\chi^2/d.o.f. = 1.0 $}) gives $\braket{r_{\pi}^2}=  {0.433(6)}\ {\rm{fm^2}}$, $F_{\pi} = 92.1(6.2)\  {\rm{MeV}}$ and $\bar{l}_6 = 16.0(1.9)$, which are consistent with the above analysis.
Our extrapolated result at the physical pion mass and continuum and infinite volume limits for the curve $f_{\pi\pi}(Q^2)$ including the systematic uncertainties from excited-state contamination, NNLO corrections, chiral extrapolation, lattice spacing and finite-volume dependence, is shown and compared with experiments in Fig.~\ref{fig:Radius_global1}; it goes through basically all the experimental data points up to $Q^2 = 1.0\, {\rm{GeV}}^2$.
Also, our results are consistent with the previous experimental analysis~\cite{Colangelo:2018mtw} and phenomenological prediction~\cite{Chen:2018rwz}.

\begin{figure}[htbp]
  \centering
  \includegraphics[width=0.45\textwidth]{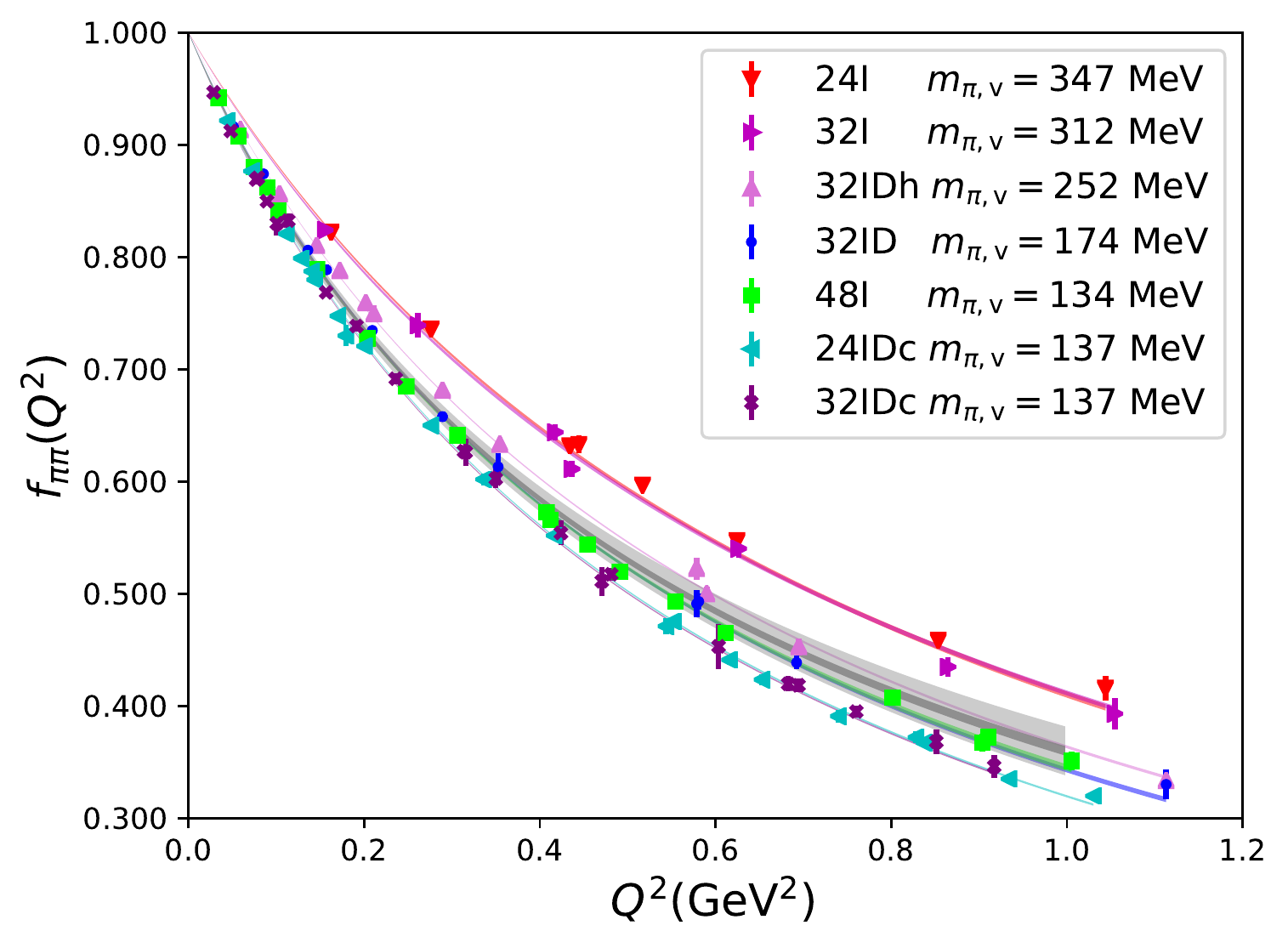}
  \caption{Pion form factor $f_{\pi \pi}(Q^2)$ on seven gauge ensembles at their unitary pion mass with the colored bands from the chiral extrapolation fit. The inner gray error band shows the fit result and statistical error extrapolated to the physical limit and the outer lighter gray band corresponds to the inclusion of the systematic uncertainties from excited-state contamination, NNLO corrections, chiral extrapolation, lattice spacing and finite-volume dependence.}
  \label{fig:ff_global}
\end{figure}

The following systematic uncertainties are included in the analysis:
\begin{itemize}
\item
{
With a variation of the fit ranges of 2pt and 3pt on 32I with pion mass $312\ {\rm MeV} $ we got the form factor at large momentum transfer $f_{\pi\pi}(Q^2 = 0.865 \ {\rm GeV^2}) = 0.4347(87)(98)$. Along with previous analysis on 32ID at small momentum transfer $f_{\pi\pi}(Q^2 = 0.051 \ {\rm GeV^2}) = 0.9158(14)(13)$, we estimate the systematic uncertainty from the excited-state contamination to be equal to the statistical uncertainty of the fitted pion form factors for all $Q^2 < 1.0 \ {\rm GeV^2}$.
}
\item
{
Since the $c_1$ and $c_2$ terms are just an estimation of the possible NNLO effects, we estimate the NNLO systematic uncertainty by setting $c_1$ and $c_2$ in Eq.~(\ref{eq:ff_SU2}) to be zero and treat the changes as the systematic uncertainty from NNLO corrections.
}
\item
{
The systematic uncertainty from the lack of a kaon log term proposed by $SU$(3) NLO ChPT is calculated with
\begin{eqnarray}\label{eq:kaon_mass}
\frac{Q^2}{12 (4 \pi F_0 )^2} \left[ {\rm{ln}} \frac{M^2_{K,{\rm{max}}}}{m^2_{K,{\rm{phys}}}} \right] ,
\end{eqnarray}
which is the difference between using $M_{K,{\rm{max}}}$ and $m_{K,{\rm{phys}}}$ in the ChPT formula. This is treated as the systematic uncertainty from chiral extrapolation.
}
\item 
{
We use the difference between the fit predictions in the continuum limit with those from the smallest lattice spacing (32I) as the systematic uncertainty of finite lattice spacing.
}
\item 
{
{The systematic uncertainty from finite-volume effects is estimated by the difference between the fit predictions 
for 24IDc with $m_\pi L \sim 3.33$ and 32IDc $m_\pi L \sim 4.45$ with both ensembles at the physical pion mass.}
}
\end{itemize}

}

\begin{figure}[htbp]
  \centering
  \includegraphics[width=0.4\textwidth]{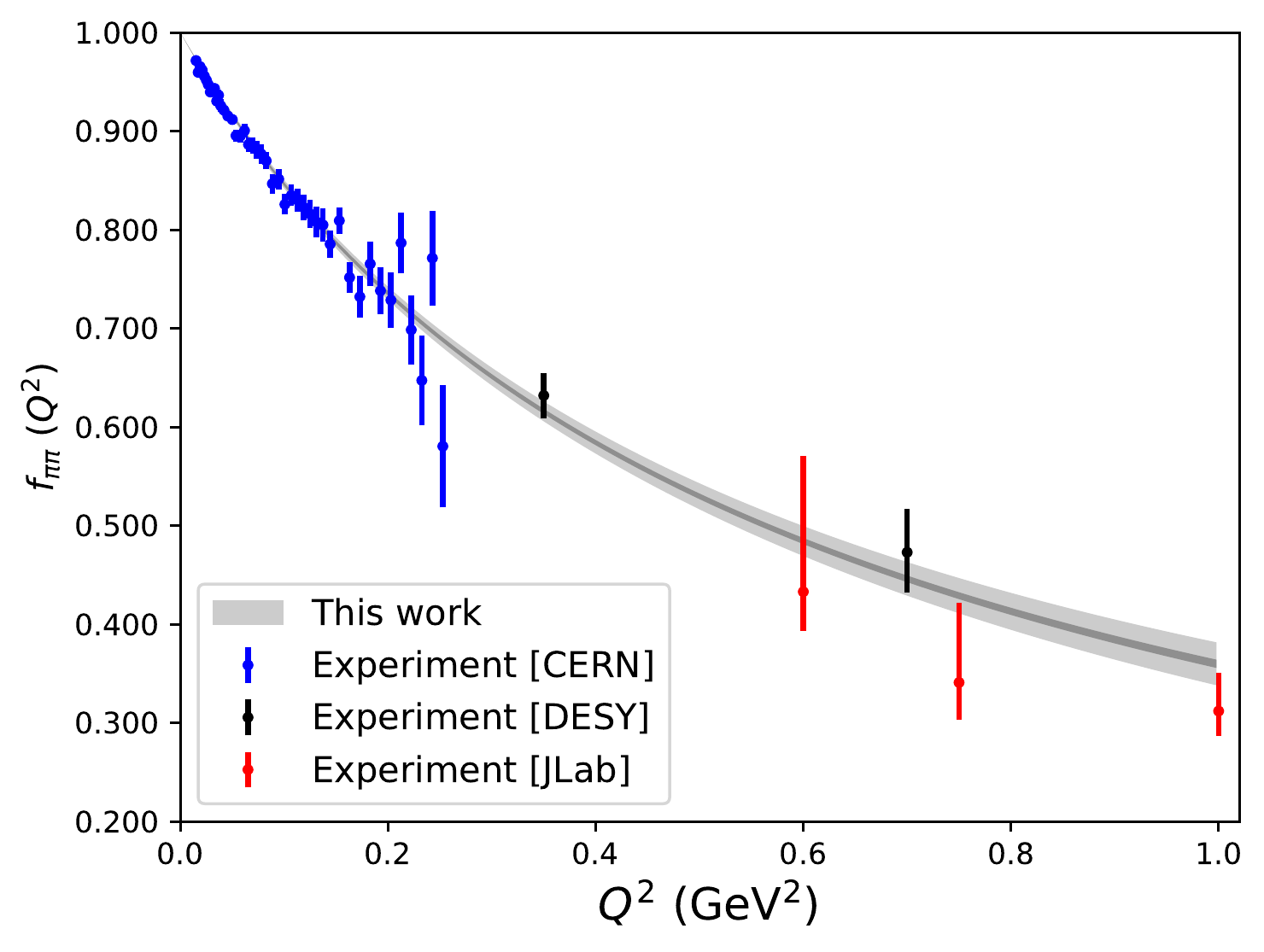}
  \caption{Comparison of the pion form factor $f_{\pi \pi}(Q^2)$ at the physical pion mass with the CERN experiment at $Q^2 < 0.25\  {\rm{GeV^2}}$~\cite{Amendolia:1986wj} and the JLab and DESY experiment data at larger $Q^2$~\cite{Huber:2008id,Blok:2008jy,Horn:2007ug,Horn:2006tm,Volmer:2000ek}. The inner gray band is the statistical error and the outer band includes the systematic uncertainties.}
  \label{fig:Radius_global1}
\end{figure}

}

{
\section{Summary}\label{sec:summary}
We have presented a calculation of the pion form factor using overlap fermions with a range of valence pion masses on seven RBC/UKQCD domain-wall ensembles including two which have the physical pion mass.
The lattice results for $\braket{r_{\pi}^2}$ in the continuum and infinite volume limits are compiled in Fig.~\ref{fig:Radius_lattice} together with that of experiment. 
Our globally fitted pion mean square charge radius is $\braket{r^2_\pi} = 0.430(5)({13}) \ {\rm{fm^2}}$, which includes systematic errors from chiral extrapolation, finite lattice spacing, finite volume, and others; it agrees with experimental value of $\braket{r^2_\pi} = 0.434(5) \ {\rm{fm^2}}$ within one sigma.  

We find that $\braket{r_{\pi}^2}$ has a strong dependence on both the valence and sea pion masses. More precisely, it depends majorly on the mass of the pion with one valence quark and one sea quark.
{A good fit of the chiral log term confirms that the pion radius diverges in the chiral limit.}
We also give the extrapolated form factor $f_{\pi \pi}(Q^2)$, and the result agrees well with the experimental data points (up to $Q^2 = 1.0 \  {\rm{GeV^2}} $).

Thus this work shows that the hadron form factor and the corresponding radius can be studied accurately and efficiently by combining LMS with the multi-mass algorithm of overlap fermions and FFT on the stochastic-sandwich method.
This raises the expectation of an efficacious investigation of the form factor of the nucleon and its pion-mass dependence with relatively small overhead on multiple quark masses and momentum transfers.
{Note that for an accurate prediction of the charge radius and form factor with 1\% overall uncertainty, calculations at smaller lattice spacing and larger source-sink separation will be essential, together with the QED and isospin breaking corrections.}

}

\begin{acknowledgments}
We thank the RBC/UKQCD Collaborations for providing their domain-wall gauge configurations and also thank L.-C. Jin and R. J. Hill for constructive discussions.
This work is supported in part by the U.S. DOE Grant No.~DE-SC0013065 and DOE Grant No.~DE-AC05-06OR23177 which is within the framework of the TMD Topical Collaboration.
Y.Y is supported by the Strategic Priority Research Program of Chinese Academy of Sciences, Grant No.~XDC01040100 and XDB34030300.
J. L. is supported by the Science and Technology Program of Guangzhou (No.~\!2019050001).
This research used resources of the Oak Ridge Leadership Computing Facility at the Oak Ridge National Laboratory, which is supported by the Office of Science of the U.S. Department of Energy under Contract No.~\!DE-AC05-00OR22725. This work used Stampede time under the Extreme Science and Engineering Discovery Environment (XSEDE), which is supported by National Science Foundation Grant No.~\!ACI-1053575.
We also thank the National Energy Research Scientific Computing Center (NERSC) for providing HPC resources that have contributed to the research results reported within this paper.
We acknowledge the facilities of the USQCD Collaboration used for this research in part, which are funded by the Office of Science of the U.S. Department of Energy.
\end{acknowledgments}

\appendix
{
\section{{Autocorrelation of measurements}}
We have chosen a set of evenly separated configurations for measurement from the full Monte Carlo evolutions available for each ensemble.
The separations are 40, 32, 10, 8, 8, 10, 32 for 24I, 32I, 48I, 24IDc, 32IDc, 32ID, 32IDh, respectively.
For the error analysis, we treat measurements from different configurations as independent and average the measurements over each configuration before analysis.

In the left panels of Fig.~\ref{fig:auto_corr_a}, we have plotted the integrated autocorrelation time on 24IDc and 48I, defined as,
\begin{eqnarray}
\begin{aligned}
&\tau_{\rm int}(\Delta_{\rm cut}) = \frac{1}{2} + \sum_{\Delta = 1}^{\Delta_{\rm cut}} C(\Delta),  \\
&C(\Delta) = \left\langle \frac{(Y(t)-\bar{Y}) (Y(t+\Delta) - \bar{Y})}{\sigma^2} \right\rangle_t.
\end{aligned}
\end{eqnarray} 
The error of $C(\Delta)$ is estimated by jackknife re-sampling of the average on $t$ and the error of the integrated autocorrelation time is estimated with simple error propagation.
The plot shows {the three-point functions with current position $\tau = t_{\textrm{f}}/2$} for the valence pion mass $137 \ {\rm{MeV}}$ and $148 \ {\rm{MeV}}$ at the smallest separation $C_{\rm 3pt}(t_{\textrm{f}}/2,t_{\textrm{f}},\vec{p}_{\textrm{i}}=0,\vec{p}_{\textrm{f}}=0)$ on 24IDc and 48I, respectively.
The integrated autocorrelation times are less than 1 within uncertainty for both ensembles which confirms the independence of the measurements on each of the configurations.

In the right panels of Fig.~\ref{fig:auto_corr_a}, we plot the central values and errors of $C_{\rm 3pt}(t_{\textrm{f}}/2,t_{\textrm{f}},\vec{p}_{\textrm{i}}=0,\vec{p}_{\textrm{f}}=0)$ on 24IDc and 48I as a function of binning size $t_{\rm bin}$. 
The statistical errors change very little as the bin size is increased for both ensembles which again confirms the measurements' independence.

}

{
\section{Dispersion relations}\label{Appendix:dispersion}

\begin{figure}[htbp]
\centering
\includegraphics[scale=0.45]{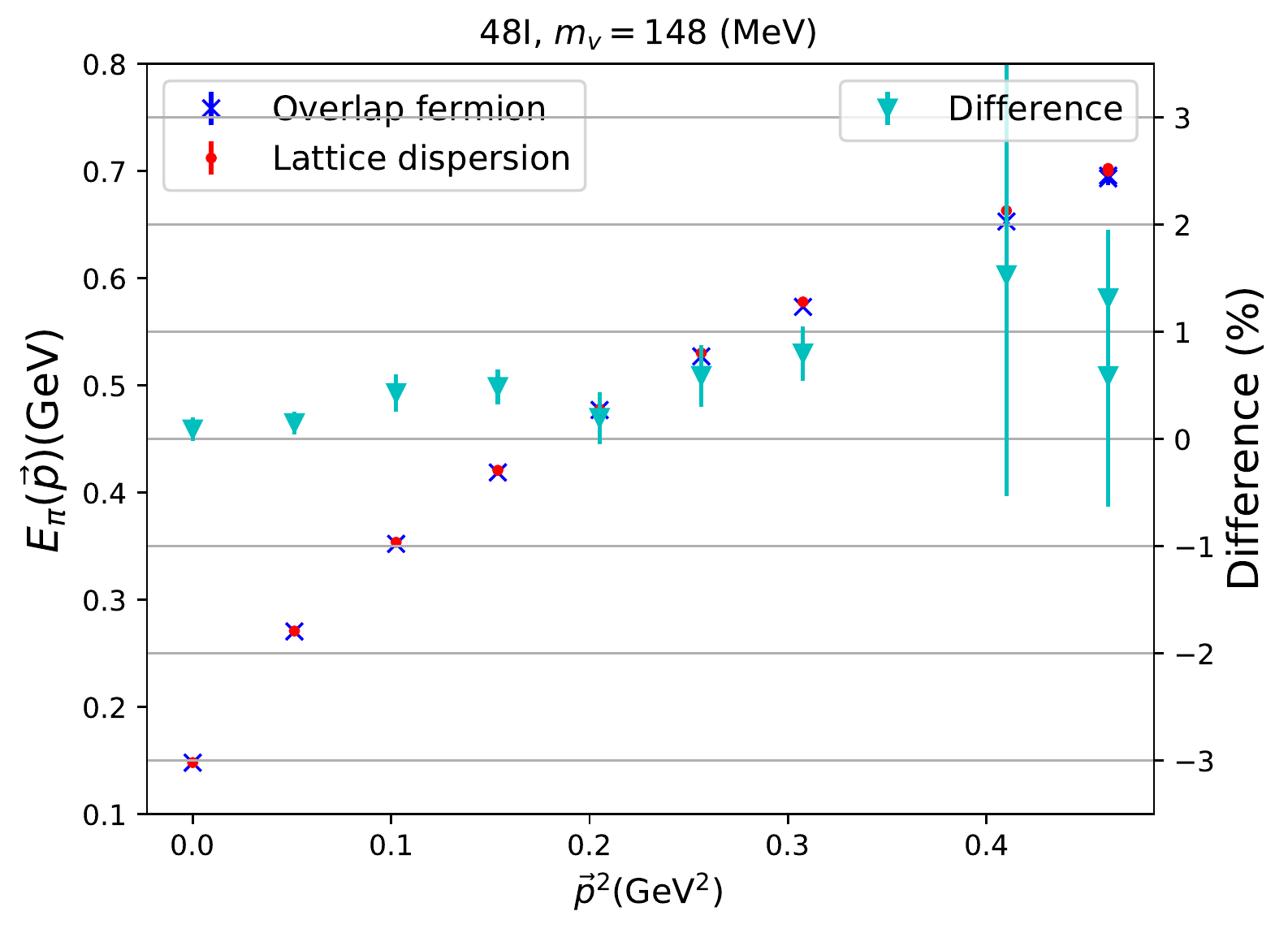}
\caption{Example plot of the pion energies as a function of $\vec{p}^2$ on 48I with pion mass $148\ {\rm{MeV}}$. The blue points correspond to the pion energies $E_{\pi}(\vec{p})$ from 2pt function fits and the red points are calculated with Eq.~\ref{Eq:app_dis} using $E_{\pi}(\vec{0})$. Their percentage differences are also shown with cyan points with the scale on the right.}
\label{fig:dia_Dis2pt}
\end{figure}

{
In Fig.~\ref{fig:dia_Dis2pt},
we compare the pion energy $E(\vec{p})$ obtained from the fitting of 2pts to the lattice dispersion relation
\begin{eqnarray}\label{Eq:app_dis}
\begin{aligned}
\hat{E}^2 = \hat{m}^2 + \sum_i {\hat{p}}_i^2
\end{aligned}
\end{eqnarray} 
where $a \hat{E} = 2 {\rm{sinh}}(a E /2)$, $a \hat{m} = 2 {\rm{sinh}}(a m /2)$ and $a \hat{p}_i = 2 {\rm{sin}}(ap_i/2)$.
As can be seen, the dispersion relation is well satisfied under the $1\%$ level for the momenta considered in this paper.
}

\begin{figure}[htbp]
\centering
\includegraphics[scale=0.45]{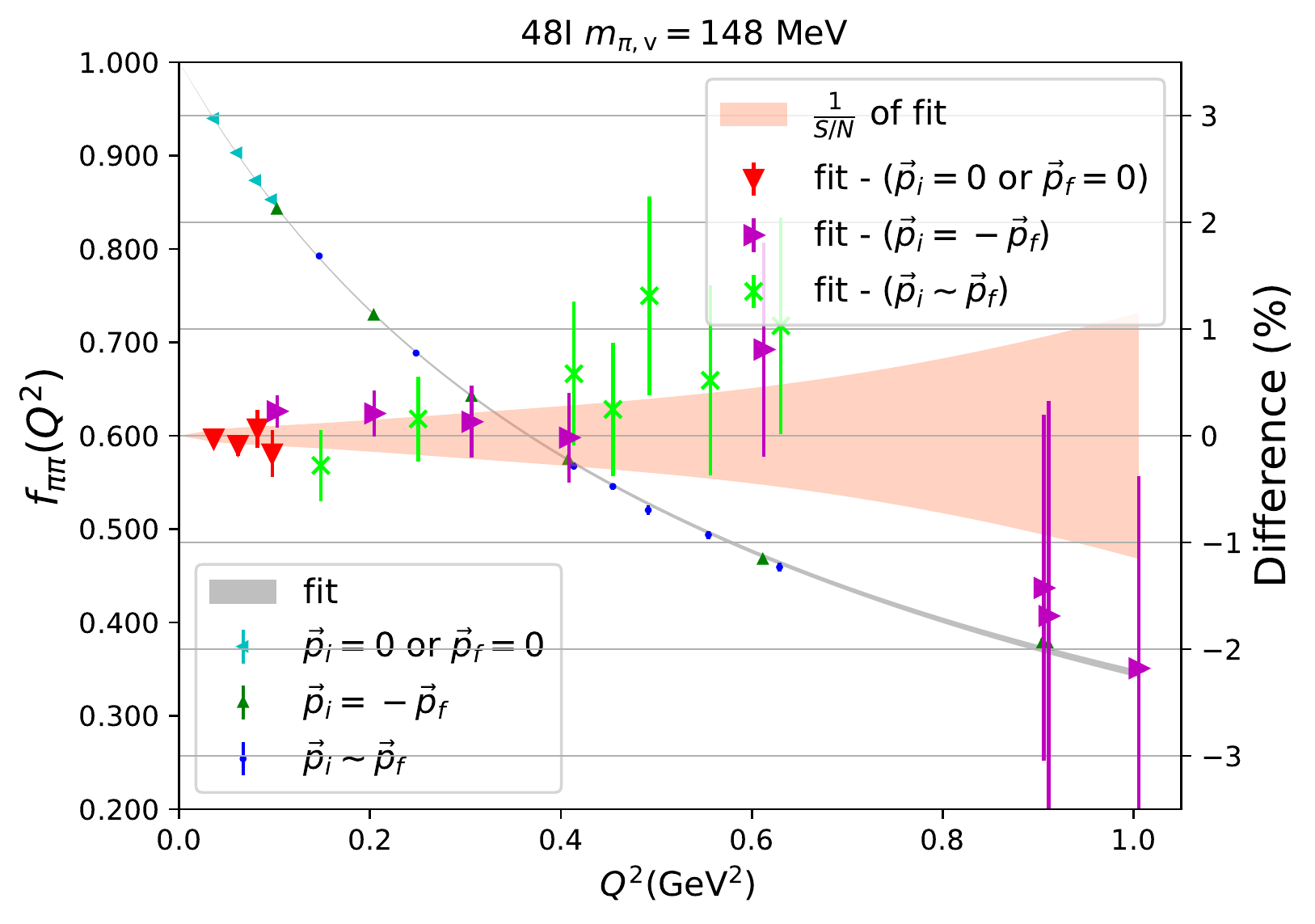}
\caption{Example plot of the pion form factor on 48I with pion mass $148\ {\rm{MeV}}$. The cyan, green and blue points correspond to the $f_{\pi \pi}(Q^2)$ from different $\vec{p}_{\rm{i}}$ and $\vec{p}_{\rm{f}}$ combinations. The gray band is the $z$-expansion fit result. The percentage differences between the data and corresponding fit results are also plotted with red, magenta and lime colors with the scale on the right. As a reference, the inverse of signal-to-noise ratio of the fit result is displayed as percentage with the coral region.}
\label{fig:dia_Dis3pt}
\end{figure}

{In Fig.~\ref{fig:dia_Dis3pt},
we have plotted the pion form factors on 48I with different $\vec{p}_{\rm{i}}$ and $\vec{p}_{\rm{f}}$ cases marked with different colors.
The values for different cases overlap with each other quite well at similar $Q^2$ at the $1\%$ level.
This confirms that the combination of $\vec{p}_{\rm{i}}$ and $\vec{p}_{\rm{f}}$ considered in this paper are consistent with each other which will lead to a well defined physical limit.}

}

{
\section{Normalization of the local vector current for overlap fermions}

The left panel of Fig.~\ref{fig:ZV_32ID_b} shows the determination of the normalization constant $Z_V$ on 32ID by fitting 
the inverse of the forward matrix element as $\frac{2 E}{\bra{\pi(p)} V_4 \ket{\pi(p)}}$ with $\vec{p}=0$.
The data points from different source-sink separations
overlap well with each other under the 0.1\% level, so we have done a simple linear fit with {$\chi^2/d.o.f.\sim 0.7$}.

As we are using overlap fermions which have exact chiral symmetry on the lattice,
the axial normalization (finite renormalization) constant is equal to the local vector current normalization constant, as confirmed in~\cite{Liu:2013yxz}.
The axial normalization constant on 32ID was calculated in~\cite{Liang:2018pis} from the Ward identity: $Z_A = \frac{2 m_q \bra{0} P \ket{\pi}}{m_\pi \bra{0} A_4 \ket{\pi}}$ with $P$ and $A_4$ the pseudo-scalar quark bilinear operator and the temporal component of the axial-vector operator, respectively.
As shown in the right panel of Fig.~\ref{fig:ZV_32ID_b}, the axial normalization constant agrees well with the local vector current normalization constant used in this paper very well at the massless limit.

}

\begin{widetext}

\begin{figure}[htbp]
  \centering
  \includegraphics[scale=0.45]{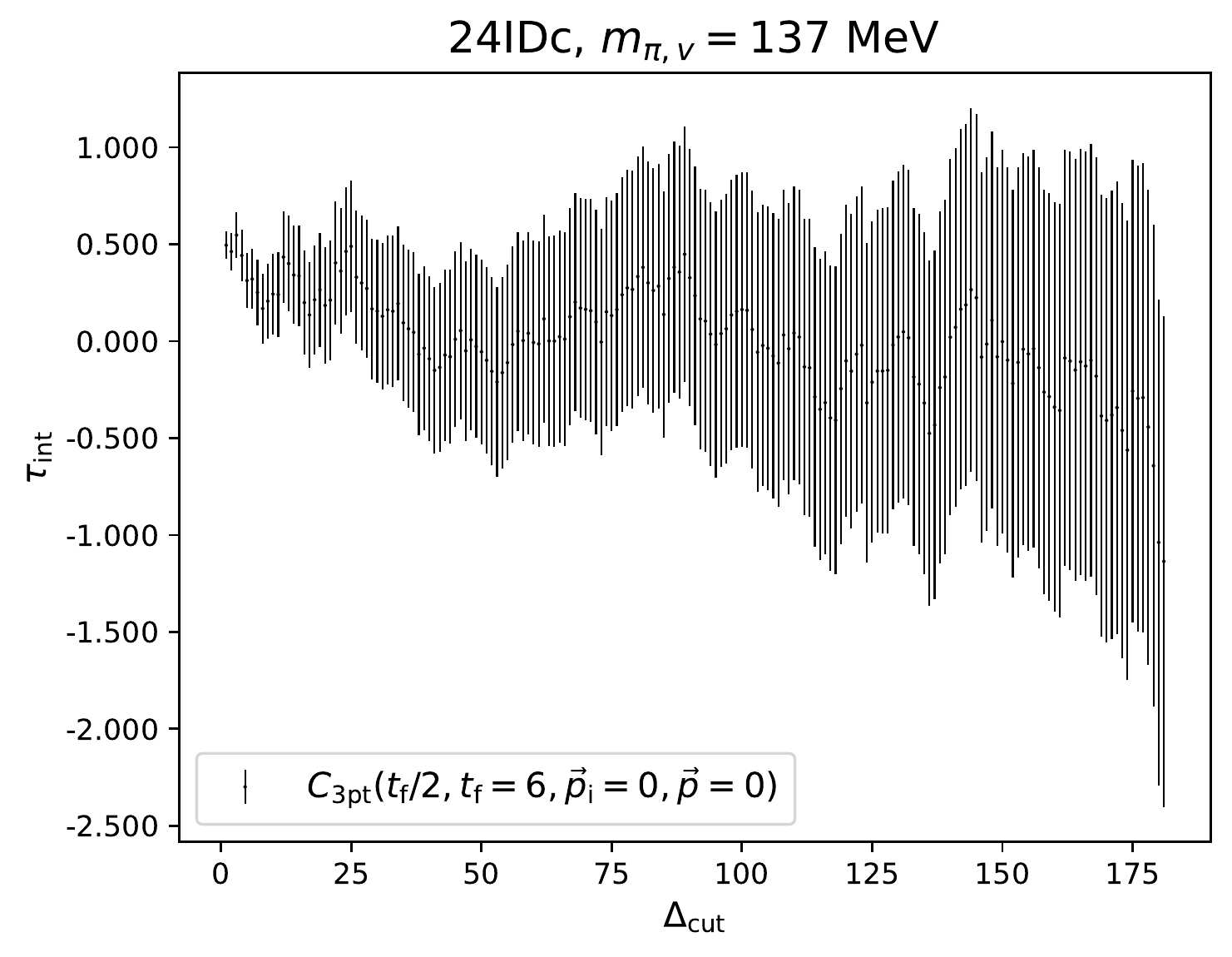}
  \includegraphics[scale=0.45]{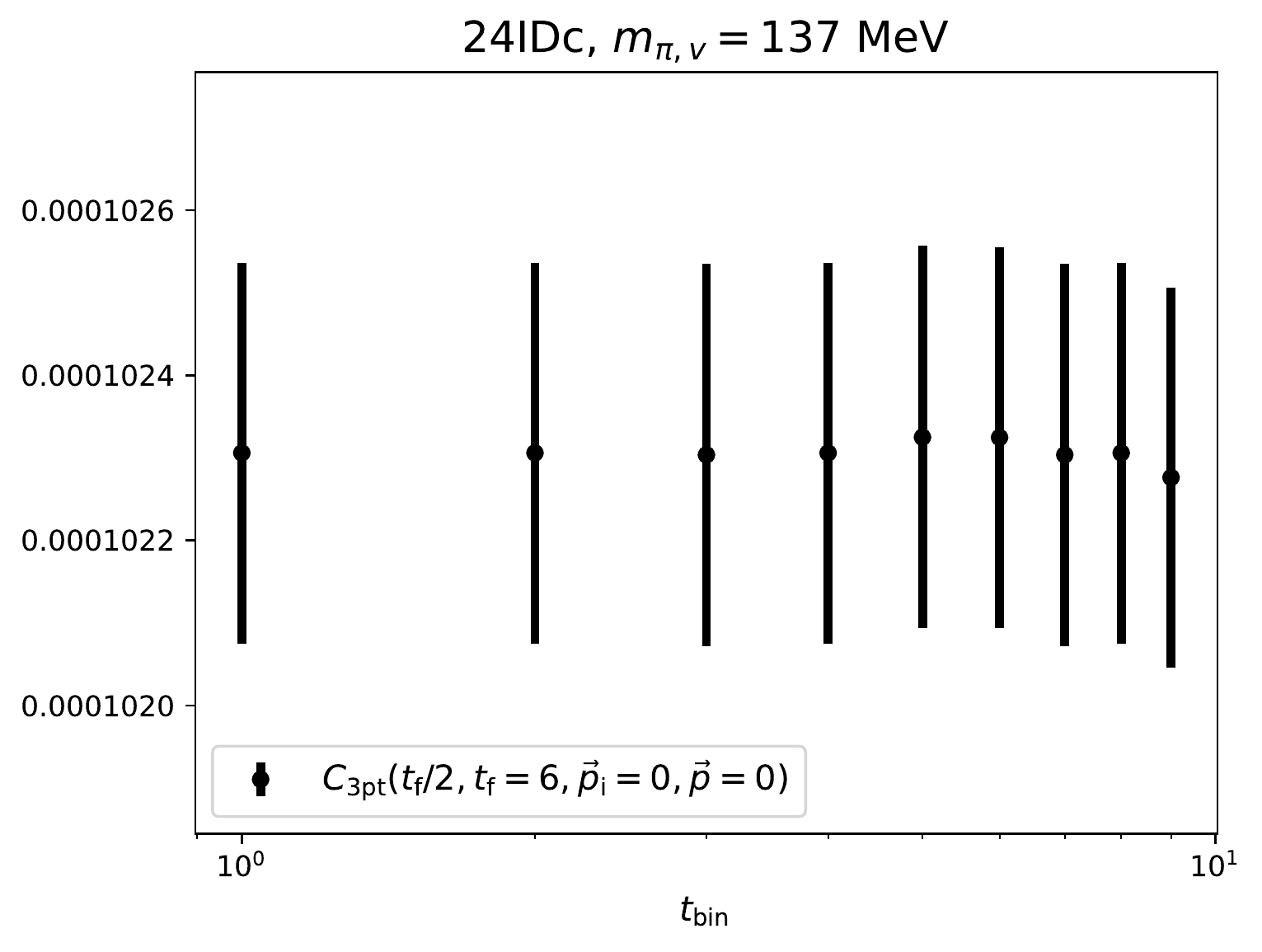}   \\
  \includegraphics[scale=0.45]{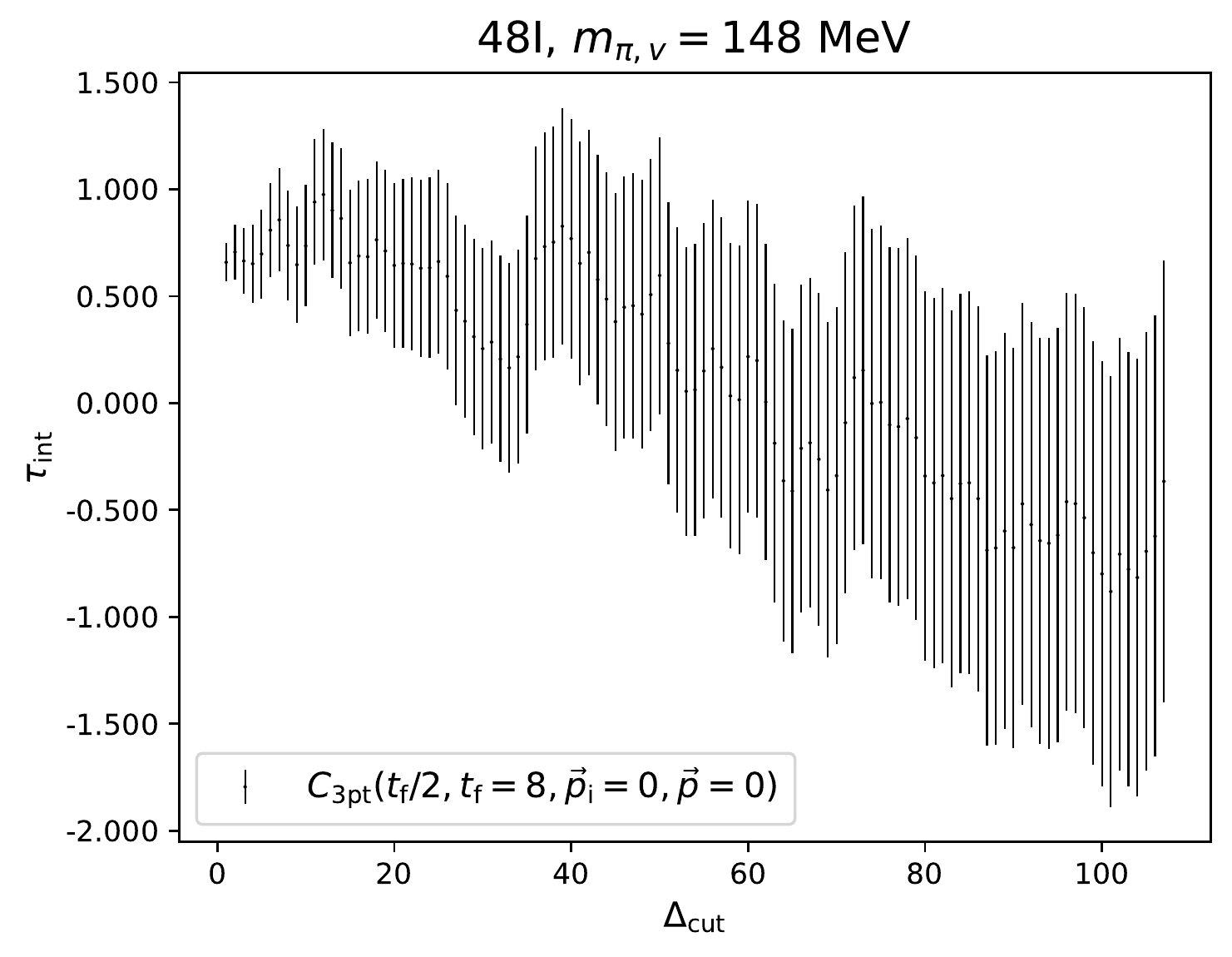}
  \includegraphics[scale=0.45]{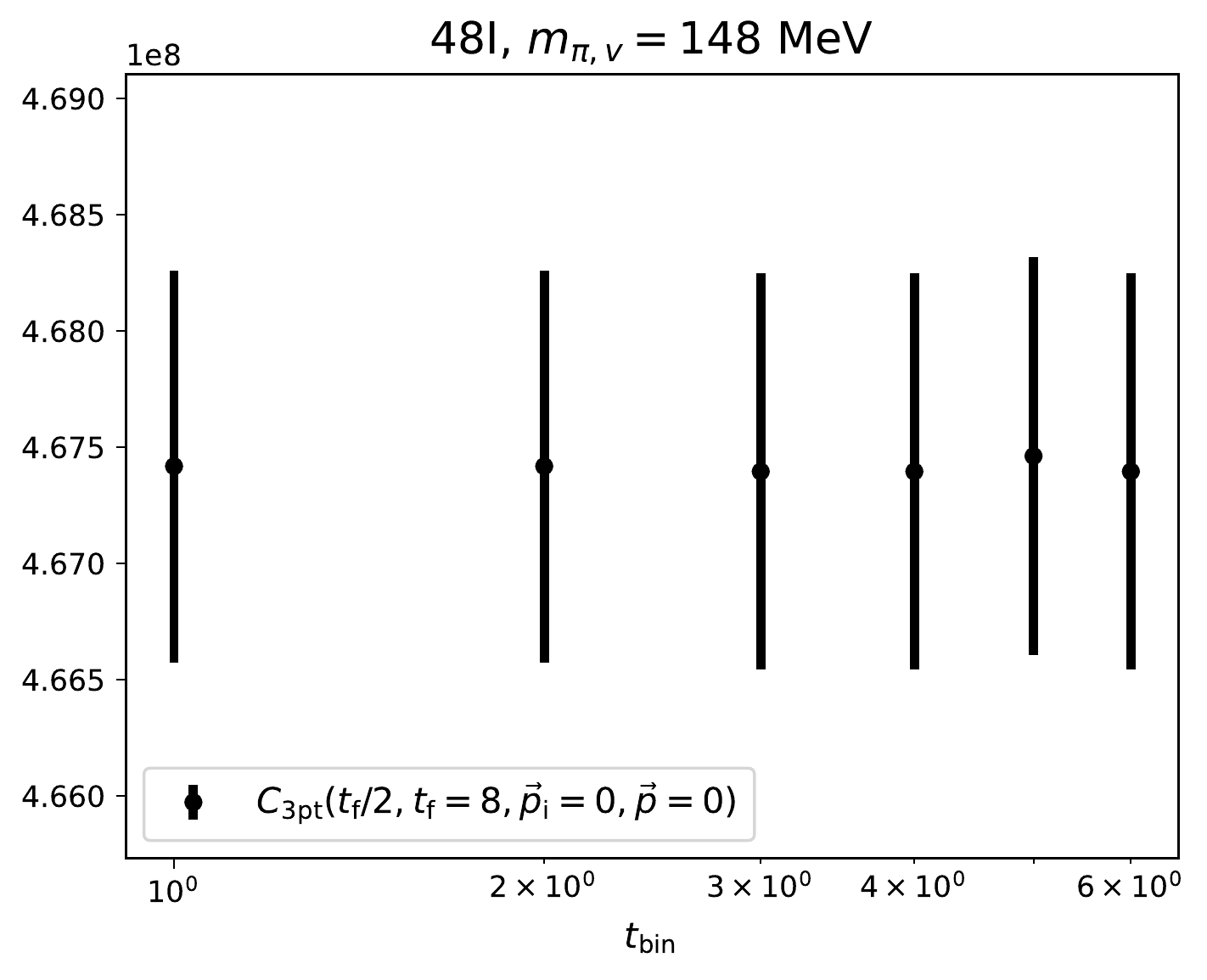}
  \caption{Plots in the left panels show the integrated autocorrelation of $C_{\rm 3pt}(t_{\textrm{f}}/2,t_{\textrm{f}},\vec{p}_{\textrm{i}}=0,\vec{p}_{\textrm{f}}=0)$ on 24IDc (top) and 48I (bottom). Plots in the right panels show the binning tests for the same quantity.}
  \label{fig:auto_corr_a}
\end{figure}

\begin{figure}[htbp]
  \centering
  \includegraphics[scale=0.45]{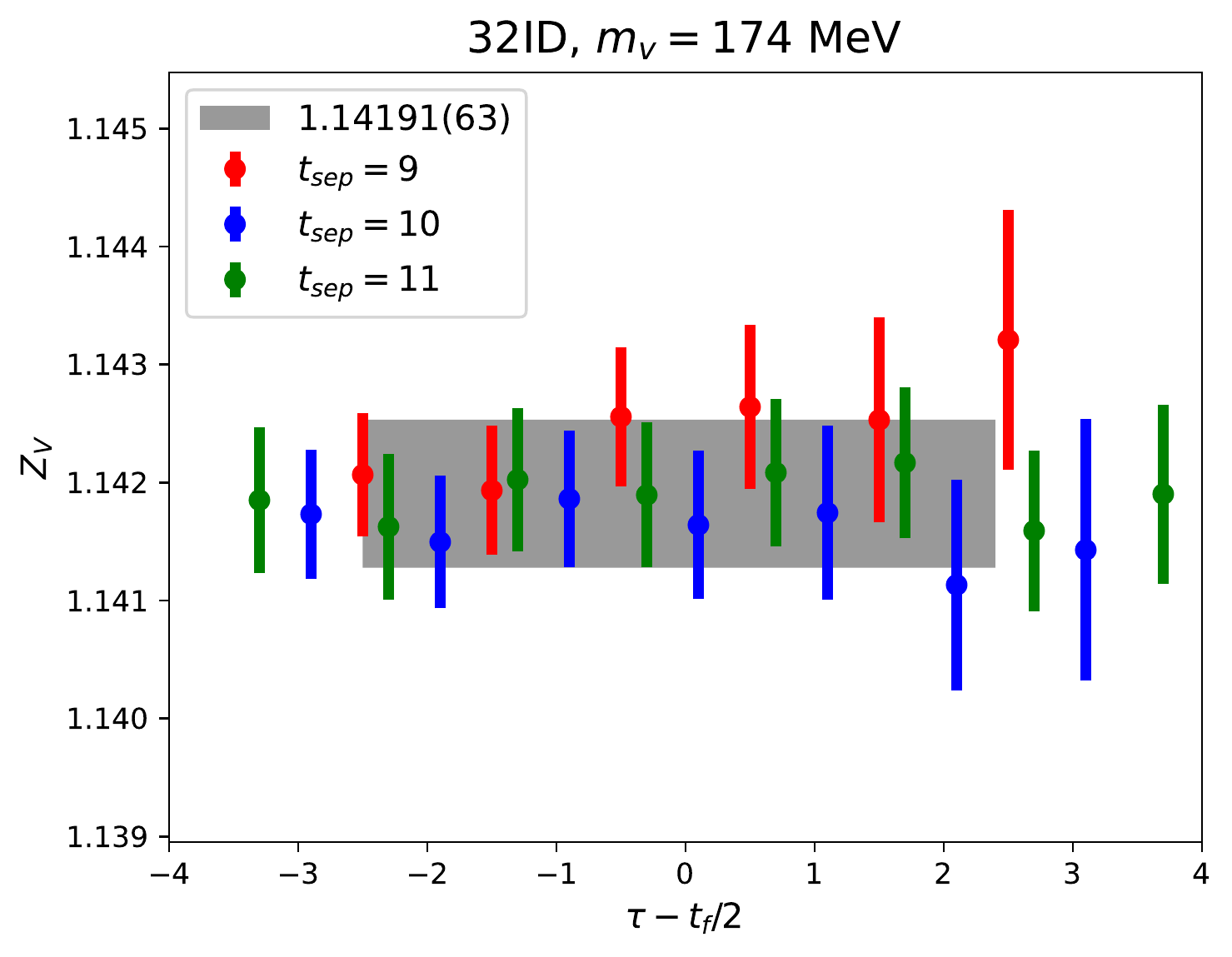}
  \includegraphics[scale=0.45]{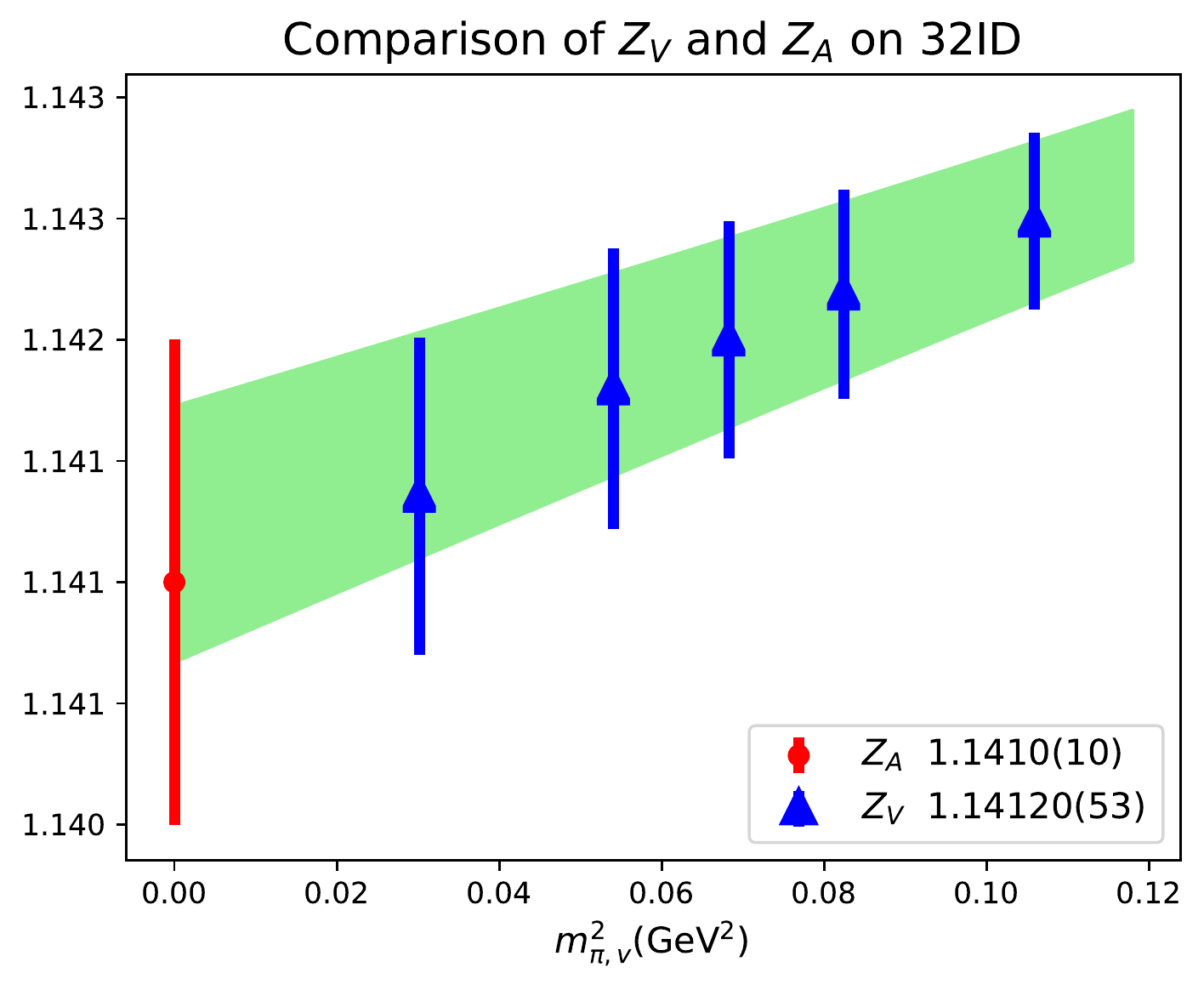}
  \caption{The left panel is an example plot of the normalization constant for the local vector determined from 
the forward matrix element as $Z_V \equiv \frac{2 E}{\bra{\pi(p)} V_4 \ket{\pi(p)}}$ with $\vec{p}=0$ on 32ID.
The gray band is a constant fit of the data points at various values of source-sink separation.
The right panel is the comparison of the renormalization of the axial vector current from the Ward identity and the local vector current normalization constant on 32ID.}
  \label{fig:ZV_32ID_b}
\end{figure}

{
\section{Correlations of fit parameters}

\begin{table}[H]
  \centering{
  \begin{tabular}{| c | c | c | c | c |}
    \hline
              & $F$ & $\bar{l}_6$ & $b_2^{ID}$  & $\bar{l}_4$    \\  \hline
Central value & 0.0908(43)   & 17.1(1.4)   & 0.0510(27)   & 4.44(26)  \\ \hline
Correlation & $F$ & $\bar{l}_6$ & $b_2^{ID}$ & $\bar{l}_4$   \\
     $F$       & 1.86e-05   & 5.90e-03   & -4.11e-07   & 9.19e-04    \\ \hline 
   $\bar{l}_6$ & 5.90e-03   & 1.91e+00   & -3.98e-04   & 3.16e-01    \\ \hline 
  $b_2^{ID}$   & -4.11e-07   & -3.98e-04   & 7.55e-06   & -6.91e-05    \\ \hline 
   $\bar{l}_4$ & 9.19e-04   & 3.16e-01   & -6.91e-05   & 6.91e-02    \\ \hline 
  \end{tabular}
  \caption{The central values and correlations of the fit parameters $F$, $\bar{l}_6$, $b_2^{ID}$ and $\bar{l}_4$ in Eq.~(\ref{eq:global_fit}).}
  \label{tab:radius_corr}
  }
\end{table}

\begin{table}[H]
 \tiny
  \centering{
  \begin{tabular}{| c | c | c | c | c | c | c | c | c | c | c | c |}
    \hline
  &  $F$ &    $\bar{l}_6$   &   $c_1$   &   $c_2$   &  $c_3^{ID}$  &   $c_3^{I}$  &  $c_4^{ID}$   & $c_4^{I}$  & $c_5$   &   $c_6$  & $\bar{l}_4$ \\  \hline
Central value & 0.0865(65)   &  16.0(1.9)   & -0.000032(56)   & 9.3(2.8)e-06   & 0.0587(33)   &  0.042(11)   &  0.276(30)   &   0.31(10)   & -0.00041(10)   & 7.2(6.9)e-06   &   4.45(27)  \\  \hline
Correlation   &  $F$ &    $\bar{l6b}$   &   $c_1$   &   $c_2$   &  $c_3^{ID}$  &   $c_3^{I}$  &  $c_4^{ID}$   & $c_4^{I}$  & $c_5$   &   $c_6$  & $\bar{l}_4$  \\  \hline
  $F$         & 4.22e-05   & 1.26e-02   & -3.21e-07   & 8.03e-09   & -1.02e-06   & 5.26e-06   & -3.54e-06   & 1.20e-05   & -1.10e-07   & 4.93e-10   & 1.34e-04    \\ \hline
  $\bar{l}_6$ & 1.26e-02   & 3.79e+00   & -9.51e-05   & 2.21e-06   & -6.96e-04   & 4.01e-04   & 1.32e-03   & 1.05e-02   & -3.88e-05   & 2.49e-07   & 5.65e-02    \\ \hline
  $c_1$       & -3.21e-07   & -9.51e-05   & 3.12e-09   & -5.35e-11   & 1.34e-08   & -2.89e-08   & -1.83e-08   & -2.83e-07   & 7.17e-10   & -8.98e-12   & 5.71e-06    \\ \hline
  $c_2$       & 8.03e-09   & 2.21e-06   & -5.35e-11   & 7.91e-12   & 4.65e-09   & 1.51e-08   & -7.43e-08   & -2.17e-07   & 7.54e-12   & -3.03e-12   & 6.42e-08    \\ \hline
  $c_3^{ID}$  & -1.02e-06   & -6.96e-04   & 1.34e-08   & 4.65e-09   & 1.06e-05   & 2.93e-05   & -5.79e-05   & -1.65e-04   & 5.39e-08   & -2.87e-09   & 1.27e-05    \\ \hline
  $c_3^{I}$   & 5.26e-06   & 4.01e-04   & -2.89e-08   & 1.51e-08   & 2.93e-05   & 1.14e-04   & -1.65e-04   & -6.43e-04   & 1.80e-09   & -1.00e-09   & -2.70e-05    \\ \hline
  $c_4^{ID}$  & -3.54e-06   & 1.32e-03   & -1.83e-08   & -7.43e-08   & -5.79e-05   & -1.65e-04   & 8.88e-04   & 2.55e-03   & -3.35e-07   & 3.12e-08   & 7.27e-05    \\ \hline
  $c_4^{I}$   & 1.20e-05   & 1.05e-02   & -2.83e-07   & -2.17e-07   & -1.65e-04   & -6.43e-04   & 2.55e-03   & 1.08e-02   & 8.64e-08   & -4.55e-08   & -1.39e-04    \\ \hline
  $c_5$       & -1.10e-07   & -3.88e-05   & 7.17e-10   & 7.54e-12   & 5.39e-08   & 1.80e-09   & -3.35e-07   & 8.64e-08   & 1.05e-08   & -3.66e-10   & -5.88e-06    \\ \hline
  $c_6$       & 4.93e-10   & 2.49e-07   & -8.98e-12   & -3.03e-12   & -2.87e-09   & -1.00e-09   & 3.12e-08   & -4.55e-08   & -3.66e-10   & 4.79e-11   & -5.50e-08    \\ \hline
  $\bar{l}_4$ & 1.34e-04   & 5.65e-02   & 5.71e-06   & 6.42e-08   & 1.27e-05   & -2.70e-05   & 7.27e-05   & -1.39e-04   & -5.88e-06   & -5.50e-08   & 7.54e-02    \\ \hline
  \end{tabular}
  \caption{The central values and correlations of the fit parameters in Eq.~(\ref{eq:ff_SU2}).}
  \label{tab:pff_NLO_corr}
  }
\end{table}

}

\end{widetext}

\bibliography{./Pion_Form_Factor.bib}

\begin{thebibliography}{63}%
\makeatletter
\providecommand \@ifxundefined [1]{%
 \@ifx{#1\undefined}
}%
\providecommand \@ifnum [1]{%
 \ifnum #1\expandafter \@firstoftwo
 \else \expandafter \@secondoftwo
 \fi
}%
\providecommand \@ifx [1]{%
 \ifx #1\expandafter \@firstoftwo
 \else \expandafter \@secondoftwo
 \fi
}%
\providecommand \natexlab [1]{#1}%
\providecommand \enquote  [1]{``#1''}%
\providecommand \bibnamefont  [1]{#1}%
\providecommand \bibfnamefont [1]{#1}%
\providecommand \citenamefont [1]{#1}%
\providecommand \href@noop [0]{\@secondoftwo}%
\providecommand \href [0]{\begingroup \@sanitize@url \@href}%
\providecommand \@href[1]{\@@startlink{#1}\@@href}%
\providecommand \@@href[1]{\endgroup#1\@@endlink}%
\providecommand \@sanitize@url [0]{\catcode `\\12\catcode `\$12\catcode
  `\&12\catcode `\#12\catcode `\^12\catcode `\_12\catcode `\%12\relax}%
\providecommand \@@startlink[1]{}%
\providecommand \@@endlink[0]{}%
\providecommand \url  [0]{\begingroup\@sanitize@url \@url }%
\providecommand \@url [1]{\endgroup\@href {#1}{\urlprefix }}%
\providecommand \urlprefix  [0]{URL }%
\providecommand \Eprint [0]{\href }%
\providecommand \doibase [0]{http://dx.doi.org/}%
\providecommand \selectlanguage [0]{\@gobble}%
\providecommand \bibinfo  [0]{\@secondoftwo}%
\providecommand \bibfield  [0]{\@secondoftwo}%
\providecommand \translation [1]{[#1]}%
\providecommand \BibitemOpen [0]{}%
\providecommand \bibitemStop [0]{}%
\providecommand \bibitemNoStop [0]{.\EOS\space}%
\providecommand \EOS [0]{\spacefactor3000\relax}%
\providecommand \BibitemShut  [1]{\csname bibitem#1\endcsname}%
\let\auto@bib@innerbib\@empty
\bibitem [{\citenamefont {Dally}\ \emph {et~al.}(1982)\citenamefont {Dally}
  \emph {et~al.}}]{Dally:1982zk}%
  \BibitemOpen
  \bibfield  {author} {\bibinfo {author} {\bibfnamefont {E.}~\bibnamefont
  {Dally}} \emph {et~al.},\ }\href {\doibase 10.1103/PhysRevLett.48.375}
  {\bibfield  {journal} {\bibinfo  {journal} {Phys. Rev. Lett.}\ }\textbf
  {\bibinfo {volume} {48}},\ \bibinfo {pages} {375} (\bibinfo {year}
  {1982})}\BibitemShut {NoStop}%
\bibitem [{\citenamefont {Amendolia}\ \emph {et~al.}(1986)\citenamefont
  {Amendolia} \emph {et~al.}}]{Amendolia:1986wj}%
  \BibitemOpen
  \bibfield  {author} {\bibinfo {author} {\bibfnamefont {S.}~\bibnamefont
  {Amendolia}} \emph {et~al.} (\bibinfo {collaboration} {NA7}),\ }\href
  {\doibase 10.1016/0550-3213(86)90437-2} {\bibfield  {journal} {\bibinfo
  {journal} {Nucl. Phys. B}\ }\textbf {\bibinfo {volume} {277}},\ \bibinfo
  {pages} {168} (\bibinfo {year} {1986})}\BibitemShut {NoStop}%
\bibitem [{\citenamefont {Gough~Eschrich}\ \emph {et~al.}(2001)\citenamefont
  {Gough~Eschrich} \emph {et~al.}}]{GoughEschrich:2001ji}%
  \BibitemOpen
  \bibfield  {author} {\bibinfo {author} {\bibfnamefont {I.~M.}\ \bibnamefont
  {Gough~Eschrich}} \emph {et~al.} (\bibinfo {collaboration} {SELEX}),\ }\href
  {\doibase 10.1016/S0370-2693(01)01285-0} {\bibfield  {journal} {\bibinfo
  {journal} {Phys. Lett. B}\ }\textbf {\bibinfo {volume} {522}},\ \bibinfo
  {pages} {233} (\bibinfo {year} {2001})},\ \Eprint
  {http://arxiv.org/abs/hep-ex/0106053} {arXiv:hep-ex/0106053} \BibitemShut
  {NoStop}%
\bibitem [{\citenamefont {Ananthanarayan}\ \emph {et~al.}(2017)\citenamefont
  {Ananthanarayan}, \citenamefont {Caprini},\ and\ \citenamefont
  {Das}}]{Ananthanarayan:2017efc}%
  \BibitemOpen
  \bibfield  {author} {\bibinfo {author} {\bibfnamefont {B.}~\bibnamefont
  {Ananthanarayan}}, \bibinfo {author} {\bibfnamefont {I.}~\bibnamefont
  {Caprini}}, \ and\ \bibinfo {author} {\bibfnamefont {D.}~\bibnamefont
  {Das}},\ }\href {\doibase 10.1103/PhysRevLett.119.132002} {\bibfield
  {journal} {\bibinfo  {journal} {Phys. Rev. Lett.}\ }\textbf {\bibinfo
  {volume} {119}},\ \bibinfo {pages} {132002} (\bibinfo {year} {2017})},\
  \Eprint {http://arxiv.org/abs/1706.04020} {arXiv:1706.04020 [hep-ph]}
  \BibitemShut {NoStop}%
\bibitem [{\citenamefont {Colangelo}\ \emph {et~al.}(2019)\citenamefont
  {Colangelo}, \citenamefont {Hoferichter},\ and\ \citenamefont
  {Stoffer}}]{Colangelo:2018mtw}%
  \BibitemOpen
  \bibfield  {author} {\bibinfo {author} {\bibfnamefont {G.}~\bibnamefont
  {Colangelo}}, \bibinfo {author} {\bibfnamefont {M.}~\bibnamefont
  {Hoferichter}}, \ and\ \bibinfo {author} {\bibfnamefont {P.}~\bibnamefont
  {Stoffer}},\ }\href {\doibase 10.1007/JHEP02(2019)006} {\bibfield  {journal}
  {\bibinfo  {journal} {JHEP}\ }\textbf {\bibinfo {volume} {02}},\ \bibinfo
  {pages} {006} (\bibinfo {year} {2019})},\ \Eprint
  {http://arxiv.org/abs/1810.00007} {arXiv:1810.00007 [hep-ph]} \BibitemShut
  {NoStop}%
\bibitem [{\citenamefont {Tanabashi}\ \emph {et~al.}(2018)\citenamefont
  {Tanabashi} \emph {et~al.}}]{Tanabashi:2018oca}%
  \BibitemOpen
  \bibfield  {author} {\bibinfo {author} {\bibfnamefont {M.}~\bibnamefont
  {Tanabashi}} \emph {et~al.} (\bibinfo {collaboration} {Particle Data
  Group}),\ }\href {\doibase 10.1103/PhysRevD.98.030001} {\bibfield  {journal}
  {\bibinfo  {journal} {Phys. Rev. D}\ }\textbf {\bibinfo {volume} {98}},\
  \bibinfo {pages} {030001} (\bibinfo {year} {2018})}\BibitemShut {NoStop}%
\bibitem [{\citenamefont {Frazer}\ and\ \citenamefont
  {Fulco}(1960)}]{Frazer:1960zzb}%
  \BibitemOpen
  \bibfield  {author} {\bibinfo {author} {\bibfnamefont {W.~R.}\ \bibnamefont
  {Frazer}}\ and\ \bibinfo {author} {\bibfnamefont {J.~R.}\ \bibnamefont
  {Fulco}},\ }\href {\doibase 10.1103/PhysRev.117.1609} {\bibfield  {journal}
  {\bibinfo  {journal} {Phys. Rev.}\ }\textbf {\bibinfo {volume} {117}},\
  \bibinfo {pages} {1609} (\bibinfo {year} {1960})}\BibitemShut {NoStop}%
\bibitem [{\citenamefont {Holladay}(1956)}]{Holladay:1956zz}%
  \BibitemOpen
  \bibfield  {author} {\bibinfo {author} {\bibfnamefont {W.}~\bibnamefont
  {Holladay}},\ }\href {\doibase 10.1103/PhysRev.101.1198} {\bibfield
  {journal} {\bibinfo  {journal} {Phys. Rev.}\ }\textbf {\bibinfo {volume}
  {101}},\ \bibinfo {pages} {1198} (\bibinfo {year} {1956})}\BibitemShut
  {NoStop}%
\bibitem [{\citenamefont {Gasser}\ and\ \citenamefont
  {Leutwyler}(1985)}]{Gasser:1984ux}%
  \BibitemOpen
  \bibfield  {author} {\bibinfo {author} {\bibfnamefont {J.}~\bibnamefont
  {Gasser}}\ and\ \bibinfo {author} {\bibfnamefont {H.}~\bibnamefont
  {Leutwyler}},\ }\href {\doibase 10.1016/0550-3213(85)90493-6} {\bibfield
  {journal} {\bibinfo  {journal} {Nucl. Phys. B}\ }\textbf {\bibinfo {volume}
  {250}},\ \bibinfo {pages} {517} (\bibinfo {year} {1985})}\BibitemShut
  {NoStop}%
\bibitem [{\citenamefont {Bijnens}\ \emph {et~al.}(1998)\citenamefont
  {Bijnens}, \citenamefont {Colangelo},\ and\ \citenamefont
  {Talavera}}]{Bijnens:1998fm}%
  \BibitemOpen
  \bibfield  {author} {\bibinfo {author} {\bibfnamefont {J.}~\bibnamefont
  {Bijnens}}, \bibinfo {author} {\bibfnamefont {G.}~\bibnamefont {Colangelo}},
  \ and\ \bibinfo {author} {\bibfnamefont {P.}~\bibnamefont {Talavera}},\
  }\href {\doibase 10.1088/1126-6708/1998/05/014} {\bibfield  {journal}
  {\bibinfo  {journal} {JHEP}\ }\textbf {\bibinfo {volume} {05}},\ \bibinfo
  {pages} {014} (\bibinfo {year} {1998})},\ \Eprint
  {http://arxiv.org/abs/hep-ph/9805389} {arXiv:hep-ph/9805389} \BibitemShut
  {NoStop}%
\bibitem [{\citenamefont {Martinelli}\ and\ \citenamefont
  {Sachrajda}(1988)}]{Martinelli:1987bh}%
  \BibitemOpen
  \bibfield  {author} {\bibinfo {author} {\bibfnamefont {G.}~\bibnamefont
  {Martinelli}}\ and\ \bibinfo {author} {\bibfnamefont {C.~T.}\ \bibnamefont
  {Sachrajda}},\ }\href {\doibase 10.1016/0550-3213(88)90445-2} {\bibfield
  {journal} {\bibinfo  {journal} {Nucl. Phys. B}\ }\textbf {\bibinfo {volume}
  {306}},\ \bibinfo {pages} {865} (\bibinfo {year} {1988})}\BibitemShut
  {NoStop}%
\bibitem [{\citenamefont {Draper}\ \emph {et~al.}(1989)\citenamefont {Draper},
  \citenamefont {Woloshyn}, \citenamefont {Wilcox},\ and\ \citenamefont
  {Liu}}]{Draper:1988bp}%
  \BibitemOpen
  \bibfield  {author} {\bibinfo {author} {\bibfnamefont {T.}~\bibnamefont
  {Draper}}, \bibinfo {author} {\bibfnamefont {R.}~\bibnamefont {Woloshyn}},
  \bibinfo {author} {\bibfnamefont {W.}~\bibnamefont {Wilcox}}, \ and\ \bibinfo
  {author} {\bibfnamefont {K.-F.}\ \bibnamefont {Liu}},\ }\href {\doibase
  10.1016/0550-3213(89)90609-3} {\bibfield  {journal} {\bibinfo  {journal}
  {Nucl. Phys. B}\ }\textbf {\bibinfo {volume} {318}},\ \bibinfo {pages} {319}
  (\bibinfo {year} {1989})}\BibitemShut {NoStop}%
\bibitem [{\citenamefont {Br{\"o}mmel}\ \emph {et~al.}(2007)\citenamefont
  {Br{\"o}mmel} \emph {et~al.}}]{Brommel:2006ww}%
  \BibitemOpen
  \bibfield  {author} {\bibinfo {author} {\bibfnamefont {D.}~\bibnamefont
  {Br{\"o}mmel}} \emph {et~al.} (\bibinfo {collaboration} {QCDSF/UKQCD}),\
  }\href {\doibase 10.1140/epjc/s10052-007-0295-6} {\bibfield  {journal}
  {\bibinfo  {journal} {Eur. Phys. J. C}\ }\textbf {\bibinfo {volume} {51}},\
  \bibinfo {pages} {335} (\bibinfo {year} {2007})},\ \Eprint
  {http://arxiv.org/abs/hep-lat/0608021} {arXiv:hep-lat/0608021} \BibitemShut
  {NoStop}%
\bibitem [{\citenamefont {Frezzotti}\ \emph {et~al.}(2009)\citenamefont
  {Frezzotti}, \citenamefont {Lubicz},\ and\ \citenamefont
  {Simula}}]{Frezzotti:2008dr}%
  \BibitemOpen
  \bibfield  {author} {\bibinfo {author} {\bibfnamefont {R.}~\bibnamefont
  {Frezzotti}}, \bibinfo {author} {\bibfnamefont {V.}~\bibnamefont {Lubicz}}, \
  and\ \bibinfo {author} {\bibfnamefont {S.}~\bibnamefont {Simula}} (\bibinfo
  {collaboration} {ETM}),\ }\href {\doibase 10.1103/PhysRevD.79.074506}
  {\bibfield  {journal} {\bibinfo  {journal} {Phys. Rev. D}\ }\textbf {\bibinfo
  {volume} {79}},\ \bibinfo {pages} {074506} (\bibinfo {year} {2009})},\
  \Eprint {http://arxiv.org/abs/0812.4042} {arXiv:0812.4042 [hep-lat]}
  \BibitemShut {NoStop}%
\bibitem [{\citenamefont {Aoki}\ \emph {et~al.}(2009)\citenamefont {Aoki} \emph
  {et~al.}}]{Aoki:2009qn}%
  \BibitemOpen
  \bibfield  {author} {\bibinfo {author} {\bibfnamefont {S.}~\bibnamefont
  {Aoki}} \emph {et~al.} (\bibinfo {collaboration} {JLQCD, TWQCD}),\ }\href
  {\doibase 10.1103/PhysRevD.80.034508} {\bibfield  {journal} {\bibinfo
  {journal} {Phys. Rev. D}\ }\textbf {\bibinfo {volume} {80}},\ \bibinfo
  {pages} {034508} (\bibinfo {year} {2009})},\ \Eprint
  {http://arxiv.org/abs/0905.2465} {arXiv:0905.2465 [hep-lat]} \BibitemShut
  {NoStop}%
\bibitem [{\citenamefont {Brandt}\ \emph {et~al.}(2013)\citenamefont {Brandt},
  \citenamefont {J{\"u}ttner},\ and\ \citenamefont {Wittig}}]{Brandt:2013dua}%
  \BibitemOpen
  \bibfield  {author} {\bibinfo {author} {\bibfnamefont {B.~B.}\ \bibnamefont
  {Brandt}}, \bibinfo {author} {\bibfnamefont {A.}~\bibnamefont {J{\"u}ttner}},
  \ and\ \bibinfo {author} {\bibfnamefont {H.}~\bibnamefont {Wittig}},\ }\href
  {\doibase 10.1007/JHEP11(2013)034} {\bibfield  {journal} {\bibinfo  {journal}
  {JHEP}\ }\textbf {\bibinfo {volume} {11}},\ \bibinfo {pages} {034} (\bibinfo
  {year} {2013})},\ \Eprint {http://arxiv.org/abs/1306.2916} {arXiv:1306.2916
  [hep-lat]} \BibitemShut {NoStop}%
\bibitem [{\citenamefont {Alexandrou}\ \emph {et~al.}(2018)\citenamefont
  {Alexandrou} \emph {et~al.}}]{Alexandrou:2017blh}%
  \BibitemOpen
  \bibfield  {author} {\bibinfo {author} {\bibfnamefont {C.}~\bibnamefont
  {Alexandrou}} \emph {et~al.} (\bibinfo {collaboration} {ETM}),\ }\href
  {\doibase 10.1103/PhysRevD.97.014508} {\bibfield  {journal} {\bibinfo
  {journal} {Phys. Rev. D}\ }\textbf {\bibinfo {volume} {97}},\ \bibinfo
  {pages} {014508} (\bibinfo {year} {2018})},\ \Eprint
  {http://arxiv.org/abs/1710.10401} {arXiv:1710.10401 [hep-lat]} \BibitemShut
  {NoStop}%
\bibitem [{\citenamefont {Bonnet}\ \emph {et~al.}(2005)\citenamefont {Bonnet},
  \citenamefont {Edwards}, \citenamefont {Fleming}, \citenamefont {Lewis},\
  and\ \citenamefont {Richards}}]{Bonnet:2004fr}%
  \BibitemOpen
  \bibfield  {author} {\bibinfo {author} {\bibfnamefont {F.~D.}\ \bibnamefont
  {Bonnet}}, \bibinfo {author} {\bibfnamefont {R.~G.}\ \bibnamefont {Edwards}},
  \bibinfo {author} {\bibfnamefont {G.~T.}\ \bibnamefont {Fleming}}, \bibinfo
  {author} {\bibfnamefont {R.}~\bibnamefont {Lewis}}, \ and\ \bibinfo {author}
  {\bibfnamefont {D.~G.}\ \bibnamefont {Richards}} (\bibinfo {collaboration}
  {Lattice Hadron Physics}),\ }\href {\doibase 10.1103/PhysRevD.72.054506}
  {\bibfield  {journal} {\bibinfo  {journal} {Phys. Rev. D}\ }\textbf {\bibinfo
  {volume} {72}},\ \bibinfo {pages} {054506} (\bibinfo {year} {2005})},\
  \Eprint {http://arxiv.org/abs/hep-lat/0411028} {arXiv:hep-lat/0411028}
  \BibitemShut {NoStop}%
\bibitem [{\citenamefont {Boyle}\ \emph {et~al.}(2008)\citenamefont {Boyle},
  \citenamefont {Flynn}, \citenamefont {J{\"u}ttner}, \citenamefont {Kelly},
  \citenamefont {de~Lima}, \citenamefont {Maynard}, \citenamefont {Sachrajda},\
  and\ \citenamefont {Zanotti}}]{Boyle:2008yd}%
  \BibitemOpen
  \bibfield  {author} {\bibinfo {author} {\bibfnamefont {P.}~\bibnamefont
  {Boyle}}, \bibinfo {author} {\bibfnamefont {J.}~\bibnamefont {Flynn}},
  \bibinfo {author} {\bibfnamefont {A.}~\bibnamefont {J{\"u}ttner}}, \bibinfo
  {author} {\bibfnamefont {C.}~\bibnamefont {Kelly}}, \bibinfo {author}
  {\bibfnamefont {H.}~\bibnamefont {de~Lima}}, \bibinfo {author} {\bibfnamefont
  {C.}~\bibnamefont {Maynard}}, \bibinfo {author} {\bibfnamefont
  {C.}~\bibnamefont {Sachrajda}}, \ and\ \bibinfo {author} {\bibfnamefont
  {J.}~\bibnamefont {Zanotti}},\ }\href {\doibase
  10.1088/1126-6708/2008/07/112} {\bibfield  {journal} {\bibinfo  {journal}
  {JHEP}\ }\textbf {\bibinfo {volume} {07}},\ \bibinfo {pages} {112} (\bibinfo
  {year} {2008})},\ \Eprint {http://arxiv.org/abs/0804.3971} {arXiv:0804.3971
  [hep-lat]} \BibitemShut {NoStop}%
\bibitem [{\citenamefont {Nguyen}\ \emph {et~al.}(2011)\citenamefont {Nguyen},
  \citenamefont {Ishikawa}, \citenamefont {Ukawa},\ and\ \citenamefont
  {Ukita}}]{Nguyen:2011ek}%
  \BibitemOpen
  \bibfield  {author} {\bibinfo {author} {\bibfnamefont {O.~H.}\ \bibnamefont
  {Nguyen}}, \bibinfo {author} {\bibfnamefont {K.-I.}\ \bibnamefont
  {Ishikawa}}, \bibinfo {author} {\bibfnamefont {A.}~\bibnamefont {Ukawa}}, \
  and\ \bibinfo {author} {\bibfnamefont {N.}~\bibnamefont {Ukita}},\ }\href
  {\doibase 10.1007/JHEP04(2011)122} {\bibfield  {journal} {\bibinfo  {journal}
  {JHEP}\ }\textbf {\bibinfo {volume} {04}},\ \bibinfo {pages} {122} (\bibinfo
  {year} {2011})},\ \Eprint {http://arxiv.org/abs/1102.3652} {arXiv:1102.3652
  [hep-lat]} \BibitemShut {NoStop}%
\bibitem [{\citenamefont {Fukaya}\ \emph {et~al.}(2014)\citenamefont {Fukaya},
  \citenamefont {Aoki}, \citenamefont {Hashimoto}, \citenamefont {Kaneko},
  \citenamefont {Matsufuru},\ and\ \citenamefont {Noaki}}]{Fukaya:2014jka}%
  \BibitemOpen
  \bibfield  {author} {\bibinfo {author} {\bibfnamefont {H.}~\bibnamefont
  {Fukaya}}, \bibinfo {author} {\bibfnamefont {S.}~\bibnamefont {Aoki}},
  \bibinfo {author} {\bibfnamefont {S.}~\bibnamefont {Hashimoto}}, \bibinfo
  {author} {\bibfnamefont {T.}~\bibnamefont {Kaneko}}, \bibinfo {author}
  {\bibfnamefont {H.}~\bibnamefont {Matsufuru}}, \ and\ \bibinfo {author}
  {\bibfnamefont {J.}~\bibnamefont {Noaki}},\ }\href {\doibase
  10.1103/PhysRevD.90.034506} {\bibfield  {journal} {\bibinfo  {journal} {Phys.
  Rev. D}\ }\textbf {\bibinfo {volume} {90}},\ \bibinfo {pages} {034506}
  (\bibinfo {year} {2014})},\ \Eprint {http://arxiv.org/abs/1405.4077}
  {arXiv:1405.4077 [hep-lat]} \BibitemShut {NoStop}%
\bibitem [{\citenamefont {Aoki}\ \emph {et~al.}(2016)\citenamefont {Aoki},
  \citenamefont {Cossu}, \citenamefont {Feng}, \citenamefont {Hashimoto},
  \citenamefont {Kaneko}, \citenamefont {Noaki},\ and\ \citenamefont
  {Onogi}}]{Aoki:2015pba}%
  \BibitemOpen
  \bibfield  {author} {\bibinfo {author} {\bibfnamefont {S.}~\bibnamefont
  {Aoki}}, \bibinfo {author} {\bibfnamefont {G.}~\bibnamefont {Cossu}},
  \bibinfo {author} {\bibfnamefont {X.}~\bibnamefont {Feng}}, \bibinfo {author}
  {\bibfnamefont {S.}~\bibnamefont {Hashimoto}}, \bibinfo {author}
  {\bibfnamefont {T.}~\bibnamefont {Kaneko}}, \bibinfo {author} {\bibfnamefont
  {J.}~\bibnamefont {Noaki}}, \ and\ \bibinfo {author} {\bibfnamefont
  {T.}~\bibnamefont {Onogi}} (\bibinfo {collaboration} {JLQCD}),\ }\href
  {\doibase 10.1103/PhysRevD.93.034504} {\bibfield  {journal} {\bibinfo
  {journal} {Phys. Rev. D}\ }\textbf {\bibinfo {volume} {93}},\ \bibinfo
  {pages} {034504} (\bibinfo {year} {2016})},\ \Eprint
  {http://arxiv.org/abs/1510.06470} {arXiv:1510.06470 [hep-lat]} \BibitemShut
  {NoStop}%
\bibitem [{\citenamefont {Feng}\ \emph {et~al.}(2020)\citenamefont {Feng},
  \citenamefont {Fu},\ and\ \citenamefont {Jin}}]{Feng:2019geu}%
  \BibitemOpen
  \bibfield  {author} {\bibinfo {author} {\bibfnamefont {X.}~\bibnamefont
  {Feng}}, \bibinfo {author} {\bibfnamefont {Y.}~\bibnamefont {Fu}}, \ and\
  \bibinfo {author} {\bibfnamefont {L.-C.}\ \bibnamefont {Jin}},\ }\href
  {\doibase 10.1103/PhysRevD.101.051502} {\bibfield  {journal} {\bibinfo
  {journal} {Phys. Rev. D}\ }\textbf {\bibinfo {volume} {101}},\ \bibinfo
  {pages} {051502} (\bibinfo {year} {2020})},\ \Eprint
  {http://arxiv.org/abs/1911.04064} {arXiv:1911.04064 [hep-lat]} \BibitemShut
  {NoStop}%
\bibitem [{\citenamefont {Koponen}\ \emph {et~al.}(2016)\citenamefont
  {Koponen}, \citenamefont {Bursa}, \citenamefont {Davies}, \citenamefont
  {Dowdall},\ and\ \citenamefont {Lepage}}]{Koponen:2015tkr}%
  \BibitemOpen
  \bibfield  {author} {\bibinfo {author} {\bibfnamefont {J.}~\bibnamefont
  {Koponen}}, \bibinfo {author} {\bibfnamefont {F.}~\bibnamefont {Bursa}},
  \bibinfo {author} {\bibfnamefont {C.}~\bibnamefont {Davies}}, \bibinfo
  {author} {\bibfnamefont {R.}~\bibnamefont {Dowdall}}, \ and\ \bibinfo
  {author} {\bibfnamefont {G.}~\bibnamefont {Lepage}},\ }\href {\doibase
  10.1103/PhysRevD.93.054503} {\bibfield  {journal} {\bibinfo  {journal} {Phys.
  Rev. D}\ }\textbf {\bibinfo {volume} {93}},\ \bibinfo {pages} {054503}
  (\bibinfo {year} {2016})},\ \Eprint {http://arxiv.org/abs/1511.07382}
  {arXiv:1511.07382 [hep-lat]} \BibitemShut {NoStop}%
\bibitem [{\citenamefont {Li}\ \emph {et~al.}(2010)\citenamefont {Li} \emph
  {et~al.}}]{Li:2010pw}%
  \BibitemOpen
  \bibfield  {author} {\bibinfo {author} {\bibfnamefont {A.}~\bibnamefont {Li}}
  \emph {et~al.} (\bibinfo {collaboration} {xQCD}),\ }\href {\doibase
  10.1103/PhysRevD.82.114501} {\bibfield  {journal} {\bibinfo  {journal} {Phys.
  Rev. D}\ }\textbf {\bibinfo {volume} {82}},\ \bibinfo {pages} {114501}
  (\bibinfo {year} {2010})},\ \Eprint {http://arxiv.org/abs/1005.5424}
  {arXiv:1005.5424 [hep-lat]} \BibitemShut {NoStop}%
\bibitem [{\citenamefont {Yang}\ \emph {et~al.}(2018)\citenamefont {Yang},
  \citenamefont {Liang}, \citenamefont {Bi}, \citenamefont {Chen},
  \citenamefont {Draper}, \citenamefont {Liu},\ and\ \citenamefont
  {Liu}}]{Yang:2018nqn}%
  \BibitemOpen
  \bibfield  {author} {\bibinfo {author} {\bibfnamefont {Y.-B.}\ \bibnamefont
  {Yang}}, \bibinfo {author} {\bibfnamefont {J.}~\bibnamefont {Liang}},
  \bibinfo {author} {\bibfnamefont {Y.-J.}\ \bibnamefont {Bi}}, \bibinfo
  {author} {\bibfnamefont {Y.}~\bibnamefont {Chen}}, \bibinfo {author}
  {\bibfnamefont {T.}~\bibnamefont {Draper}}, \bibinfo {author} {\bibfnamefont
  {K.-F.}\ \bibnamefont {Liu}}, \ and\ \bibinfo {author} {\bibfnamefont
  {Z.}~\bibnamefont {Liu}},\ }\href {\doibase 10.1103/PhysRevLett.121.212001}
  {\bibfield  {journal} {\bibinfo  {journal} {Phys. Rev. Lett.}\ }\textbf
  {\bibinfo {volume} {121}},\ \bibinfo {pages} {212001} (\bibinfo {year}
  {2018})},\ \Eprint {http://arxiv.org/abs/1808.08677} {arXiv:1808.08677
  [hep-lat]} \BibitemShut {NoStop}%
\bibitem [{\citenamefont {Sufian}\ \emph {et~al.}(2017)\citenamefont {Sufian},
  \citenamefont {Yang}, \citenamefont {Alexandru}, \citenamefont {Draper},
  \citenamefont {Liang},\ and\ \citenamefont {Liu}}]{Sufian:2016pex}%
  \BibitemOpen
  \bibfield  {author} {\bibinfo {author} {\bibfnamefont {R.~S.}\ \bibnamefont
  {Sufian}}, \bibinfo {author} {\bibfnamefont {Y.-B.}\ \bibnamefont {Yang}},
  \bibinfo {author} {\bibfnamefont {A.}~\bibnamefont {Alexandru}}, \bibinfo
  {author} {\bibfnamefont {T.}~\bibnamefont {Draper}}, \bibinfo {author}
  {\bibfnamefont {J.}~\bibnamefont {Liang}}, \ and\ \bibinfo {author}
  {\bibfnamefont {K.-F.}\ \bibnamefont {Liu}},\ }\href {\doibase
  10.1103/PhysRevLett.118.042001} {\bibfield  {journal} {\bibinfo  {journal}
  {Phys. Rev. Lett.}\ }\textbf {\bibinfo {volume} {118}},\ \bibinfo {pages}
  {042001} (\bibinfo {year} {2017})},\ \Eprint
  {http://arxiv.org/abs/1606.07075} {arXiv:1606.07075 [hep-ph]} \BibitemShut
  {NoStop}%
\bibitem [{\citenamefont {Yang}\ \emph {et~al.}(2017)\citenamefont {Yang},
  \citenamefont {Sufian}, \citenamefont {Alexandru}, \citenamefont {Draper},
  \citenamefont {Glatzmaier}, \citenamefont {Liu},\ and\ \citenamefont
  {Zhao}}]{Yang:2016plb}%
  \BibitemOpen
  \bibfield  {author} {\bibinfo {author} {\bibfnamefont {Y.-B.}\ \bibnamefont
  {Yang}}, \bibinfo {author} {\bibfnamefont {R.~S.}\ \bibnamefont {Sufian}},
  \bibinfo {author} {\bibfnamefont {A.}~\bibnamefont {Alexandru}}, \bibinfo
  {author} {\bibfnamefont {T.}~\bibnamefont {Draper}}, \bibinfo {author}
  {\bibfnamefont {M.~J.}\ \bibnamefont {Glatzmaier}}, \bibinfo {author}
  {\bibfnamefont {K.-F.}\ \bibnamefont {Liu}}, \ and\ \bibinfo {author}
  {\bibfnamefont {Y.}~\bibnamefont {Zhao}},\ }\href {\doibase
  10.1103/PhysRevLett.118.102001} {\bibfield  {journal} {\bibinfo  {journal}
  {Phys. Rev. Lett.}\ }\textbf {\bibinfo {volume} {118}},\ \bibinfo {pages}
  {102001} (\bibinfo {year} {2017})},\ \Eprint
  {http://arxiv.org/abs/1609.05937} {arXiv:1609.05937 [hep-ph]} \BibitemShut
  {NoStop}%
\bibitem [{\citenamefont {Cooley}\ and\ \citenamefont
  {Tukey}(1965)}]{cooleyAlgorithmMachineCalculation1965}%
  \BibitemOpen
  \bibfield  {author} {\bibinfo {author} {\bibfnamefont {J.~W.}\ \bibnamefont
  {Cooley}}\ and\ \bibinfo {author} {\bibfnamefont {J.~W.}\ \bibnamefont
  {Tukey}},\ }\href {\doibase 10.1090/S0025-5718-1965-0178586-1} {\bibfield
  {journal} {\bibinfo  {journal} {Mathematics of Computation}\ }\textbf
  {\bibinfo {volume} {19}},\ \bibinfo {pages} {297} (\bibinfo {year}
  {1965})}\BibitemShut {NoStop}%
\bibitem [{\citenamefont {Wang}(2020)}]{Wang:2020yqv}%
  \BibitemOpen
  \bibfield  {author} {\bibinfo {author} {\bibfnamefont {G.}~\bibnamefont
  {Wang}},\ }\emph {\bibinfo {title} {{The Pion Form Factor and Momentum and
  Angular Momentum Fractions of the Proton in Lattice QCD}}},\ \href {\doibase
  10.13023/etd.2020.406} {Ph.D. thesis},\ \bibinfo  {school} {Kentucky U.}
  (\bibinfo {year} {2020})\BibitemShut {NoStop}%
\bibitem [{\citenamefont {Aoki}\ \emph {et~al.}(2011)\citenamefont {Aoki} \emph
  {et~al.}}]{Aoki:2010dy}%
  \BibitemOpen
  \bibfield  {author} {\bibinfo {author} {\bibfnamefont {Y.}~\bibnamefont
  {Aoki}} \emph {et~al.} (\bibinfo {collaboration} {RBC, UKQCD}),\ }\href
  {\doibase 10.1103/PhysRevD.83.074508} {\bibfield  {journal} {\bibinfo
  {journal} {Phys. Rev. D}\ }\textbf {\bibinfo {volume} {83}},\ \bibinfo
  {pages} {074508} (\bibinfo {year} {2011})},\ \Eprint
  {http://arxiv.org/abs/1011.0892} {arXiv:1011.0892 [hep-lat]} \BibitemShut
  {NoStop}%
\bibitem [{\citenamefont {Blum}\ \emph {et~al.}(2016)\citenamefont {Blum} \emph
  {et~al.}}]{Blum:2014tka}%
  \BibitemOpen
  \bibfield  {author} {\bibinfo {author} {\bibfnamefont {T.}~\bibnamefont
  {Blum}} \emph {et~al.} (\bibinfo {collaboration} {RBC, UKQCD}),\ }\href
  {\doibase 10.1103/PhysRevD.93.074505} {\bibfield  {journal} {\bibinfo
  {journal} {Phys. Rev. D}\ }\textbf {\bibinfo {volume} {93}},\ \bibinfo
  {pages} {074505} (\bibinfo {year} {2016})},\ \Eprint
  {http://arxiv.org/abs/1411.7017} {arXiv:1411.7017 [hep-lat]} \BibitemShut
  {NoStop}%
\bibitem [{\citenamefont {Arthur}\ \emph {et~al.}(2013)\citenamefont {Arthur}
  \emph {et~al.}}]{Arthur:2012opa}%
  \BibitemOpen
  \bibfield  {author} {\bibinfo {author} {\bibfnamefont {R.}~\bibnamefont
  {Arthur}} \emph {et~al.} (\bibinfo {collaboration} {RBC, UKQCD}),\ }\href
  {\doibase 10.1103/PhysRevD.87.094514} {\bibfield  {journal} {\bibinfo
  {journal} {Phys. Rev. D}\ }\textbf {\bibinfo {volume} {87}},\ \bibinfo
  {pages} {094514} (\bibinfo {year} {2013})},\ \Eprint
  {http://arxiv.org/abs/1208.4412} {arXiv:1208.4412 [hep-lat]} \BibitemShut
  {NoStop}%
\bibitem [{\citenamefont {Boyle}\ \emph {et~al.}(2016)\citenamefont {Boyle}
  \emph {et~al.}}]{Boyle:2015exm}%
  \BibitemOpen
  \bibfield  {author} {\bibinfo {author} {\bibfnamefont {P.}~\bibnamefont
  {Boyle}} \emph {et~al.},\ }\href {\doibase 10.1103/PhysRevD.93.054502}
  {\bibfield  {journal} {\bibinfo  {journal} {Phys. Rev. D}\ }\textbf {\bibinfo
  {volume} {93}},\ \bibinfo {pages} {054502} (\bibinfo {year} {2016})},\
  \Eprint {http://arxiv.org/abs/1511.01950} {arXiv:1511.01950 [hep-lat]}
  \BibitemShut {NoStop}%
\bibitem [{\citenamefont {Chiu}(1999)}]{Chiu:1998eu}%
  \BibitemOpen
  \bibfield  {author} {\bibinfo {author} {\bibfnamefont {T.-W.}\ \bibnamefont
  {Chiu}},\ }\href {\doibase 10.1103/PhysRevD.60.034503} {\bibfield  {journal}
  {\bibinfo  {journal} {Phys. Rev. D}\ }\textbf {\bibinfo {volume} {60}},\
  \bibinfo {pages} {034503} (\bibinfo {year} {1999})},\ \Eprint
  {http://arxiv.org/abs/hep-lat/9810052} {arXiv:hep-lat/9810052} \BibitemShut
  {NoStop}%
\bibitem [{\citenamefont {Liu}(2005)}]{Liu:2002qu}%
  \BibitemOpen
  \bibfield  {author} {\bibinfo {author} {\bibfnamefont {K.-F.}\ \bibnamefont
  {Liu}},\ }\href {\doibase 10.1142/S0217751X05022366} {\bibfield  {journal}
  {\bibinfo  {journal} {Int. J. Mod. Phys. A}\ }\textbf {\bibinfo {volume}
  {20}},\ \bibinfo {pages} {7241} (\bibinfo {year} {2005})},\ \Eprint
  {http://arxiv.org/abs/hep-lat/0206002} {arXiv:hep-lat/0206002} \BibitemShut
  {NoStop}%
\bibitem [{\citenamefont {Chiu}\ and\ \citenamefont
  {Zenkin}(1999)}]{Chiu:1998gp}%
  \BibitemOpen
  \bibfield  {author} {\bibinfo {author} {\bibfnamefont {T.-W.}\ \bibnamefont
  {Chiu}}\ and\ \bibinfo {author} {\bibfnamefont {S.~V.}\ \bibnamefont
  {Zenkin}},\ }\href {\doibase 10.1103/PhysRevD.59.074501} {\bibfield
  {journal} {\bibinfo  {journal} {Phys. Rev. D}\ }\textbf {\bibinfo {volume}
  {59}},\ \bibinfo {pages} {074501} (\bibinfo {year} {1999})},\ \Eprint
  {http://arxiv.org/abs/hep-lat/9806019} {arXiv:hep-lat/9806019} \BibitemShut
  {NoStop}%
\bibitem [{\citenamefont {DeGrand}\ and\ \citenamefont
  {Loft}(1991)}]{DeGrand:1990dz}%
  \BibitemOpen
  \bibfield  {author} {\bibinfo {author} {\bibfnamefont {T.~A.}\ \bibnamefont
  {DeGrand}}\ and\ \bibinfo {author} {\bibfnamefont {R.~D.}\ \bibnamefont
  {Loft}},\ }\href {\doibase 10.1016/0010-4655(91)90158-H} {\bibfield
  {journal} {\bibinfo  {journal} {Comput. Phys. Commun.}\ }\textbf {\bibinfo
  {volume} {65}},\ \bibinfo {pages} {84} (\bibinfo {year} {1991})}\BibitemShut
  {NoStop}%
\bibitem [{\citenamefont {Allton}\ \emph {et~al.}(1991)\citenamefont {Allton},
  \citenamefont {Sachrajda}, \citenamefont {Lubicz}, \citenamefont {Maiani},\
  and\ \citenamefont {Martinelli}}]{Allton:1990qg}%
  \BibitemOpen
  \bibfield  {author} {\bibinfo {author} {\bibfnamefont {C.}~\bibnamefont
  {Allton}}, \bibinfo {author} {\bibfnamefont {C.~T.}\ \bibnamefont
  {Sachrajda}}, \bibinfo {author} {\bibfnamefont {V.}~\bibnamefont {Lubicz}},
  \bibinfo {author} {\bibfnamefont {L.}~\bibnamefont {Maiani}}, \ and\ \bibinfo
  {author} {\bibfnamefont {G.}~\bibnamefont {Martinelli}},\ }\href {\doibase
  10.1016/0550-3213(91)90337-W} {\bibfield  {journal} {\bibinfo  {journal}
  {Nucl. Phys. B}\ }\textbf {\bibinfo {volume} {349}},\ \bibinfo {pages} {598}
  (\bibinfo {year} {1991})}\BibitemShut {NoStop}%
\bibitem [{\citenamefont {Liang}\ \emph {et~al.}(2017)\citenamefont {Liang},
  \citenamefont {Yang}, \citenamefont {Liu}, \citenamefont {Alexandru},
  \citenamefont {Draper},\ and\ \citenamefont {Sufian}}]{Liang:2016fgy}%
  \BibitemOpen
  \bibfield  {author} {\bibinfo {author} {\bibfnamefont {J.}~\bibnamefont
  {Liang}}, \bibinfo {author} {\bibfnamefont {Y.-B.}\ \bibnamefont {Yang}},
  \bibinfo {author} {\bibfnamefont {K.-F.}\ \bibnamefont {Liu}}, \bibinfo
  {author} {\bibfnamefont {A.}~\bibnamefont {Alexandru}}, \bibinfo {author}
  {\bibfnamefont {T.}~\bibnamefont {Draper}}, \ and\ \bibinfo {author}
  {\bibfnamefont {R.~S.}\ \bibnamefont {Sufian}},\ }\href {\doibase
  10.1103/PhysRevD.96.034519} {\bibfield  {journal} {\bibinfo  {journal} {Phys.
  Rev. D}\ }\textbf {\bibinfo {volume} {96}},\ \bibinfo {pages} {034519}
  (\bibinfo {year} {2017})},\ \Eprint {http://arxiv.org/abs/1612.04388}
  {arXiv:1612.04388 [hep-lat]} \BibitemShut {NoStop}%
\bibitem [{\citenamefont {Dong}\ and\ \citenamefont {Liu}(1994)}]{Dong:1993pk}%
  \BibitemOpen
  \bibfield  {author} {\bibinfo {author} {\bibfnamefont {S.-J.}\ \bibnamefont
  {Dong}}\ and\ \bibinfo {author} {\bibfnamefont {K.-F.}\ \bibnamefont {Liu}},\
  }\href {\doibase 10.1016/0370-2693(94)90440-5} {\bibfield  {journal}
  {\bibinfo  {journal} {Phys. Lett. B}\ }\textbf {\bibinfo {volume} {328}},\
  \bibinfo {pages} {130} (\bibinfo {year} {1994})},\ \Eprint
  {http://arxiv.org/abs/hep-lat/9308015} {arXiv:hep-lat/9308015} \BibitemShut
  {NoStop}%
\bibitem [{\citenamefont {Bernard}\ \emph {et~al.}(1985)\citenamefont
  {Bernard}, \citenamefont {Draper}, \citenamefont {Hockney}, \citenamefont
  {Rushton},\ and\ \citenamefont {Soni}}]{Bernard:1985tm}%
  \BibitemOpen
  \bibfield  {author} {\bibinfo {author} {\bibfnamefont {C.~W.}\ \bibnamefont
  {Bernard}}, \bibinfo {author} {\bibfnamefont {T.}~\bibnamefont {Draper}},
  \bibinfo {author} {\bibfnamefont {G.}~\bibnamefont {Hockney}}, \bibinfo
  {author} {\bibfnamefont {A.}~\bibnamefont {Rushton}}, \ and\ \bibinfo
  {author} {\bibfnamefont {A.}~\bibnamefont {Soni}},\ }\href {\doibase
  10.1103/PhysRevLett.55.2770} {\bibfield  {journal} {\bibinfo  {journal}
  {Phys. Rev. Lett.}\ }\textbf {\bibinfo {volume} {55}},\ \bibinfo {pages}
  {2770} (\bibinfo {year} {1985})}\BibitemShut {NoStop}%
\bibitem [{\citenamefont {Martinelli}\ and\ \citenamefont
  {Sachrajda}(1989)}]{Martinelli:1988rr}%
  \BibitemOpen
  \bibfield  {author} {\bibinfo {author} {\bibfnamefont {G.}~\bibnamefont
  {Martinelli}}\ and\ \bibinfo {author} {\bibfnamefont {C.~T.}\ \bibnamefont
  {Sachrajda}},\ }\href {\doibase 10.1016/0550-3213(89)90035-7} {\bibfield
  {journal} {\bibinfo  {journal} {Nucl. Phys. B}\ }\textbf {\bibinfo {volume}
  {316}},\ \bibinfo {pages} {355} (\bibinfo {year} {1989})}\BibitemShut
  {NoStop}%
\bibitem [{\citenamefont {Yang}\ \emph {et~al.}(2016)\citenamefont {Yang},
  \citenamefont {Alexandru}, \citenamefont {Draper}, \citenamefont {Gong},\
  and\ \citenamefont {Liu}}]{Yang:2015zja}%
  \BibitemOpen
  \bibfield  {author} {\bibinfo {author} {\bibfnamefont {Y.-B.}\ \bibnamefont
  {Yang}}, \bibinfo {author} {\bibfnamefont {A.}~\bibnamefont {Alexandru}},
  \bibinfo {author} {\bibfnamefont {T.}~\bibnamefont {Draper}}, \bibinfo
  {author} {\bibfnamefont {M.}~\bibnamefont {Gong}}, \ and\ \bibinfo {author}
  {\bibfnamefont {K.-F.}\ \bibnamefont {Liu}},\ }\href {\doibase
  10.1103/PhysRevD.93.034503} {\bibfield  {journal} {\bibinfo  {journal} {Phys.
  Rev. D}\ }\textbf {\bibinfo {volume} {93}},\ \bibinfo {pages} {034503}
  (\bibinfo {year} {2016})},\ \Eprint {http://arxiv.org/abs/1509.04616}
  {arXiv:1509.04616 [hep-lat]} \BibitemShut {NoStop}%
\bibitem [{\citenamefont {Liang}\ \emph {et~al.}(2018)\citenamefont {Liang},
  \citenamefont {Yang}, \citenamefont {Draper}, \citenamefont {Gong},\ and\
  \citenamefont {Liu}}]{Liang:2018pis}%
  \BibitemOpen
  \bibfield  {author} {\bibinfo {author} {\bibfnamefont {J.}~\bibnamefont
  {Liang}}, \bibinfo {author} {\bibfnamefont {Y.-B.}\ \bibnamefont {Yang}},
  \bibinfo {author} {\bibfnamefont {T.}~\bibnamefont {Draper}}, \bibinfo
  {author} {\bibfnamefont {M.}~\bibnamefont {Gong}}, \ and\ \bibinfo {author}
  {\bibfnamefont {K.-F.}\ \bibnamefont {Liu}},\ }\href {\doibase
  10.1103/PhysRevD.98.074505} {\bibfield  {journal} {\bibinfo  {journal} {Phys.
  Rev. D}\ }\textbf {\bibinfo {volume} {98}},\ \bibinfo {pages} {074505}
  (\bibinfo {year} {2018})},\ \Eprint {http://arxiv.org/abs/1806.08366}
  {arXiv:1806.08366 [hep-ph]} \BibitemShut {NoStop}%
\bibitem [{\citenamefont {Lee}\ \emph {et~al.}(2015)\citenamefont {Lee},
  \citenamefont {Arrington},\ and\ \citenamefont {Hill}}]{Lee:2015jqa}%
  \BibitemOpen
  \bibfield  {author} {\bibinfo {author} {\bibfnamefont {G.}~\bibnamefont
  {Lee}}, \bibinfo {author} {\bibfnamefont {J.~R.}\ \bibnamefont {Arrington}},
  \ and\ \bibinfo {author} {\bibfnamefont {R.~J.}\ \bibnamefont {Hill}},\
  }\href {\doibase 10.1103/PhysRevD.92.013013} {\bibfield  {journal} {\bibinfo
  {journal} {Phys. Rev. D}\ }\textbf {\bibinfo {volume} {92}},\ \bibinfo
  {pages} {013013} (\bibinfo {year} {2015})},\ \Eprint
  {http://arxiv.org/abs/1505.01489} {arXiv:1505.01489 [hep-ph]} \BibitemShut
  {NoStop}%
\bibitem [{\citenamefont {Lujan}\ \emph {et~al.}(2012)\citenamefont {Lujan},
  \citenamefont {Alexandru}, \citenamefont {Chen}, \citenamefont {Draper},
  \citenamefont {Freeman}, \citenamefont {Gong}, \citenamefont {Lee},
  \citenamefont {Li}, \citenamefont {Liu},\ and\ \citenamefont
  {Mathur}}]{Lujan:2012wg}%
  \BibitemOpen
  \bibfield  {author} {\bibinfo {author} {\bibfnamefont {M.}~\bibnamefont
  {Lujan}}, \bibinfo {author} {\bibfnamefont {A.}~\bibnamefont {Alexandru}},
  \bibinfo {author} {\bibfnamefont {Y.}~\bibnamefont {Chen}}, \bibinfo {author}
  {\bibfnamefont {T.}~\bibnamefont {Draper}}, \bibinfo {author} {\bibfnamefont
  {W.}~\bibnamefont {Freeman}}, \bibinfo {author} {\bibfnamefont
  {M.}~\bibnamefont {Gong}}, \bibinfo {author} {\bibfnamefont {F.}~\bibnamefont
  {Lee}}, \bibinfo {author} {\bibfnamefont {A.}~\bibnamefont {Li}}, \bibinfo
  {author} {\bibfnamefont {K.}~\bibnamefont {Liu}}, \ and\ \bibinfo {author}
  {\bibfnamefont {N.}~\bibnamefont {Mathur}},\ }\href {\doibase
  10.1103/PhysRevD.86.014501} {\bibfield  {journal} {\bibinfo  {journal} {Phys.
  Rev. D}\ }\textbf {\bibinfo {volume} {86}},\ \bibinfo {pages} {014501}
  (\bibinfo {year} {2012})},\ \Eprint {http://arxiv.org/abs/1204.6256}
  {arXiv:1204.6256 [hep-lat]} \BibitemShut {NoStop}%
\bibitem [{\citenamefont {Wang}\ \emph {et~al.}(2021)\citenamefont {Wang},
  \citenamefont {Jin}, \citenamefont {Yang},\ and\ \citenamefont
  {Zhao}}]{Deltamix_paper}%
  \BibitemOpen
  \bibfield  {author} {\bibinfo {author} {\bibfnamefont {G.}~\bibnamefont
  {Wang}}, \bibinfo {author} {\bibfnamefont {L.-C.~J.}\ \bibnamefont {Jin}},
  \bibinfo {author} {\bibfnamefont {Y.-B.}\ \bibnamefont {Yang}}, \ and\
  \bibinfo {author} {\bibfnamefont {D.-J.}\ \bibnamefont {Zhao}},\ }\href@noop
  {} {\bibfield  {journal} {\bibinfo  {journal} {{in preparation}}\ } (\bibinfo
  {year} {2021})}\BibitemShut {NoStop}%
\bibitem [{\citenamefont {Lepage}\ and\ \citenamefont
  {Brodsky}(1979)}]{Lepage:1979zb}%
  \BibitemOpen
  \bibfield  {author} {\bibinfo {author} {\bibfnamefont {G.}~\bibnamefont
  {Lepage}}\ and\ \bibinfo {author} {\bibfnamefont {S.~J.}\ \bibnamefont
  {Brodsky}},\ }\href {\doibase 10.1016/0370-2693(79)90554-9} {\bibfield
  {journal} {\bibinfo  {journal} {Phys. Lett. B}\ }\textbf {\bibinfo {volume}
  {87}},\ \bibinfo {pages} {359} (\bibinfo {year} {1979})}\BibitemShut
  {NoStop}%
\bibitem [{\citenamefont {Farrar}\ and\ \citenamefont
  {Jackson}(1979)}]{Farrar:1979aw}%
  \BibitemOpen
  \bibfield  {author} {\bibinfo {author} {\bibfnamefont {G.~R.}\ \bibnamefont
  {Farrar}}\ and\ \bibinfo {author} {\bibfnamefont {D.~R.}\ \bibnamefont
  {Jackson}},\ }\href {\doibase 10.1103/PhysRevLett.43.246} {\bibfield
  {journal} {\bibinfo  {journal} {Phys. Rev. Lett.}\ }\textbf {\bibinfo
  {volume} {43}},\ \bibinfo {pages} {246} (\bibinfo {year} {1979})}\BibitemShut
  {NoStop}%
\bibitem [{\citenamefont {Arndt}\ and\ \citenamefont
  {Tiburzi}(2003)}]{Arndt:2003ww}%
  \BibitemOpen
  \bibfield  {author} {\bibinfo {author} {\bibfnamefont {D.}~\bibnamefont
  {Arndt}}\ and\ \bibinfo {author} {\bibfnamefont {B.~C.}\ \bibnamefont
  {Tiburzi}},\ }\href {\doibase 10.1103/PhysRevD.68.094501} {\bibfield
  {journal} {\bibinfo  {journal} {Phys. Rev. D}\ }\textbf {\bibinfo {volume}
  {68}},\ \bibinfo {pages} {094501} (\bibinfo {year} {2003})},\ \Eprint
  {http://arxiv.org/abs/hep-lat/0307003} {arXiv:hep-lat/0307003} \BibitemShut
  {NoStop}%
\bibitem [{\citenamefont {Golterman}\ and\ \citenamefont
  {Leung}(1998)}]{Golterman:1997st}%
  \BibitemOpen
  \bibfield  {author} {\bibinfo {author} {\bibfnamefont {M.~F.}\ \bibnamefont
  {Golterman}}\ and\ \bibinfo {author} {\bibfnamefont {K.-C.}\ \bibnamefont
  {Leung}},\ }\href {\doibase 10.1103/PhysRevD.57.5703} {\bibfield  {journal}
  {\bibinfo  {journal} {Phys. Rev. D}\ }\textbf {\bibinfo {volume} {57}},\
  \bibinfo {pages} {5703} (\bibinfo {year} {1998})},\ \Eprint
  {http://arxiv.org/abs/hep-lat/9711033} {arXiv:hep-lat/9711033} \BibitemShut
  {NoStop}%
\bibitem [{\citenamefont {Bunton}\ \emph {et~al.}(2006)\citenamefont {Bunton},
  \citenamefont {Jiang},\ and\ \citenamefont {Tiburzi}}]{Bunton:2006va}%
  \BibitemOpen
  \bibfield  {author} {\bibinfo {author} {\bibfnamefont {T.}~\bibnamefont
  {Bunton}}, \bibinfo {author} {\bibfnamefont {F.-J.}\ \bibnamefont {Jiang}}, \
  and\ \bibinfo {author} {\bibfnamefont {B.}~\bibnamefont {Tiburzi}},\ }\href
  {\doibase 10.1103/PhysRevD.74.099902} {\bibfield  {journal} {\bibinfo
  {journal} {Phys. Rev. D}\ }\textbf {\bibinfo {volume} {74}},\ \bibinfo
  {pages} {034514} (\bibinfo {year} {2006})},\ \bibinfo {note} {[Erratum:
  Phys.Rev.D 74, 099902 (2006)]},\ \Eprint
  {http://arxiv.org/abs/hep-lat/0607001} {arXiv:hep-lat/0607001} \BibitemShut
  {NoStop}%
\bibitem [{\citenamefont {Jiang}\ and\ \citenamefont
  {Tiburzi}(2007)}]{Jiang:2006gna}%
  \BibitemOpen
  \bibfield  {author} {\bibinfo {author} {\bibfnamefont {F.-J.}\ \bibnamefont
  {Jiang}}\ and\ \bibinfo {author} {\bibfnamefont {B.}~\bibnamefont
  {Tiburzi}},\ }\href {\doibase 10.1016/j.physletb.2006.12.041} {\bibfield
  {journal} {\bibinfo  {journal} {Phys. Lett. B}\ }\textbf {\bibinfo {volume}
  {645}},\ \bibinfo {pages} {314} (\bibinfo {year} {2007})},\ \Eprint
  {http://arxiv.org/abs/hep-lat/0610103} {arXiv:hep-lat/0610103} \BibitemShut
  {NoStop}%
\bibitem [{\citenamefont {Colangelo}\ and\ \citenamefont
  {Vaghi}()}]{Colangelo:2016wgs}%
  \BibitemOpen
  \bibfield  {author} {\bibinfo {author} {\bibfnamefont {G.}~\bibnamefont
  {Colangelo}}\ and\ \bibinfo {author} {\bibfnamefont {A.}~\bibnamefont
  {Vaghi}},\ }\href {\doibase 10.1007/JHEP07(2016)134} {\bibfield  {journal}
  {\bibinfo  {journal} {JHEP}\ }\textbf {\bibinfo {volume} {07}},\ \bibinfo
  {pages} {134}},\ \Eprint {http://arxiv.org/abs/1607.00916} {arXiv:1607.00916
  [hep-lat]} \BibitemShut {NoStop}%
\bibitem [{\citenamefont {Aoki}\ \emph {et~al.}(2020)\citenamefont {Aoki} \emph
  {et~al.}}]{Aoki:2019cca}%
  \BibitemOpen
  \bibfield  {author} {\bibinfo {author} {\bibfnamefont {S.}~\bibnamefont
  {Aoki}} \emph {et~al.} (\bibinfo {collaboration} {Flavour Lattice Averaging
  Group}),\ }\href {\doibase 10.1140/epjc/s10052-019-7354-7} {\bibfield
  {journal} {\bibinfo  {journal} {Eur. Phys. J. C}\ }\textbf {\bibinfo {volume}
  {80}},\ \bibinfo {pages} {113} (\bibinfo {year} {2020})},\ \Eprint
  {http://arxiv.org/abs/1902.08191} {arXiv:1902.08191 [hep-lat]} \BibitemShut
  {NoStop}%
\bibitem [{\citenamefont {Chen}\ \emph {et~al.}(2018)\citenamefont {Chen},
  \citenamefont {Ding}, \citenamefont {Chang},\ and\ \citenamefont
  {Roberts}}]{Chen:2018rwz}%
  \BibitemOpen
  \bibfield  {author} {\bibinfo {author} {\bibfnamefont {M.}~\bibnamefont
  {Chen}}, \bibinfo {author} {\bibfnamefont {M.}~\bibnamefont {Ding}}, \bibinfo
  {author} {\bibfnamefont {L.}~\bibnamefont {Chang}}, \ and\ \bibinfo {author}
  {\bibfnamefont {C.~D.}\ \bibnamefont {Roberts}},\ }\href {\doibase
  10.1103/PhysRevD.98.091505} {\bibfield  {journal} {\bibinfo  {journal} {Phys.
  Rev. D}\ }\textbf {\bibinfo {volume} {98}},\ \bibinfo {pages} {091505}
  (\bibinfo {year} {2018})},\ \Eprint {http://arxiv.org/abs/1808.09461}
  {arXiv:1808.09461 [nucl-th]} \BibitemShut {NoStop}%
\bibitem [{\citenamefont {Huber}\ \emph {et~al.}(2008)\citenamefont {Huber}
  \emph {et~al.}}]{Huber:2008id}%
  \BibitemOpen
  \bibfield  {author} {\bibinfo {author} {\bibfnamefont {G.}~\bibnamefont
  {Huber}} \emph {et~al.} (\bibinfo {collaboration} {Jefferson Lab}),\ }\href
  {\doibase 10.1103/PhysRevC.78.045203} {\bibfield  {journal} {\bibinfo
  {journal} {Phys. Rev. C}\ }\textbf {\bibinfo {volume} {78}},\ \bibinfo
  {pages} {045203} (\bibinfo {year} {2008})},\ \Eprint
  {http://arxiv.org/abs/0809.3052} {arXiv:0809.3052 [nucl-ex]} \BibitemShut
  {NoStop}%
\bibitem [{\citenamefont {Blok}\ \emph {et~al.}(2008)\citenamefont {Blok} \emph
  {et~al.}}]{Blok:2008jy}%
  \BibitemOpen
  \bibfield  {author} {\bibinfo {author} {\bibfnamefont {H.}~\bibnamefont
  {Blok}} \emph {et~al.} (\bibinfo {collaboration} {Jefferson Lab}),\ }\href
  {\doibase 10.1103/PhysRevC.78.045202} {\bibfield  {journal} {\bibinfo
  {journal} {Phys. Rev. C}\ }\textbf {\bibinfo {volume} {78}},\ \bibinfo
  {pages} {045202} (\bibinfo {year} {2008})},\ \Eprint
  {http://arxiv.org/abs/0809.3161} {arXiv:0809.3161 [nucl-ex]} \BibitemShut
  {NoStop}%
\bibitem [{\citenamefont {Horn}\ \emph {et~al.}(2008)\citenamefont {Horn} \emph
  {et~al.}}]{Horn:2007ug}%
  \BibitemOpen
  \bibfield  {author} {\bibinfo {author} {\bibfnamefont {T.}~\bibnamefont
  {Horn}} \emph {et~al.},\ }\href {\doibase 10.1103/PhysRevC.78.058201}
  {\bibfield  {journal} {\bibinfo  {journal} {Phys. Rev. C}\ }\textbf {\bibinfo
  {volume} {78}},\ \bibinfo {pages} {058201} (\bibinfo {year} {2008})},\
  \Eprint {http://arxiv.org/abs/0707.1794} {arXiv:0707.1794 [nucl-ex]}
  \BibitemShut {NoStop}%
\bibitem [{\citenamefont {Horn}\ \emph {et~al.}(2006)\citenamefont {Horn} \emph
  {et~al.}}]{Horn:2006tm}%
  \BibitemOpen
  \bibfield  {author} {\bibinfo {author} {\bibfnamefont {T.}~\bibnamefont
  {Horn}} \emph {et~al.} (\bibinfo {collaboration} {Jefferson Lab F(pi)-2}),\
  }\href {\doibase 10.1103/PhysRevLett.97.192001} {\bibfield  {journal}
  {\bibinfo  {journal} {Phys. Rev. Lett.}\ }\textbf {\bibinfo {volume} {97}},\
  \bibinfo {pages} {192001} (\bibinfo {year} {2006})},\ \Eprint
  {http://arxiv.org/abs/nucl-ex/0607005} {arXiv:nucl-ex/0607005} \BibitemShut
  {NoStop}%
\bibitem [{\citenamefont {Volmer}\ \emph {et~al.}(2001)\citenamefont {Volmer}
  \emph {et~al.}}]{Volmer:2000ek}%
  \BibitemOpen
  \bibfield  {author} {\bibinfo {author} {\bibfnamefont {J.}~\bibnamefont
  {Volmer}} \emph {et~al.} (\bibinfo {collaboration} {Jefferson Lab F(pi)}),\
  }\href {\doibase 10.1103/PhysRevLett.86.1713} {\bibfield  {journal} {\bibinfo
   {journal} {Phys. Rev. Lett.}\ }\textbf {\bibinfo {volume} {86}},\ \bibinfo
  {pages} {1713} (\bibinfo {year} {2001})},\ \Eprint
  {http://arxiv.org/abs/nucl-ex/0010009} {arXiv:nucl-ex/0010009} \BibitemShut
  {NoStop}%
\bibitem [{\citenamefont {Liu}\ \emph {et~al.}(2014)\citenamefont {Liu},
  \citenamefont {Chen}, \citenamefont {Dong}, \citenamefont {Glatzmaier},
  \citenamefont {Gong}, \citenamefont {Li}, \citenamefont {Liu}, \citenamefont
  {Yang},\ and\ \citenamefont {Zhang}}]{Liu:2013yxz}%
  \BibitemOpen
  \bibfield  {author} {\bibinfo {author} {\bibfnamefont {Z.}~\bibnamefont
  {Liu}}, \bibinfo {author} {\bibfnamefont {Y.}~\bibnamefont {Chen}}, \bibinfo
  {author} {\bibfnamefont {S.-J.}\ \bibnamefont {Dong}}, \bibinfo {author}
  {\bibfnamefont {M.}~\bibnamefont {Glatzmaier}}, \bibinfo {author}
  {\bibfnamefont {M.}~\bibnamefont {Gong}}, \bibinfo {author} {\bibfnamefont
  {A.}~\bibnamefont {Li}}, \bibinfo {author} {\bibfnamefont {K.-F.}\
  \bibnamefont {Liu}}, \bibinfo {author} {\bibfnamefont {Y.-B.}\ \bibnamefont
  {Yang}}, \ and\ \bibinfo {author} {\bibfnamefont {J.-B.}\ \bibnamefont
  {Zhang}} (\bibinfo {collaboration} {chiQCD}),\ }\href {\doibase
  10.1103/PhysRevD.90.034505} {\bibfield  {journal} {\bibinfo  {journal} {Phys.
  Rev. D}\ }\textbf {\bibinfo {volume} {90}},\ \bibinfo {pages} {034505}
  (\bibinfo {year} {2014})},\ \Eprint {http://arxiv.org/abs/1312.7628}
  {arXiv:1312.7628 [hep-lat]} \BibitemShut {NoStop}%
\end{thebibliography}%

\end{document}